\documentclass[aps,prb,onecolumn,nofootinbib,citeautoscript,10pt]{revtex4-2}

\usepackage{amsmath,amssymb} 
\usepackage{comment}

\usepackage[tight]{subfigure} 

\usepackage[dvipsnames]{xcolor} 
\usepackage[papersize={8.5in,11in}]{geometry}
\usepackage[colorlinks=true]{hyperref}
\hypersetup{
    bookmarks=true,         
    unicode=false,          
    pdftoolbar=true,        
    pdfmenubar=true,        
    pdffitwindow=false,     
    pdfstartview={FitH},    
    pdfkeywords={keyword1} {key2} {key3}, 
    pdfnewwindow=true,      
    colorlinks=true,       
    linkcolor=magenta, 
    citecolor=blue,        
    filecolor=magenta,      
    urlcolor=blue          
} 

\usepackage{dcolumn}
\usepackage{color}
\usepackage{amssymb,amsmath}
\usepackage{tabularx,graphicx}
\usepackage{epstopdf}
\usepackage{latexsym}
\usepackage{colortbl}
\usepackage{psfrag}
\usepackage{bbm,bm,array,physics}
\usepackage{dsfont}
\usepackage{float, mathrsfs}
\newcommand{\bs}[1]{\boldsymbol{#1}}

\def \nn{\nonumber \\}
\def\*#1{\boldsymbol{#1}} 

\begin{document}

\title{Unconventional magnetoelectric conductivity and electrochemical response from dipole-like sources of Berry curvature}

\author{Ipsita Mandal}
\email{ipsita.mandal@snu.edu.in}

\affiliation{Department of Physics, Shiv Nadar Institution of Eminence (SNIoE), Gautam Buddha Nagar, Uttar Pradesh 201314, India}

\begin{abstract} 
We investigate the longitudinal magnetoelectric conductivity and nonlinear electrochemical response (ECR) of three-dimensional nodal-ring semimetals, including vortex nodal-rings (VNRs) and $\mathcal{PT}$-symmetric nodal-rings (PTNRs), together with three-band Hopf semimetals (HSMs), using the semiclassical Boltzmann formalism. The nodal-rings of the VNRs and PTNRs lie in the $k_z=0$ plane, whereas the nodal points of HSMs host isolated Berry-curvature (BC) dipoles with axes along the $k_z$ direction. In VNRs, such dipoles may be viewed as forming a continuous distribution along the nodal-ring, connecting the two classes of systems. For collinear electric and magnetic ($\mathcal{B}$) fields along the anisotropy axis $\boldsymbol {\hat{z}}$, the linear response is computed exactly and displays odd powers of $B_z$ for a single ring or node. For a two-node HSM with oppositely-oriented BC-dipole axes, these odd terms cancel in the sum. We analyse the nonlinear ECR, described by third-rank response tensors that relate the current to the combined action of an electric field and a chemical-potential gradient. The surviving tensor components are determined by the tilt direction of the bandstructure, and we obtain explicit results for simultaneous in-plane and axial tilts. A comparison of the three systems reveals a close correspondence between the responses of VNRs and HSMs, while PTNRs behave qualitatively differently. We attribute these similarities and differences to the topology of the BC field, namely dipole-like sources in VNRs and HSMs versus the ring-confined BC structure of PTNRs.
\end{abstract}

\maketitle
\tableofcontents

\section{Introduction}

\begin{figure*}[t]
\subfigure[]{\includegraphics[width=0.2 \textwidth]{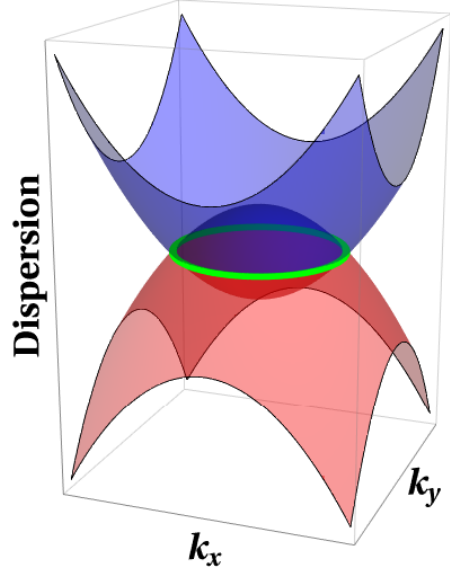} \hspace{ 1 cm}
\includegraphics[width=0.5 \textwidth]{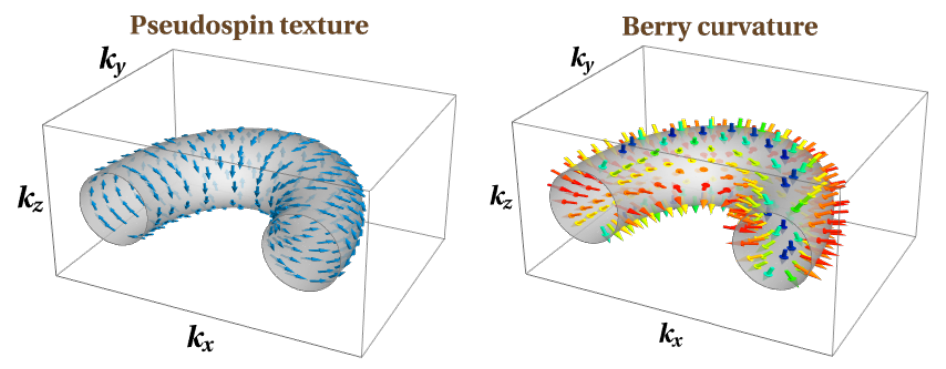}  }
\subfigure[]{\includegraphics[width=0.2 \textwidth]{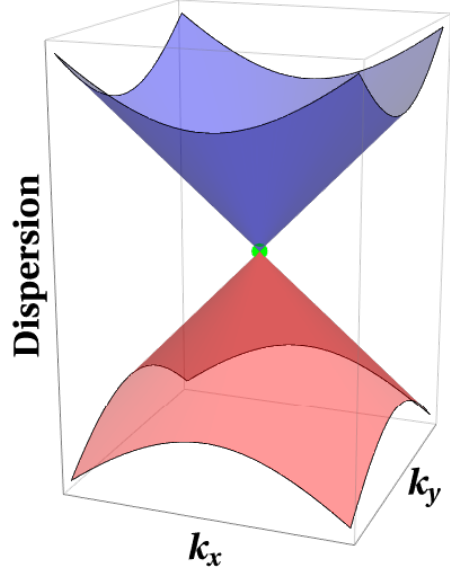} \hspace{ 1 cm}
\includegraphics[width=0.5 \textwidth]{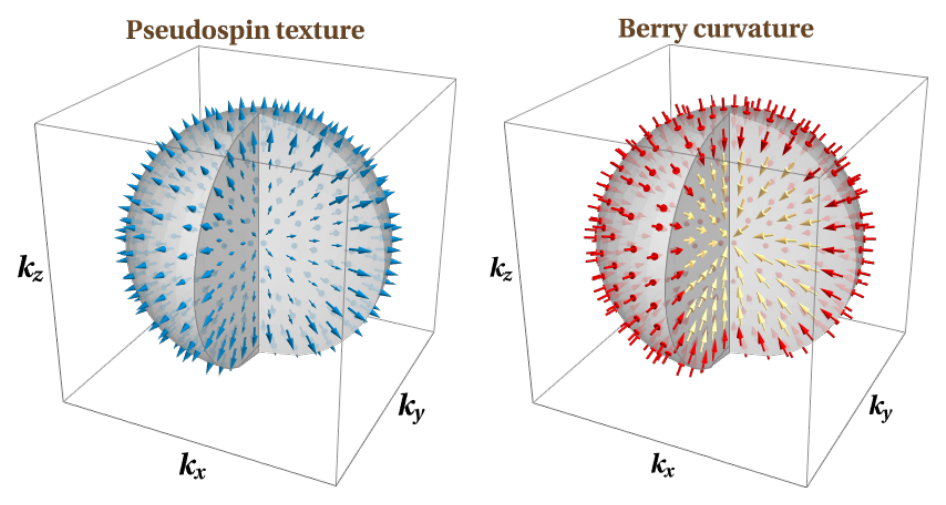}  }
\caption{\label{figfs} Distinctions between a VNR and a WSM: (a) We show the unique features of a VNR in terms of its dispersion against the $k_x k_y $-plane showing two bands crossing along a circle (highlighted in green), setting $k_z = 0 $, and the profile of the vector fields representing the pseudospin and BC. (b) For the sake of comparison with a Weyl node, we show its dispersion (with the green dot denoting the nodal-point), and the pseudospin and BC distributions. The vector fields for each case have been drawn in the vicinity of a section of the Fermi surface for the positive-energy band. The BC of the VNR forms a dipole in every cross-section of the nodal-ring, whereas the BC of a Weyl node is a monopole.}
\end{figure*}

Investigations of transport properties in three-dimensional (3d) semimetals have catapulted into prominence as they often reflect the underlying topological features of the Brillouin zone (BZ). When the BZ harbours symmetry-protected band-crossings in the momentum space (denoted here by $\bs k \equiv \lbrace k_x, k_y, k_z \rbrace $), the resulting nodal points~\cite{burkov11_Weyl, yan17_topological, bernevig, grushin-multifold} or curves~\cite{balents-nodal, vortex-nrsm} can be characterised by some topological invariant or index. Our efforts are mainly directed at identifying meaningful response which will capture this invariant unambiguously. The topological properties often arise as a nontrivial Berry curvature (BC)~\cite{xiao_review, sundaram99_wavepacket,graf-Nband, graf_thesis}, sourced by the Berry phases of the associated Bloch bands meeting at the nodes of a semimetal. Starting with the pioneering example of the Weyl semimetals (WSMs), the most important telltale signatures of the BC are the intrinsic anomalous-Hall effect~\cite{haldane04_berry,goswami13_axionic, burkov14_anomolous} and the appearance of longitudinal and transverse conductivity in planar-Hall configurations \cite{zhang16_linear, chen16_thermoelectric, nandy_2017_chiral, ips-mwsm-floquet, amit_magneto, das20_thermal, ips-kush-review, ips-ruiz, ips-tilted, ips-rsw-ph, ips-shreya, ips-spin1-ph, timm, ips-exact-spin1, ips-exact-rsw, ips-exact-kwn, li2023_planar, ips-nlsm-ph, ips-gnr-strain}. Here, we will focus on a two-band system containing a nodal-loop in the form of a circle [see Fig.~\ref{figfs}(a)], thus carrying a representation of the pseudospin-1/2 quantum number (just like the WSMs). Generically, their effective Hamiltonian can be represented as $\boldsymbol{d} (\bm k) \cdot \boldsymbol \sigma$, where $ \boldsymbol{\sigma} = \left\lbrace \sigma_x,\sigma_y,\sigma_z\right \rbrace $ is the vector comprising the three Pauli matrices as its three components and defining the $2\times 2$ matrix-operators in the pseudospin space. The 3-component vector $\bs d$ encodes the pseudospin structure of the two-component spinors representing the two Bloch bands. For a typical nodal-loop protected by $\mathcal P \mathcal T$-symmetry (where $\mathcal P$ and $\mathcal T$ represent the operators implementing the inversion and time-reversal symmetries, respectively), which we will refer to as a $\mathcal P \mathcal T$-symmetric nodal-ring (PTNR), the BC is nonzero only along the loop itself, vanishing identically elsewhere.\footnote{Adding a mass term of magnitude $\Delta $, for example generated by spin-orbit coupling, makes it gapped. This induces a nonvanishing BC $\propto \Delta$ which happens to have only in-plane components \cite{schnyder_nodal, yang1, yang_review_nlsm, chen_nlsm, ips-magnus, flores, li2023_planar, claudia_nlsm, enke, ips-nlsm-ph, ips-gnr-strain}.}
This singular-natured BC is characterised by a nonvanishing Zak phase \cite{schnyder_nodal, biao_nodal, zak_nodal, ips-nlsm-ph}. However, in the absence of both these symmetries, a different type of nodal-ring may exist, dubbed as the vortex nodal-ring (VNR) \cite{vortex-nrsm}, which does harbour a nonvanishing nonsingular BC away from the nodal-loop. The nomenclature is related to the fact that the vector field representing $\boldsymbol d (\bf k)$ takes a vortex-like or smoke-ring texture in the momentum space [cf. Fig.~\ref{figfs}(a)].

\begin{figure*}[t!]
\includegraphics[width= 0.8 \textwidth]{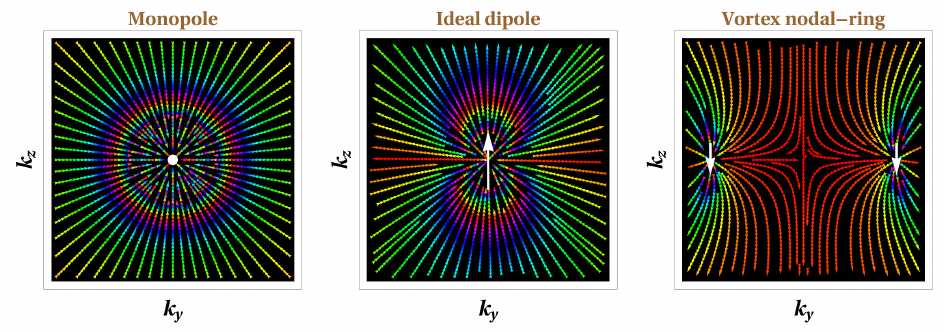}
\caption{\label{figbc} Profile of the BC-flux distribution for a monopole (e.g., Weyl node), an ideal dipole (e.g., three-band HSM) [cf. Eq.~\eqref{eqbcommhopf} with $s=2$], and a VNR [cf. Eq.~\eqref{eqbcomm} with $s=2$], projected on the $k_y k_z $-plane at $k_x =0 $. While the white dot denotes the monopole singularity, the white arrows denote the dipole-like singularities of the BC. Panels (a), (b) and (c) show the monopole, the HSM and the VNR, respectively. The dipoles are isolated for the HSM and continuously distributed along the ring for the VNR.}
\end{figure*}

The magnitude of the BC-monopole carried by a nodal point is synonymous with the Chern number of any closed 2d surface (in momentum space) containing the node inside, giving us the associated topological invariant in terms of the BC-monopole charge. Beyond this simple picture, we are also now aware of more possibilities of the behaviour of the BC where the nodes act as sources of higher-order poles \cite{graf-hopf, hopf2, hopf3,  ips-fa, ips-fa-lett}, which include ideal dipoles, quadrupoles, and so on. They can be thought of as the momentum-space analogues of such objects familiar from multipole expansions of electromagnetism. Here, we will consider three-band massless Hopf semimetals (HSMs), which host BC-dipoles. In an HSM, such a dipole sits as an isolated object at each nodal point, whereas in a VNR the dipoles are not confined to a point but can instead be visualised as a continuous distribution spread along the nodal-ring itself. 
The connection between the two dipolar systems is made explicit in Fig.~\ref{figbc}. In the plane spanned by $\kappa_\perp$ and $k_z$ at fixed $\phi$, the BC of the VNR has the angular structure $k_z\,\bs k$ of the HSM, with the radial dependence of the HSM replaced by the one set by the distance from the ring. The VNR can therefore be viewed as a one-parameter family, labelled by $\phi$, of such dipoles. Since the dispersions of the VNR and the PTNR coincide (cf. Fig.~\ref{figcyclide}), the similarity of the response-characteristics of the VNR and the HSM, as opposed to the PTNR, can be traced to the BC texture and not to the shapes of the Fermi surfaces.

While a nonzero BC-vector field is the first important topological property to affect transport (see Fig.~\ref{figbc} for a comprehensive illustration), it is accompanied by another vector field called the orbital magnetic moment (OMM), sourced by the same Berry phase \cite{xiao_review, sundaram99_wavepacket}, and also affects transport involving a magnetic field \cite{ips-ruiz, ips-tilted, ips-rsw-ph, ips-shreya, ips-spin1-ph,timm, ips-exact-spin1, ips-exact-rsw, ips-exact-kwn,  ips-nlsm-ph}. While topology-induced response has been studied quite widely for the PTNRs \cite{schnyder_nodal, yang1, yang_review_nlsm, chen_nlsm, ips-magnus, flores, li2023_planar, claudia_nlsm, enke, ips-nlsm-ph}, such characteristics have not been explored for the VNRs and BC-dipole nodal points, which we undertake in this paper. In particular, we will perform an analytical computation of the linear magnetoelectric conductivity in planar-Hall set-ups, similar to what was studied in our earlier work involving PTNRs \cite{ips-nlsm-ph}. Such set-ups refer to the application of static and uniform electric ($\bs E $) and magnetic ($\bs B$) fields, where $\bs B $ is not necessarily perpendicular to $\bs E $. Rather than resorting to the relaxation-time approximation (RTA) used in that earlier work, we solve the linearised Boltzmann equation for the collinear configuration exactly, and find that the resulting magnetoconductivity for the VNR and the HSM contains an additional linear-in-$B_z$ term (here $B_z$ is the component of $\bs B$ along $\hat{\bs z}$), absent for BC-monopole and PTNR nodes, which we trace directly to the absence of standalone time-reversal symmetry in these two dipolar models. We extend this analysis to a two-node HSM, whose conjugate nodes carry oppositely-oriented BC-dipole axes. The RTA treatment of the VNR and the HSM, which also finds the odd-in-$B_z$ terms and the vanishing of the internode conductivity of the HSM, is presented in Ref.~\cite{ips-firdous-vnr}. The exact numerical solution shows that the odd-in-$B_z$ terms of the two nodes cancel in the sum, and that the internode scattering acts as an additional channel that only modifies the slope of the response of each node. The net response is even in $B_z$ and, for the chosen parameters, much smaller than that of a single node.

Nonlinear transport under the effect of an external electric field (or temperature gradient) and the gradient of the chemical potential constitutes a response of electrochemical nature \cite{ips-shreya}. Here, we dub the associated electrical conductivity tensor as the electrochemical response (ECR). Since the Chern number is zero for the systems under consideration in this paper, several of these coefficients vanish, in contrast to nodal points harbouring BC-monopoles. Even so, we find that nonzero coefficients emerge in the tilted cases, offering a nontrivial characterisation of the underlying topology. We further show that the tilt direction alone dictates which components of the ECR survive, and that this selection rule sets the PTNR apart from the VNR and the HSM: since its BC is confined entirely to the nodal-ring, a tilt along the ring's own axis leaves the ring pinned at zero energy and yields no response whatsoever, in contrast to the VNR and the HSM, where the same axial tilt activates a distinct set of diagonal ECR components. For simultaneous in-plane and axial tilts, we derive explicit expressions for the VNR and the HSM, which contain the single-tilt results as limiting cases and show that the two tilts interfere only at third order.

The paper is organised as follows. Sec.~\ref{secmodels} describes the effective models of the three systems. Sec.~\ref{secexact} derives the exact longitudinal magnetoelectric conductivity for a single VNR, a single BC-dipole nodal point, and a two-node HSM with oppositely-oriented dipole axes. Sec.~\ref{secnonlinear} treats the nonlinear ECR, and Sec.~\ref{secsum} gives a summary and outlook. Appendix~\ref{appcyclides} covers the integrals for tilted nodal-rings, Appendix~\ref{appcomb} the combined in-plane and axial tilts, Appendix~\ref{appsing} the singularity of the collision matrix, and Appendix~\ref{appcont} the internode transfer rate.

All our expressions in the paper are written using the natural units. Basically, this implies that the reduced Planck's constant ($\hbar $), the speed of light ($c$), and the Boltzmann constant ($k_B $) are each set to unity. The magnitude of electric charge, $e$, has no units and also equals unity in the natural units. However, for tracking the appearance of the electric charge, we will retain $e$ in our expressions. 

\section{Models}
\label{secmodels}

In the following three subsections, we describe the three distinct systems of interest in this paper.

\subsection{Vortex nodal-ring}
\label{secvrn}

The minimal model of a VNR, comprising two bands and a circular nodal-loop lying in the $k_x k_y$-plane at $k_z = 0$, is captured by \cite{vortex-nrsm}
\begin{align}
& \mathcal{H}_{0} (\bs k ) = {\bs d}_0 (\bs k) \cdot \boldsymbol{\sigma} 
+ v_0 \,{\bs \eta} \cdot {\bs k} \;\,\mathbb{I}_{2 \times 2}\,,
\quad {\bs d}_0 (\bs k)
= \left \lbrace  \frac{-\,k_x\,k_z} {m_z}, \,\frac{-\,k_y\, k_z} {m_z},
\, \frac{ k_\perp^2-k_0^2 - k_z^2} {2\,m_\perp}  \right \rbrace, \quad
k_\perp = \sqrt{k_x^2 + k_y^2 }\,,\nn &
\bs \eta = \lbrace \eta_x ,\, \eta_y, \, \eta_z \rbrace\,.
\end{align}
where $k_0$, $m_z$, and $m_\perp $ are material-dependent parameters. At the $ k_z = 0$ plane, the two bands cross at $k_\perp = k_0  $, defining a nodal-ring of radius $ k_0 $. Therefore, $v_z \equiv k_0/m_z$ and $v_\perp \equiv k_0/m_\perp $ encode the values of the Fermi velocities perpendicular and parallel to the nodal plane, respectively. In the following, we will set $v_z = v_\perp  = v_0 $ for the sake of simplicity. We have also included a generic tilting term, $v_0 \,{\bs \eta} \cdot {\bs k}$.

For low-energy excitations confined in the vicinity of the resulting Fermi surface, the dominant transport-signatures of the low-energy quasiparticles can be captured by a linearised Hamiltonian, $\mathcal{H}$. It comprises terms which deviate only at linear-order-in-momentum from the location of the nodal line \cite{linearize-nlsm, ips-nlsm-ph}. This can be accomplished by implementing a transformation to the cylindrical coordinates as follows:
\begin{align}
\label{eqtrs}
k_x = k_\perp \cos \phi \,, \quad
k_y = k_\perp \sin \phi \,,\quad k_\perp =  k_0 + \kappa_\perp \,.
\end{align}
Clearly, $ \kappa_\perp \in [-\, k_0,\,\infty )$. However, for the low-energy processes we are interested in, we must have $|\kappa_\perp | \ll k_0 $, as $k_0$ acts as an ultraviolet-energy scale.
Subsequently, linearising in the small-valued variables (viz. $\kappa_\perp $ and $k_z$), we end up with
\begin{align}
\label{eqham}
& {\mathcal{H}}_{vn} (\bs k) = {\bs d}_{vn}  ( \bs k) \cdot \boldsymbol{\sigma}
+ v_0 \,{\bs \eta} \cdot {\bs k}\; \mathbb{I}_{2 \times 2} \,,
\quad
{\bs d}_{vn}  ( \bs k) = v_0 \left \lbrace  - \,k_z \cos \phi,\,
-\,k_z \sin \phi,\, \kappa_\perp \right \rbrace .
\end{align}
For the linearised Hamiltonian $\mathcal H$, the eigenvalues of the two bands are obtained as
\begin{align}
\label{eqev}
\varepsilon_s  ({ \bs k}) &= (-1)^s\, \epsilon_k  + v_0 \,{\bs \eta} \cdot {\bs k} \,, \quad 
 \epsilon_k = v_0\, \sqrt{ \kappa_\perp^2 +  k_z^2} \,,
\quad s \in \lbrace 1,2 \rbrace.
\end{align}
The band-velocities of the quasiparticles are given by
\begin{align}
\label{eqv}
{\boldsymbol v}^{(0,s)} ( \bs{k}) 
\equiv \nabla_{\bs{k}} \varepsilon_s  (\bs{k}) 
= \frac{ (-1)^s\, v_0^2}{\epsilon_k }  \left\lbrace 
\kappa_\perp \cos \phi, \, \kappa_\perp \sin\phi , \,k_z \right\rbrace
+  
v_0 \,{\bs \eta}  .
\end{align}
The Berry curvature (BC) and the orbital magnetic moment (OMM), associated with the $s^{\rm{th}}$ band, turn out to be
\begin{align}
\label{eqbcomm}
\bs \Omega_s ({ \bs k})= 
\frac{(-1)^{s+1} \, v_0^3\,k_z} 
{2\, (k_0 + \kappa_\perp )\, \epsilon_k^3}
\left\lbrace \kappa_\perp \cos \phi  , \, \kappa_\perp \sin \phi,\, k_z \right\rbrace, \quad 
{\boldsymbol{m}} ({ \bs k}) 
= \frac{ -\,e \, v_0^3\,k_z} 
{2\, (k_0 + \kappa_\perp )\, \epsilon_k^2}
\left\lbrace \kappa_\perp \cos \phi  , \, \kappa_\perp \sin \phi,\, k_z \right\rbrace.
\end{align}
See Fig.~\ref{figbc}(c) for a depiction of the cross-section of the BC profile in the $k_x = 0$ plane.
The OMM behaves exactly like the electron's spin because, on applying a magnetic field ($\bs B$), it couples to it through
a Zeeman-like term. Therefore, the shift in energy caused by the OMM is captured by
\begin{align}
\varepsilon_m ({ \bs k}) 
= \frac{e \,v_0^3 \,k_z 
\left( B_x \,\kappa_\perp  \cos \phi + B_y \,\kappa_\perp  \sin \phi 
+ B_z k_z \right)}
{2 \,\epsilon_k^2 \,(k_0 + \kappa_\perp )}\,.
\end{align}
The OMM-induced parts are the same for both the bands and, hence, we have not attached any $s$ index to them.
The BC can be imagined to be caused by a continuous distribution of ideal dipoles along the nodal-ring, as seen from Figs.~\ref{figbc}(c) and \ref{figfs}(a).

We will work in the $T \rightarrow 0 $ limit, such that the Fermi-Dirac-distribution function, denoted as $f_0 (\mathcal E) $ for a band with dispersion $E$, goes to $\Theta (\mu- \mathcal E)$. Hence, its derivative with respect to $\mathcal E$ goes as $\partial_\mathcal E f_0 \rightarrow -\, \delta (\mathcal E - \mu )$. We will use the shorthand-notation of $\int_k \cdots \equiv \int \frac{ d^3 \bs k}{(2\, \pi)^3 }  \cdots $ for the $\bs k$-integrals. While performing these integrals, we will resort to using the toroidal coordinates, defined as follows:
\begin{align}
\label{eqtrsf}
k_x = \left( k_0+ \kappa_\perp\right) \cos \phi  \,, \quad
k_y = \left ( k_0 +  \kappa_\perp \right ) \sin \phi \,,\quad
k_z = \kappa \sin \gamma \,, \quad  \kappa_\perp \equiv \kappa \cos \gamma  = k_\perp - k_0\,.
\end{align}
The Jacobian of the transformation is given by
\begin{align}
J =   \kappa \, k_\perp \,.
\end{align}
In terms of the toroidal coordinates, $k_0$ represents the major radius (i.e., the distance between a point on the nodal-ring and the centre of the torus), $\kappa $ denotes the minor radius (i.e., the radius of the cross-section of the torus). The angular coordinates, $\phi $ and $\gamma $, span over $  [0, 2 \pi)$, representing rotation around the torus's axis of revolution and rotation around the centre of a cross-section, respectively.
In our calculations for conductivity, we will consider applying a non-negative value of the chemical potential, $\mu$. Therefore, for $\eta = 0$, only the $s=2$ band will contribute.

\begin{figure*}[t!]
\subfigure{\includegraphics[width= 0.75 \textwidth]{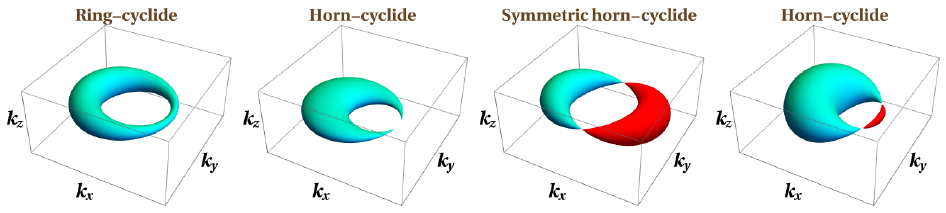}}
\subfigure{\includegraphics[width= 0.75 \textwidth]{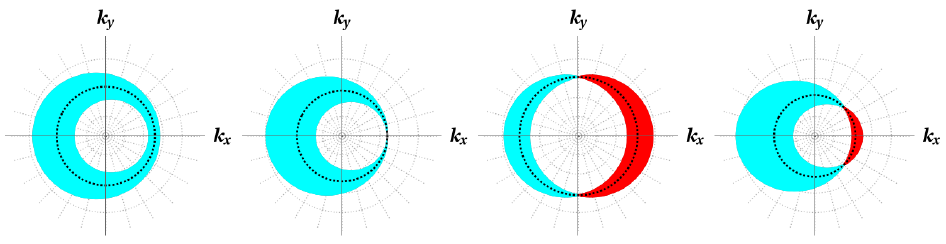}}
\caption{\label{figcyclide} On introducing tilt, the Fermi surfaces of nodal-rings take the forms of different kinds of cyclides, which depend on the values of $\mu$ and $\tilde \eta = v_0\, k_0\,\eta_x $. The red and cyan colours depict the $s=1$ and $s=2$ bands, respectively. While the upper panel illustrates the full 2d Fermi surfaces, the lower panel shows the corresponding projections on the $k_x k_y$-plane for $k_z = 0 $. The dotted black circle represents the original PTNR or VNR at $\mu = \eta_x = 0$ for the sake of reference. The same cyclides arise for the VNR and the PTNR, since the two share the same dispersion.}
\end{figure*}

For the tilted case, some extra care needs to be taken. When the tilting is with respect to the $k_x k_y $-plane, the ring-toroid structure can get drastically modified. Let us choose this in-plane tilting with respect to the $k_x$-axis, without any loss of generality. Depending on the relative values of $\mu $ and $\tilde \eta \equiv  v_0\, k_0\, \eta_x$, there are several possibilities \cite{ahn-tilted-nl} corresponding to (a) ring-cyclide: $ 0< \tilde \eta < \mu $; (b) horn-cyclide formed from the $s=2 $ band only: $\mu> 0 $ and $ \tilde \eta = \mu $; (c) symmetric horn-cyclide formed by both the bands: $\mu =0 $ and $ \tilde \eta >0 $; and (d) (asymmetric) horn-cyclide formed by both the bands: $ 0 < \mu < \tilde \eta $. For visual aid, Fig.~\ref{figcyclide} depicts the $\phi$-dependence of these distinct-shaped cyclides.
In the language of Eqs.~\eqref{eqtrs} and \eqref{eqtrsf}, we have $\epsilon_k =  v_0 \, \kappa $ and
\begin{align} 
&\delta (\varepsilon_s -\mu)  \equiv 
\delta \Big (  (-1)^s\,v_0 \,\kappa + \eta_x \,v_0\, (k_0 + \kappa \cos \gamma) \cos \phi -\mu \Big )
= \frac{\delta \Big ( \kappa - \frac{\mu - \tilde \eta \cos \phi } { v_0\,\zeta_s } \Big )}
{ v_0 \,|\zeta_s| } \,,
\nn & \text{where }
\zeta_s = (-1)^s  + \eta_x  \cos \gamma \cos \phi
\text{ and } \tilde \eta = \eta_x \,v_0\,k_0\,.
\end{align}
Depending on the relative values of $\mu$ and $ \tilde \eta $, both the bands may contribute, and the valid range of $\phi $ is determined from the points where the Fermi-momentum vanishes, viz. $\kappa = 0 \Rightarrow  \cos \phi  =  \mu /\tilde \eta $. More details can be found in Appendix~\ref{appcyclides}.

The Hamiltonian of Eq.~\eqref{eqham} possesses three symmetries that will be central to interpreting the response below. We first consider the unitary spatial symmetries. At $\bs\eta=0$, the Hamiltonian is covariant under continuous $C_{\infty z}$ rotations about the $k_z$ axis,
\begin{align}
\mathcal H_{vn}(\mathcal R_z\bs k)
=U(\varphi)\,\mathcal H_{vn}(\bs k)\,U^\dagger(\varphi)\,, \quad
U(\varphi)=e^{-i\,\varphi\,\sigma_z/2}\,.
\end{align}
This is because the azimuthal rotation of $(k_x,k_y)$ is exactly compensated by the corresponding rotation of $(d_x,d_y)$, while $d_z$ is unchanged. The Hamiltonian is also invariant under reflection through the $k_z=0$ plane,
\begin{align}
\mathcal H_{vn}(k_x,k_y,-k_z)=\sigma_z\,\mathcal H_{vn}(k_x,k_y,k_z)\,\sigma_z\,,
\end{align}
because $\sigma_z$ leaves $d_z$ unchanged while reversing the signs of $d_x$ and $d_y$, compensating $k_z\to-k_z$.

This mirror symmetry immediately fixes the parity of the BC and OMM. Since $\bs\Omega_s$ and $\boldsymbol m$ in Eq.~\eqref{eqbcomm} carry an overall factor of $k_z$, their in-plane components are odd under $k_z\to-k_z$, whereas their $z$ components are even. Thus, on a mirror-symmetric Fermi surface, any integrand containing a single power of $\Omega_s^{x,y}$ or $m^{x,y}$ vanishes upon integration, while terms involving $\Omega_s^z$ or $m^z$ can survive. This parity structure underlies both the odd-in-$B_z$ contributions to the diagonal conductivity-components and the vanishing of the out-of-plane response for an untilted VNR.

The Hamiltonian does not possess ordinary time-reversal symmetry. For a two-band Hamiltonian $H(\bs k)=\bs d(\bs k)\cdot\bs\sigma$, an antiunitary time reversal $T=\mathcal U_T\, \mathcal K$ requires
$T\, H(\bs k)\,T^{-1}=H(-\bs k)$.
The two relevant choices are the orbital (sublattice) operation $\mathcal U_T=\mathbb I$, with $T^2=+1$, which requires
\begin{align}
d_a^*(\bs k)=(-1)^{\delta_{ay}}\, d_a(-\bs k)\,,
\end{align}
and the spin-$1/2$ operation $\mathcal U_T=i\sigma_y$ (with $T^2=-1$), which requires
\begin{align}
\bs d^*(\bs k)=-\bs d(-\bs k)\,.
\end{align}
For the VNR Hamiltonian, $\bs d$ is real and even under $\bs k\to-\bs k$. In particular, $k_z\cos\phi$ and $k_z\sin\phi$ are both invariant, because $\phi\to\phi+\pi$ simultaneously with $k_z\to-k_z$. Hence the spin-$1/2$ condition is violated noting $\bs d^*=\bs d\neq-\bs d=-\bs d(-\bs k)$, while the orbital condition fails for the $y$ component, which does not acquire the required sign. Thus the VNR is not invariant under ordinary time-reversal. Equivalently, the intrinsic dipolar Berry geometry does not have the momentum-odd transformation of a conventional Berry-curvature monopole: $\bs\Omega_s$ and $\boldsymbol m$ do not simply reverse under $\bs k\to-\bs k$. A different antiunitary symmetry nevertheless survives. Let
\begin{align}
\Xi=T \,M_y\,,
\end{align}
where $M_y$ reflects $k_y\to-k_y$ and acts on the pseudospin as
\begin{align}
\sigma_x\to-\sigma_x \,,\quad \sigma_y\to\sigma_y\,,\quad \sigma_z\to-\sigma_z \,.
\end{align}
For the spin-$1/2$ representation, the unitary parts of $T=i\,\sigma_y \,\mathcal K$ and $M_y$ combine, up to an overall phase, to the identity. Consequently, $\Xi$ acts as complex conjugation in the pseudospin sector and satisfies
\begin{align}
\label{eq:xi_invariance_vnr}
\mathcal K\,\mathcal H_{vn}(k_x,k_y,k_z)\,\mathcal K^{-1} = \mathcal H_{vn}(k_x,-k_y,k_z)\,.
\end{align}
Thus $\Xi$ relates the Hamiltonian at $(k_x,k_y,k_z)$ to that at $(k_x,-k_y,k_z)$ even though ordinary $T$ is broken. It is important to distinguish this symmetry statement from the Onsager--Casimir relation used for transport. The latter is not a claim that the VNR Hamiltonian is time-reversal invariant. Rather, the Boltzmann response obeys the generalised parameter-reversal relation
\begin{align}
\label{eq:generalized_onsager}
\sigma_{ij}\bigl(\bs B,\bs\eta,\bs\Omega_s,\bs m_s\bigr)
= \sigma_{ji} \,\bigl(-\bs B,-\bs\eta,-\bs\Omega_s,-\bs m_s\bigr)\,,
\end{align}
where $\bs B$, the tilt-vector $\bs\eta$, and the intrinsically time-reversal-odd quantities $\bs\Omega_s$ and $\bs m_s$ are all reversed. This is the appropriate microreversibility relation for the transport coefficients, which should not be interpreted as an ordinary $T$ symmetry of the fixed-parameter Hamiltonian. For the VNR, the surviving antiunitary symmetry $\Xi$ gives a corresponding covariance relation at fixed parameter values. Because $\Xi$ reverses only the $y$ component of momentum, the intrinsic dipolar quantities transform by a partial reversal rather than by the full $\bs k\to-\bs k$ operation. The response therefore obeys
\begin{align}
\label{eq:theta_covariance_vnr}
\sigma_{ij}(\bs B,\bs\eta)
=R_{ik}\, R_{jl}\,\sigma_{kl}(\Xi\,\bs B,\Xi \,\bs\eta)\,, \quad
\Xi\,\bs B=(B_x,-B_y,B_z)\,, \quad \Xi \, \bs\eta=(\eta_x,-\eta_y,\eta_z)\,,
\end{align}
with $R$ determined by the transformation of the current operator under $\Xi$. Unlike the ordinary Onsager--Casimir relation, which involves interchange of indices together with full reversal of the time-reversal-odd parameters, Eq.~\eqref{eq:theta_covariance_vnr} is a spatial covariance relation associated with the surviving partial-reversal symmetry. This distinction is essential since the absence of ordinary $T$ does not invalidate the generalised parameter-reversal relation. In fact, the weaker $\Xi$ covariance permits response structures, such as the odd-in-$B_z$ contribution to $\sigma_{zz}$ and the zero-field tilt-driven ECR discussed in Secs.~\ref{secexact} and \ref{secnonlinear}, that would be constrained differently in a fully time-reversal-symmetric system.

To summarise the logical relation between the two notions, the time-reversal symmetry of the bare Hamiltonian and the parameter-reversal relation obeyed by the transport coefficients are distinct statements. The former is a property of a fixed Hamiltonian, namely the existence of an antiunitary operator $T$ with $T\,\mathcal H(\bs k)\,T^{-1}=\mathcal H(-\bs k)$. The latter follows from the microscopic reversibility of the full dynamics and relates the conductivity of a system to that of its time-reversed partner, in which every time-reversal-odd parameter is reversed. In the Boltzmann response, these parameters are $\bs B$, $\bs\eta$, $\bs\Omega_s$ and $\bs m_s$, which gives Eq.~\eqref{eq:generalized_onsager}. When $\mathcal H$ is itself time-reversal invariant, as for the PTNR, the reversal of $\bs\Omega_s$ and $\bs m_s$ is equivalent to the original system, and Eq.~\eqref{eq:generalized_onsager} closes on a single system. For the diagonal component it then reduces to the familiar Onsager--Casimir form $\sigma_{zz}(B_z)=\sigma_{zz}(-B_z)$, which forbids odd powers of $B_z$. In a single VNR or a single HSM node, the reversal of $\bs\Omega_s$ and $\bs m_s$ flips the orientation of the BC dipole and therefore produces a different system. Eq.~\eqref{eq:generalized_onsager} then relates $\sigma_{zz}$ of the system to that of its dipole-reversed partner and not to itself, so that no constraint on the parity of $\sigma_{zz}$ in $B_z$ results. The surviving antiunitary symmetry $\Xi$ yields only the covariance relation of Eq.~\eqref{eq:theta_covariance_vnr}, which reduces to an identity for $\bs E\parallel\bs B\parallel\hat{\bs z}$. Whether odd powers of $B_z$ actually appear is then decided by the parity of the integrand under $k_z\to-k_z$, as discussed in Sec.~\ref{secexact}. For a pair of nodes with oppositely-oriented dipoles, the reversal of $\bs\Omega_s$ and $\bs m_s$ exchanges the two nodes and maps the pair onto itself, so that the net $\sigma_{zz}$ is again even in $B_z$.

Finally, a finite tilt modifies the unitary spatial symmetries. The $C_{\infty z}$ symmetry survives only for $\bs\eta\parallel\hat{\bs z}$, whereas the corresponding $v_0\eta_z k_z$ term breaks the $k_z\to-k_z$ mirror. Conversely, a tilt confined to the $k_xk_y$-plane preserves the $k_z$ mirror but breaks $C_{\infty z}$.


\subsection{${\mathcal P} {\mathcal T}$-symmetric nodal-ring}
\label{secptnr}

The minimal model of a two-band ideal nodal-ring of the PTNR variety, lying in the $k_x k_y$-plane, and linearised in momenta about the nodal-circle, is captured by \cite{balents-nodal, yang1, ips-nlsm-ph, ips-gnr-strain}
\begin{align}
\mathcal{H}_{pt} (\bs k ) = {\bs d}_{pt} (\bs k) \cdot \boldsymbol{\sigma}  
+  v_0 \,{\bs \eta} \cdot {\bs k}\; \mathbb{I}_{2\times 2}\,,
\quad {\bs d}_{pt} (\bs k) 
=
\left \lbrace v_0  \left( k_\perp - k_0 \right) , \, v_z \, k_z ,\, 0 \right \rbrace .
\end{align}
For the sake of simplicity, we will set $v_z = v_0$. The eigenenergies and the band-velocities take the same forms as the VNR, viz. Eqs.~\eqref{eqev} and \eqref{eqv}, and the associated $\bs k$-integrals are likewise evaluated in the same toroidal coordinates, Eq.~\eqref{eqtrsf}.

Unlike the VNR model, the PTNR has a vanishing BC everywhere for each band, except at the nodal-ring. It can be parametrised as \cite{schnyder_nodal, fuchs-bc-graphene}
\begin{align}
\label{eqbcpnr}
\bs \Omega_s ({ \bs k})
= (-1)^{s+1}\, \pi \, \delta(k_z)\,\delta(k_\perp-k_0)\,{\boldsymbol{\hat \phi}}
= (-1)^{s+1} \,\pi \, \delta(k_z)\,\delta(\kappa_\perp)
\left \lbrace - \sin \phi\,, \cos \phi\,, \, 0 \right \rbrace ,
\end{align}
where it becomes singular. Here, ${\boldsymbol{\hat \phi}}$ denotes the unit vector along the $\phi$-coordinate.
Similarly, the OMM is singular at the nodal-ring, expressed as
\begin{align} 
\bs{m} ( \bs k)  = e\,(-1)\, v_0\, \kappa \, \pi\, \delta(\kappa \sin \gamma)\,\delta(\kappa  \cos \gamma)
\,{\boldsymbol{\hat \phi}}
=-\, e\, v_0\, \kappa \, \pi\, \frac{\delta(\gamma)} {\kappa \,|\cos \gamma|}
\,\frac{\delta(\kappa )} {|\cos \gamma|}
\,{\boldsymbol{\hat \phi}}
= -\, e\,   v_0  \,\pi \, \delta(\kappa)\,\delta(\gamma) \,{\boldsymbol{\hat \phi}}\,.
\end{align}
For evaluating some of the conductivity expressions, we will need to use the expression of the cross-product of a vector, $\bs A$, with $\bs \Omega_s$. This turns out to be
\begin{align}
\label{eqcross}
 \bs A \times \bs \Omega_s ({ \bs k})
 & =  (-1)^{s+1} \,\pi \, \delta(k_z)\,\delta(k_\perp-k_0)\,
\bs A \times {\boldsymbol{\hat \phi}}
\nn & = (-1)^s\, \pi \, \delta(k_z)\,\delta(k_\perp-k_0)
\left \lbrace  A_z \cos \phi\,,  A_z \sin \phi\,, \, 
-A_x \cos \phi - A_y \sin \phi \right \rbrace .
\end{align}

The linearised Hamiltonian of the PTNR displays a distinct symmetry structure
from the VNR and the HSM, since the breaking of time-reversal symmetry in those models is absent here. At $\bs \eta = 0$, it is
invariant under the continuous $C_{\infty z}$ rotation about the $k_z$-axis,
\begin{align}
\mathcal{H}_{pt}(\mathcal R_z \, \bs k) = \mathcal{H}_{pt}(\bs k)\,,
\end{align}
since $\bs d_{pt}$ depends only on $k_\perp$ and $k_z$, and not on $\phi$. It is
also invariant under reflection through the $k_z=0$ plane,
\begin{align}
\mathcal{H}_{pt}(k_x,k_y,-k_z)
=\sigma_x \, \mathcal{H}_{pt}(k_x,k_y,k_z) \, \sigma_x\,,
\end{align}
because $\sigma_x$ leaves $d_x \propto (k_\perp-k_0)$ fixed while flipping the
sign of $d_y \propto k_z$. Unlike the VNR and the HSM, the BC and OMM
of the PTNR are $\delta$-function supported on the nodal-ring, whereas the
Fermi surface at $\mu > 0$ lies away from the ring. The two supports therefore
do not overlap, making the geometric contributions to $\sigma_{zz}$ vanish
identically at $\mu > 0$. Consequently, the presence of odd-in-$B_z$ terms identified
for the VNR and the HSM later in Sec.~\ref{secexact} has no geometric ingredient to
act on in the PTNR. The mirror parity of $\bs\Omega_s$ and $\bs m$, which are
even under $k_z \rightarrow -k_z$ (because they are proportional to
$\delta(k_z)$), is therefore not sampled at the Fermi surface. The Drude piece,
proportional to $v_z \propto k_z$, is odd under the mirror and integrates to
zero over the mirror-symmetric Fermi surface. Unlike the VNR and the HSM, the PTNR retains ordinary time-reversal symmetry. It realises the orbital branch of the general time-reversal condition
introduced in Sec.~\ref{secvrn}. Since $d_x^{\rm pt}=v_0(k_\perp-k_0)$ is even
and $d_y^{\rm pt}=v_z\,k_z$ is odd under $\bs k\to-\bs k$, the component-wise
condition, $ d_a^*(\bs k)= (-1)^{\delta_{ay}}\,d_a(-\bs k)$, is satisfied. Thus the antiunitary operator $T=\mathcal K$, with
$T^2=+1$, gives
\begin{align}
T\,\mathcal{H}_{pt}(\bs k)\,T^{-1} = \mathcal{H}_{pt}(-\bs k)\,.
\end{align}
Here the pseudospin labels an orbital (sublattice) degree of freedom rather
than a real spin. Thus the appropriate time-reversal operator is $T = \mathcal K$
with $T^2 = +1$, rather than the spin-$1/2$ operator
$ i\,\sigma_y\, \mathcal K$ which squares to $ -1$. The latter would have reversed
the sign of $d_x$, which would not have left $\mathcal{H}_{pt}$ invariant.
The PTNR thus retains both its unitary mirror symmetry and ordinary time-reversal, in contrast with the VNR (which breaks plain time-reversal) and the
mass-gapped nodal-ring model (which retains time-reversal but does not have the
same PTNR mirror structure). Switching on the tilt $v_0\,\bs \eta \cdot \bs k$ preserves the
$C_{\infty z}$ symmetry only for $\bs \eta \parallel \hat{\bs z}$, where the
mirror symmetry is broken instead, since $v_0\,\eta_z\,k_z$ is odd under
$k_z \rightarrow -k_z$. A tilt confined to the $k_x k_y$-plane preserves the
mirror at the cost of the $C_{\infty z}$ symmetry.

\subsection{Three-band Hopf semimetal}
\label{sechopf}

HSMs have been introduced in Ref.~\cite{graf-hopf}, which host BC-dipoles and possess linearly-dispersing bands. We consider the simplest case with a threefold-degenerate node, captured by the effective continuum Hamiltonian \cite{graf-hopf, hopf2, hopf3},
\begin{align}
\label{eqbcd2}
H_{h} =  v_0 \,(k_x \,\lambda_1 + k_y \,\lambda_2+ k_z\,\lambda_5 +{\bs \eta} \cdot {\bs k} \;\,\mathbb{I}_{3 \times 3})\,,
\end{align}
where
\begin{align}
\lambda_1 =  \begin{bmatrix}
 0 & 1 & 0 \\
 1 & 0 & 0 \\
 0 & 0 & 0 \\
\end{bmatrix}, \quad
\lambda_2 = \begin{bmatrix}
 0 & -i & 0 \\
 i & 0 & 0 \\
 0 & 0 & 0 \\
\end{bmatrix}, \quad
\lambda_5 = \begin{bmatrix}
 0 & 0 & -i \\
 0 & 0 & 0 \\
 i & 0 & 0 \\
\end{bmatrix} .
\end{align}
We have added a tilt term to see its effect in nonlinear response. The system comprises three bands, with eigenvalues
\begin{align}
\label{eqevhopf}
\varepsilon_ s  ({ \boldsymbol k})=  \begin{cases}
(-1)^s \, v_0 \,k + v_0 \,{\bs \eta} \cdot {\bs k} & \text{ for } s \in \lbrace 1,2 \rbrace
\\ v_0 \,{\bs \eta} \cdot {\bs k} & \text{ for } s=0
\end{cases}\,.
\end{align}
The value $s=0$ represents a nondispersive flat-band for no tilt, which does not contribute to transport. For a nonzero tilted node, there is a possibility of getting nonzero response. Using an orthonormal set of eigenvectors, the BC and the OMM are obtained as
\begin{align} 
\label{eqbcommhopf}
& {\boldsymbol \Omega}_s( \boldsymbol k)  = \begin{cases}
\frac{k_z\,\boldsymbol k  } {k^4} &\text{ for } s=1, 2\\
\frac{  - \,2\, k_z\,\boldsymbol k  } {k^4} & \text{ for } s=0
\end{cases}
\text{ and }
{\boldsymbol m}_s( \boldsymbol k)  = \begin{cases}
\frac{(-1)^{s+1} \,e\, v_0\, k_z\,\boldsymbol k  } {2\, k^3} &\text{ for } s=1, 2\\
0 &\text{ for } s=0
\end{cases} ,
\end{align}
respectively.
See Fig.~\ref{figbc}(b) for a depiction of the cross-section of the BC profile in the $k_x = 0$ plane.

While performing various integrals for the BC-dipole system, we will resort to using the usual spherical-polar coordinates, defined as follows:
\begin{align}
\label{eqtrs2}
k_x = k \sin \gamma \cos \phi  \,, \quad
k_y = k \sin \gamma  \sin \phi \,,\quad
k_z = k \cos \gamma \,.
\end{align}
We note that, here, $\gamma \in [0, \,\pi]$, unlike the toroidal coordinates.

The Hamiltonian of Eq.~\eqref{eqbcd2} shares the same two symmetries that
organise the VNR response, together with an additional chiral symmetry. At
$\bs \eta = 0$, it is covariant under the continuous $C_{\infty z}$ rotation
about the $k_z$-axis,
\begin{align}
H_h (\mathcal R_z \, \bs k)=U(\varphi) \, H_h (\bs k) \, U^\dagger(\varphi)\,,
\quad U(\varphi) = \text{diag} (1, e^{i \, \varphi}, 1)\,,
\end{align}
and it is invariant under reflection through the $k_z=0$ plane,
\begin{align}
H_h (k_x, k_y, -k_z)= U_\sigma \, H_h (k_x,k_y,k_z) \, U_\sigma^\dagger,
\quad U_\sigma = \text{diag}(1,1,-1)\,.
\end{align}
Since $\bs \Omega_s$ and $\bs m_s$ in Eq.~\eqref{eqbcommhopf} carry an
explicit overall factor of $k_z$ (equivalently, $\cos \gamma$), their in-plane
components are odd under this mirror, while their $z$-components are even,
exactly as for the VNR. Consequently, any integrand built from a single power
of $\Omega_s^{x,y}$ or $m^{x,y}$ is odd in $k_z$ and integrates to zero over a
mirror-symmetric Fermi surface, whereas integrands involving $\Omega_s^z$ or
$m_s^z$ are even and survive. In addition, $H_h$ anticommutes with the unitary Hermitian operator
\begin{align}
\mathcal S = \text{diag} (1,-1,-1),
\end{align}
which pins the flat-band at $\varepsilon = 0$ and forces the dispersive-band
eigenvalues to occur in $\pm \varepsilon$ pairs.
As in the VNR, a standalone time-reversal symmetry is absent. An analogue of
the antiunitary combination $\Xi \equiv T\,M_y$ nevertheless survives this
breaking: it acts on the pseudospin with
\begin{align}
\mathcal U = \mathrm{diag}(1, -1, 1),
\end{align}
and on the momentum as $(k_x, k_y, k_z) \to(-k_x, k_y, -k_z)$.
Because $\Xi$ acts on the HSM by a partial momentum-reversal rather than the
full $\bs k \to -\bs k$ that underlies the standard Onsager--Casimir relation,
the conductivity relations take the covariance form shown in
Eq.~\eqref{eq:theta_covariance_vnr}, with the appropriate $R$ matrix for the
HSM pseudospin action. This covariance relation should be distinguished from
the generalised parameter-reversal Onsager--Casimir relation: it is the
weaker constraint associated with the surviving antiunitary $\Xi$, rather
than the familiar $\sigma_{ij}(\bs B)=\sigma_{ji}(-\bs B)$, that permits the
odd-in-$B_z$ behaviour of $\sigma_{zz}$ identified in Sec.~\ref{secexact}.
Switching on the tilt $v_0\,\bs \eta \cdot \bs k \, \mathbb{I}_{3 \times 3}$
preserves the $C_{\infty z}$ symmetry only for $\bs \eta \parallel \hat{\bs z}$,
where the mirror is broken instead. A tilt confined to the $k_x k_y$-plane
preserves the mirror at the cost of the $C_{\infty z}$ symmetry.

\section{Linear magnetoconductivity for collinear electromagnetic fields } 
\label{secexact}

In this section, we will compute the magnetoelectric conductivity adopting the methodology of semiclassical Boltzmann formalism \cite{mermin, sundaram99_wavepacket, li2023_planar, ips-kush-review, ips-rsw-ph, ips-shreya, timm}, considering collinear electric ($ \boldsymbol E  =  E\, \boldsymbol{\hat{z}}$) and magnetic ($\boldsymbol B  =  B\, \boldsymbol{\hat{z}}$) fields. In particular, we will not limit ourselves to the RTA and use the exact solutions, which have been applied already for various nodal-point semimetals successfully \cite{timm, ips-exact-spin1, ips-exact-kwn, ips-exact-rsw}. Hence, we do not repeat the derivation of the main formula, of which one can find very detailed explanations in those earlier works. Instead, we start with the linearised Boltzmann equation,
\begin{align}
\label{eqvec}
&  w^z_{\chi, s} ({\boldsymbol k})
+ e \, B \left [
{\boldsymbol \Omega}_{\chi, s} ({\boldsymbol k})
 \cdot {\boldsymbol{w}}_{\chi, s} ({\boldsymbol k})  \right ]
-\, 
e \, B \left [ {\boldsymbol{w}}_{\chi, s} ({\boldsymbol k}) \cross   
\boldsymbol{\hat z}   \right ] 
\cdot \nabla_{\boldsymbol k} {\Lambda}^z_{\chi, s} (\boldsymbol {k} )  
 = {\mathcal D}^{-1}_{\chi, s}  ({\boldsymbol k})
\sum \limits_{\tilde \chi, \tilde s}
\int_{k'}
\mathcal{M}^{\chi, \tilde \chi}_{ s, \tilde s} (\boldsymbol k,\boldsymbol k^\prime)
\left[ 
 {\Lambda}^z_{ \tilde \chi, \tilde s} (\boldsymbol {k}^\prime ) 
 -  {\Lambda}^z_{\chi, s} (\boldsymbol {k} )  
 \right ],
\end{align}
whose solution gives us the conductivity. The symbol $\int_k \equiv \int d^{3}  \boldsymbol k  \, {\mathcal D}^{-1}_{\tilde \chi, \tilde s} ({\boldsymbol k} )/ (2\,\pi)^3 $ stands for the three-dimensional integral in the momentum space, containing the modified phase-space factor due to the BC. Here, an index $\chi$ has been included to take into account systems with nodal-points with a dependence on the chirality. The various symbols appearing in the equation are explained below:
\begin{enumerate}

\item The phase-space volume element for the quasiparticles occupying a Bloch band gets modified by a nonzero BC via the factor of $
\left [{\mathcal D}_{ \chi, s }  (\boldsymbol k)\right]^{-1}$, where
\begin{align}
{\mathcal D}_{ \chi, s }  (\boldsymbol k) = \left [1 
+ e \,  \left \lbrace 
{\boldsymbol B} \cdot \boldsymbol{\Omega }_{ \chi, s }  (\boldsymbol k)
\right \rbrace  \right ]^{-1}.
\end{align}

\item A nonzero OMM gives rise to a Zeeman-like correction to the bare dispersion \cite{xiao_review} in the presence of a magnetic field, modifying it to
\begin{align}
\label{eqmodi}
& \xi_{ \chi, s } (\boldsymbol k) 
= \varepsilon_ s  (\boldsymbol k) + \varepsilon_{\chi, s}^{ (m) }  (\boldsymbol k) \, ,
\quad 
\varepsilon_{\chi, s}^{(m)}   (\boldsymbol k) 
= - \,{\boldsymbol B} \cdot \boldsymbol{m }_{\chi, s}  (\boldsymbol k) \,.
\end{align}
This, in turn, modifies the group-velocity as
\begin{align}
{\boldsymbol   w}_{ \chi, s } ({\boldsymbol k} ) \equiv 
 \nabla_{{\boldsymbol k}}   \xi_{ \chi, s } ({\boldsymbol k})
 = {\boldsymbol   v}_s ({\boldsymbol k} ) + {\boldsymbol  v}^{(m)}_{\chi, s} ({\boldsymbol k} ) \,,
\quad {\boldsymbol v}^{(m)}_{\chi, s} ({\boldsymbol k} )
= \nabla_{{\boldsymbol k}} \varepsilon_{\chi, s}^{(m)}   (\boldsymbol k) \,.
\end{align}
The effects of OMM show up via the modified energy appearing in the equilibrium Fermi-Dirac distribution,
$ f_0 \big (\xi_{\chi, s} (\boldsymbol k) , \mu, T \big )
= \left [ 1 + \exp \lbrace \, 
\left(  \xi_{\chi, s} (\boldsymbol k)-\mu \right) /T  \rbrace \right ]^{-1}$,
where $T $ is the temperature.
While using $f_0$ in various equations, we will be suppressing its $\mu$- and $ T $-dependence for uncluttering of notations. Moreover, in what follows, we will restrict our calculations to the $ T = 0$ limit.

\item $ I_{\text{coll}} [f_{\chi, s} ({\boldsymbol k}) ] $ symbolises the so-called collision integral, which comprises the relevant scattering processes trying to relax $f_{\chi, s} ({\boldsymbol k})$ towards $f_0 (\xi_{\chi, s}({\boldsymbol k}))$. 
For point-scattering mechanisms,
\begin{align}
I_{\text{coll}} [f_{\chi, s}({\boldsymbol k})]
 = \sum \limits_{\tilde \chi, \tilde s}
\int_{k'}
\mathcal{M}^{\chi, \tilde \chi}_{ s, \tilde s} (\boldsymbol k,\boldsymbol k^\prime)
\left[ f_{\tilde \chi , \tilde s} ( \boldsymbol k^\prime )  - f_{\chi, s} (\boldsymbol k) \right ].
\end{align}
Furthermore, for elastic and pseudospin-independent scattering, we have the simple expression of
\begin{align}
\mathcal{M}^{\chi, \tilde \chi}_{ s, \tilde s} (\boldsymbol k,\boldsymbol k^\prime) 
= \frac{2\, \pi \, \rho_{\rm imp} 
\, |{\mathcal V} ^{\chi, \tilde \chi}_{ s, \tilde s} |^2} 
{V} \,
\Big \vert \left \lbrace  \psi_{ \tilde \chi, \tilde s }({ \boldsymbol k^\prime }) \right \rbrace^\dagger 
\; \psi_{ \chi, s }({ \boldsymbol k}) \Big \vert^2 \,
 \delta \Big( \xi_{\tilde \chi , \tilde s } (\boldsymbol k^\prime)
 - \xi_{  \chi , s } (\boldsymbol k ) \Big) \,.
\end{align}
Here, $ \rho_{\rm imp}$ represents the impurity-concentration (acting as the scattering centres), $V$ denotes the system's volume, and
$ |{\mathcal V} ^{\chi, \tilde \chi}_{ s, \tilde s} |^2 $ stands for the scattering-strength (which in general parametrises intraband, interband, intranode, and internode processes).

\item We have parametrised the deviation of the quasiparticle-distribution from the equilibrium as
$$\delta f_{\chi, s} (\boldsymbol {k}) = -\,	 e\, 	\frac{\partial  f_0 (\xi_{\chi, s}({\boldsymbol k})) }
 {\partial \xi_{\chi, s} ({\boldsymbol k})} 
	\,   E \, {\Lambda}^z_{\chi, s} ( \boldsymbol {k} ) \,,$$
where $ \bm{\Lambda}_{\chi, s} (\boldsymbol {k} ) $ is the vectorial mean-free path. For our configuration, we get a nontrivial equation only for the $z$-component of $ \bm{\Lambda}_{\chi, s} ( \boldsymbol {k} ) $, viz. $ {\Lambda}^z_{   \chi,  s} (\boldsymbol {k} )   $.

\end{enumerate}
To solve the integro-differential equation represented by Eq.~\eqref{eqvec}, we take the self-consistent ansatz \cite{timm} that 
$ {\Lambda}^z_{\chi, s} \equiv {\Lambda}^z_{\chi, s} ( \mu, \gamma)$ at an energy $\mu$, which only depends on the polar angle, $\gamma$, and the chemical potential, $\mu$. This is because, for elastic scatterings, the integral over the full momentum space can be replaced by an integral over the Fermi surface at energy $\xi_{\chi , s} (\boldsymbol k) = \mu $, and the dependence on $\phi$ should not be there because of the rotational symmetry of the entire system in the $k_x k_y$-plane. 
Consequently, the momentum-space integrals reduce to the respective Fermi surfaces at energy $\xi_{\chi , s} 
(k_F^{\chi, s} , \gamma)= \mu $ with $T$ set to zero, denoted by the set of Fermi momenta, $ \lbrace k_F^{\chi, s} (\gamma) \rbrace $. The self-consistency can be easily checked from the fact that $\left [ {\boldsymbol \Omega}^\chi_{s} (\boldsymbol k) \cdot
 {\boldsymbol{w}}_{\chi, s} (\boldsymbol k)  \right ] $ is $\phi$-independent and $ \left [  {\boldsymbol{w}}_{\chi, s} 
 (\boldsymbol k)  \cross   \boldsymbol{\hat z}  \right ] 
\cdot \nabla_{\boldsymbol k} {\Lambda}^z_{\chi, s} (\mu, \gamma ) $ evaluates to zero. Thus, although the spinor overlaps might depend on $\phi$ and $\phi^\prime $, they will drop out on performing the integrations over these angles. Hence, we will use a $ \phi$- and $\phi^\prime$-integrated
overlap function, ${\mathcal T }^{\chi, \tilde \chi}_{ s, \tilde s} (\gamma, \gamma^\prime)$. Defining 
\begin{align}
\label{eq_lambda_mu}
\tau^{-1}_{\chi, s}(\mu, \gamma)
= \sum_{\tilde \chi , \tilde s} V \int_{k^\prime }
\mathcal{M}^{\chi, \tilde \chi}_{ s, \tilde s} (\boldsymbol k,\boldsymbol k^\prime) \,, \quad
h_{\chi, s} (\mu, \gamma) = {\mathcal D}_{\chi, s} ({\boldsymbol k})
 \left[ w^z_{\chi, s} ({\boldsymbol k})
+ e \, B \left \lbrace
{\boldsymbol \Omega}_{\chi, s} ({\boldsymbol k}) \cdot {\boldsymbol{w}}_{\chi, s} ({\boldsymbol k})
  \right \rbrace \right],	
\end{align}
Eq.~\eqref{eqvec} further reduces to
\begin{align}
\label{eqlambdamu}
& h_{\chi, s}(\mu, \gamma) + \frac{ {\Lambda}^z_{\chi, s}  (\mu, \gamma)} 
{\tau_{\chi, s}(\mu, \gamma)} =
\sum_{\tilde \chi , \tilde s}  
\frac{ \rho_{\rm imp} 
\, |{\mathcal V} ^{\chi, \tilde \chi}_{ s, \tilde s} |^2 }
{ 2\, \pi }
\int d\gamma^\prime \, \frac{\mathcal J \left (k^\prime \right )
\, {\mathcal D}^{-1}_{\tilde \chi, \tilde s} ({\boldsymbol k}^\prime)}
{  \big | \partial \xi_{\chi, s}({\boldsymbol k}^\prime) /
 \partial \mathfrak{R}^\prime  \big | }
\, {\mathcal T }^{\chi, \tilde \chi}_{ s, \tilde s} (\gamma, \gamma^\prime)
\, {\Lambda}^z_{ \tilde \chi, \tilde s} (\mu, \gamma^\prime ) 
\Big \vert_{ k^\prime = k_F^{ \tilde \chi, \tilde s} } \,.
\end{align}
The factor $ \mathcal J $ arises as the Jacobian for switching to the toroidal or spherical-polar coordinates, depending on whether we are dealing with the nodal-rings or BC-dipole nodal-point. Here, $\mathfrak{R}^\prime $ denotes whichever radial coordinate is being held fixed while $\gamma^\prime $ varies along the Fermi surface, viz. $\kappa^\prime $ for the nodal-rings and $k^\prime $ for the BC-dipole nodal points, so that the derivative $\big | \partial \xi_{\chi, s}(\boldsymbol k^\prime) / \partial \mathfrak{R}^\prime \big |^{-1} $ arises from converting $ \delta \Big( \xi_{\tilde \chi , \tilde s } (\boldsymbol k^\prime) -\mu \Big) $ to
$\delta (\kappa^\prime - k_F^{ \tilde \chi, \tilde s})$ (for the nodal-rings) or $\delta (k^\prime - k_F^{ \tilde \chi, \tilde s})$ (for the nodal points), with $k_F^{  \chi, s}$ parametrising the radius of the Fermi surface as measured from the gapless ring (i.e., the locus of the band-touching curve) or the band-touching point. The final step in solving the still complicated-looking integro-differential equation is to parametrise ${\Lambda}^z_{\chi, s} (\mu,\gamma )$ by a meaningful ansatz. Since Eq.~\eqref{eqlambdamu} is linear in $ {\Lambda}^z_{\tilde \chi, \tilde s} (\mu,\gamma^\prime )$, the ansatz must be chosen such that the $\gamma^\prime$-integral of $ {\mathcal T }^{\chi, \tilde \chi}_{ s, \tilde s} (\gamma, \gamma^\prime) \, {\Lambda}^z_{\tilde \chi, \tilde s} (\mu,\gamma^\prime )$ reproduces only the harmonics already present in the ansatz, which will depend on the eigenspinors of the respective systems. This condition alone leaves an overall constant undetermined, since $\tau^{-1}_{\chi,s} (\mu,\gamma)$ itself lies within the span of these harmonics, so that a constant shift of $ {\Lambda}^z_{\chi, s} $ solves the homogeneous version of Eq.~\eqref{eqlambdamu}. This remaining freedom is precisely what the electron-number conservation, $\int_k \delta f_{\chi, s} (\boldsymbol k)=0$, fixes.
The conductivity is then determined by the expression,
\begin{align}
\label{eqsigmazzexact}
\sigma_{zz} =  -\,\frac{e^2 } { V } 
\sum_{\chi,s}
\int \frac{d^3 {\boldsymbol k}} {(2\, \pi)^3} 
	\left[   w^z_{\chi, s} (\boldsymbol k)
+ e \, B  \left \lbrace \boldsymbol \Omega_{\chi, s} (\boldsymbol k)
\cdot  \boldsymbol{w}_{\chi, s} (\boldsymbol k) \right  \rbrace   \right] 
\delta \big (\xi_{\chi, s} (\boldsymbol k)-\mu \big) \, {\Lambda}^z_{\chi, s} (\mu, \gamma )	\,. 
\end{align} 
That $\sigma_{zz}$ is the only symmetry-allowed component of the linear
response follows from the $C_{\infty z}$ symmetry about $\hat{\bs z}$
identified in Sec.~\ref{secmodels} for both the untilted VNR and the
untilted HSM. Since $\bs E$ and $\bs B$ are both directed along
$\hat{\bs z}$, the only axis singled out by the applied fields coincides
with the rotational axis of the Hamiltonian. Any component mixing $z$
with $x$ or $y$ is forbidden, by the same argument that removes such
mixed components from the ECR tensor in Sec.~\ref{secnonlinear}.

In the following subsections, we will discuss the features of the quantity, 
\begin{align}
\delta \sigma_{zz}  \equiv \sigma_{zz} (B) / \sigma_{zz}(B=0)  -1 \,,
\end{align} 
obtained through numerics. We will consider the features for a single VNR and a single BC-dipole, which feature nonzero BC and OMM without introducing a gap (unlike the PTNR). Thus, we will drop the index $\chi $ in what follows. For each case, on applying a chemical potential $\mu > 0$, only the $s=2$ band will participate in transport, which means only $|{\mathcal V}_{ 2,2} |^2$ matters (caused by intraband-scattering processes), which we parametrise as
\begin{align}
 |{\mathcal V}_{ 2,2} |^2 =  \frac{ 2\,\pi} {\rho_{\rm imp}} \, \beta_{\rm intr} \,.
\end{align} 

Numerical solutions to the final equations will confirm the coexistence of odd powers of $B_z$ alongside the usual even powers of $B_z$ in the net conductivity. This is predictable based on physical arguments resting on two separate ingredients. The first is that the ordinary Onsager--Casimir relation
$\sigma_{zz}(\bs B)=\sigma_{zz}(-\bs B)$, which would forbid the odd
term, does not hold here, because standalone time-reversal symmetry is
absent in both models. It is replaced by the weaker $\Xi$-covariant
relation identified in Sec.~\ref{secmodels}. For the collinear
configuration considered here, with $\bs B$ and $\bs\eta$ both directed
along $\hat{\bs z}$, this relation reduces to an identity and imposes no
constraint on the $B_z$-dependence of $\sigma_{zz}$. Odd powers of $B_z$
are therefore symmetry-allowed. The second is that whether such a
term actually appears is decided separately, by the parity of the
momentum-space integrand under $k_z\to-k_z$. To establish the nonvanishing odd-in-$B_z$ response, it is sufficient to examine the parity of the integrand under the mirror operation $k_z\to -k_z$. For the collinear configuration dealt with here, both the Hamiltonian and the magnetic field are invariant under this transformation, ensuring that only terms that are even in $k_z$ survive the momentum integration.
For a VNR or HSM, both $w^z_{\chi,s}\propto k_z$ and $(\bs\Omega_{\chi,s}\!\cdot\!\bs w_{\chi,s})$ are odd under $k_z\to -k_z$, the latter owing to the explicit $k_z$ factor in the BC [cf. Eqs.~\eqref{eqbcomm} and \eqref{eqbcommhopf}]. Their product is therefore even and contributes to $\sigma_{zz}$. The same parity structure persists for the OMM-induced terms, since $\varepsilon_m\propto B_z k_z^2$ is even in $k_z$. Consequently, the coefficients of odd powers of $B_z$ are symmetry-allowed and remain finite after angular integration. For a BC monopole, the situation is reversed. The isotropic BC does not supply the additional $k_z$ factor and $w^z_{\chi,s}(\bs\Omega_s\!\cdot\!\bs w_s)$ remains odd under $k_z\to -k_z$, vanishing upon integration over the mirror-symmetric Fermi surface. Likewise, the OMM correction $\varepsilon_m\propto B_z k_z/k^2$ is odd in $k_z$, causing the corresponding odd-in-$B_z$ contributions to disappear. The emergence of odd magnetoconductivity in the VNR and HSM thus follows directly from the dipolar $k_z$ texture of their BC and OMM, which restores the even parity required for a nonzero angular average.

\subsection{Vortex nodal-ring}
\label{secrank}

Using the eigenspinors of the tiltless Hamiltonian of Eq.~\eqref{eqham}, the overlap-function for a single node turns out to be
\begin{align}
\label{eqoverlap-vnr}
{\mathcal T }_{ 2,2} (\gamma, \gamma^\prime) & = 
\frac{ 1 +  \cos \gamma \cos \gamma^\prime} {2}\,.
\end{align}
This dictates the ansatz,
\begin{align}
 {\Lambda}^z_{2} (\mu,\gamma ) & = 
\tau_{2}(\mu,\gamma)
 \left [ - h_{2} (\mu, \gamma) +\lambda + a \, \cos \gamma  \right ],
\end{align}
corresponding to the positive-energy band. This leaves us with the problem of solving for the 2 unknown
coefficients, $\lbrace \lambda , \, a    \rbrace $. 
The terms $\lambda+a\,\cos\gamma$ in the ansatz constitute the vertex correction, and its meaning is best understood by separating the roles of the two parts of the collision integral. The current vertex is the effective velocity carried by a quasiparticle at $\bs k$ in response to the electric field. Its bare part is $-\,\tau_2\,h_2$, which is built from the driving term $h_2$ alone and contains the band velocity, the anomalous velocity arising from the BC, and the phase-space factor $\mathcal D_2$. The out-scattering rate $1/\tau_2$ is determined by the self-energy due to the impurities, and it measures how quickly a quasiparticle at $\bs k$ is scattered away from its state. The impurities, however, also scatter quasiparticles from $\bs k^\prime$ into $\bs k$. These carry with them the non-equilibrium shift of their distribution at $\bs k^\prime$, which is encoded in $\Lambda^z_2(\mu,\gamma^\prime)$. This in-scattering term of Eq.~\eqref{eqlambdamu} feeds the distribution at $\bs k^\prime$ back into that at $\bs k$, and the resulting addition to the bare vertex is the vertex correction.

In the diagrammatic language of the Kubo formalism, the bare vertex is dressed by a ladder of impurity lines. Within the Born approximation, each rung of the ladder is a single impurity line connecting $\bs k^\prime$ to $\bs k$, with the weight $\mathcal T_{2,2}(\gamma,\gamma^\prime)$. Summing the ladder to all orders yields the self-consistent equation for the dressed vertex, which is precisely Eq.~\eqref{eqlambdamu}. The overlap function plays the central role here, since it is the squared overlap of the spinors, and it encodes the matrix element of the impurity potential between the Bloch states at $\bs k$ and $\bs k^\prime$. It makes the scattering anisotropic, and the dependence on $\gamma$ and $\gamma^\prime$ through the harmonics of $\mathcal T_{2,2}$ determines which harmonics of $\Lambda^z_2$ are corrected. An overlap function that is even under the mirror operation $\gamma\to-\gamma$, which corresponds to $k_z\to-k_z$, gives no correction to a vertex that is odd under it, since the in-scattering integral of an odd function over a weight that is even then vanishes.

\begin{figure*}[t!]
\subfigure[]{\includegraphics[width= 0.41 \textwidth]{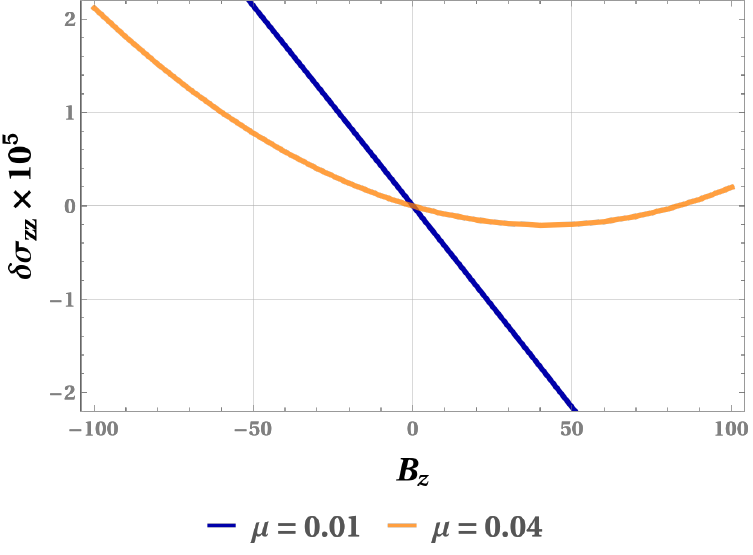}} \hspace{ 1 cm }
\subfigure[]{\includegraphics[width= 0.4 \textwidth]{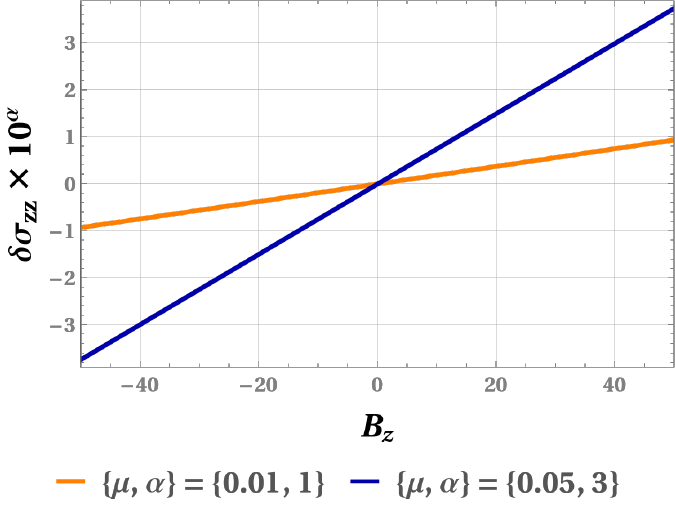}}
\caption{\label{figexact} Behaviour of $\delta \sigma_{zz}$ as a function of $B_z = \pm |\bs B|$ (in eV$^2$), setting $v_0 = 0.0004$ and $\beta_{\rm intr}=1$, and $k_0=200$ eV for the VNR. While subfigure (a) represents the characteristics of a VNR, subfigure (b) depicts the response of a single-node case of an HSM. The curves in subfigure (b) are multiplied by $10^\alpha$, with the value of $\alpha$ given in the legend.}
\end{figure*}

The vertex correction therefore describes the fraction of the momentum, imparted by the electric field to the quasiparticles, that is retained through the scattering events, instead of being lost at each of them. In the RTA, the in-scattering term is dropped altogether, which amounts to assuming that the momentum is lost completely at every event. This approximation is exact only if the correction vanishes. The analogous distinction in the simplest case of an isotropic band is the one between the quantum lifetime and the transport time, where the factor $1-\cos\theta$ in the scattering rate accounts for the retained forward momentum. Plugging in the ansatz in Eq.~\eqref{eqlambdamu} furnishes
the linear equations, which can be written as a matrix equation of the form
\begin{align}
\label{eqmatrix0}
\mathcal A \, \mathcal C = \mathcal H \,, \text{ where }
\mathcal C =\begin{bmatrix}
\lambda & a 
\end{bmatrix}^{\rm T}\,.
\end{align}
Additionally, the electron-number conservation furnishes the constraint of
$\int_k  \delta f_{2} (\boldsymbol {k}) = 0 $, which fixes the amplitude of a zero mode left undetermined by Eq.~\eqref{eqmatrix0} on its own. 
The matrix $\mathcal A$ is singular because the collision operator conserves the electron number, as discussed in Appendix~\ref{appsing} for all the cases of this section. For the VNR, the matrix $\mathcal A$ has rank 1 out of 2, with a one-dimensional kernel, and the number constraint fixes the amplitude of the zero mode. The solvability condition of Eq.~\eqref{eqmatrix0} is that the pumping rate $\int d\gamma\,\mathcal W\,h_2$ vanishes. For the VNR, the mirror symmetry $k_z\to-k_z$ makes $\mathcal W$, $\tau_2$ and the harmonics $1$ and $\cos\gamma$ even under $\gamma\to-\gamma$, whereas $h_2$ is odd. Both components of the source in Eq.~\eqref{eqmatrix0} therefore vanish, and the number constraint then sets $\lambda=a=0$. The vertex correction is thus absent for the VNR in collinear fields, which leaves $\Lambda^z_2=-\,\tau_2\,h_2$ with the anisotropic $\tau_2(\mu,\gamma)$.
Fig.~\ref{figexact}(a) illustrates the conductivity behaviour setting $v_0 = 0.0004$, $\beta_{\rm intr} = 1 $, and $k_0=200$ eV. We vary $\mu$ and show the behaviour for $\mu=0.01$ and $\mu=0.04$. The linear-in-$B_z$ term dominates for $\mu=0.01$. For $\mu=0.04$, the linear and quadratic terms are comparable over the plotted range, and the quadratic term takes over at the largest values of $|B_z|$.

\subsection{Single isolated node of Hopf semimetal}
\label{secsinglehopf}

Using the eigenspinors of the tiltless Hamiltonian of Eq.~\eqref{eqbcd2}, the overlap-function for a single node turns out to be
\begin{align}
\label{eqoverlap-bcd}
{\mathcal T }_{ 2,2} (\gamma, \gamma^\prime) & = 
\frac{ 3 + 4 \cos \gamma \cos \gamma^\prime + \cos (2 \gamma ) \cos (2 \gamma^\prime)  } {8}\,.
\end{align}
This dictates the ansatz,
\begin{align}
 {\Lambda}^z_{2} (\mu,\gamma ) & = 
\tau_{2}(\mu,\gamma)
 \left [ - h_{2} (\mu, \gamma) +\lambda + a \, \cos \gamma + b \,\cos(2\gamma)  \right ],
\end{align}
corresponding to the positive-energy band. This leaves us with the problem of solving for the 3 unknown
coefficients, $\lbrace \lambda , \, a  , \, b  \rbrace $. Plugging in the ansatz in Eq.~\eqref{eqlambdamu} furnishes
the linear equations, which can be written as a matrix equation of the form
\begin{align}
\label{eqmatrix}
\mathcal A \, \mathcal C = \mathcal H \,, \text{ where }
\mathcal C =\begin{bmatrix}
\lambda & a  & b 
\end{bmatrix}^{\rm T}\,.
\end{align}
Additionally, the electron-number conservation furnishes the constraint of
$\int_k  \delta f_{2} (\boldsymbol {k}) = 0 $, which fixes the amplitude of a zero mode left undetermined by Eq.~\eqref{eqmatrix} on its own. 
The matrix $\mathcal A$ of Eq.~\eqref{eqmatrix} is singular, with a one-dimensional kernel and rank 2, for the reasons discussed in Appendix~\ref{appsing}. The solvability condition that the pumping rate $\int d\gamma\,\mathcal W\,h_2$ vanishes is satisfied by parity.

The number constraint can be understood from parity. The quantities $\mathcal D_2$, $\varepsilon_m$, $\bs\Omega_2$ and $k_F(\gamma)$, which depends only on $\cos^2\gamma$, are even under $\bs k\to-\bs k$, whereas the band velocity $\bs v_2$ and the OMM-induced velocity $\bs v^{(m)}_2=\nabla_{\bs k}\varepsilon_m$ are odd. Hence $w^z_2$ and $\bs\Omega_2\cdot\bs w_2$ are odd, and $h_2$ is odd under $\gamma\to\pi-\gamma$. This holds for the full expression and no expansion in $B_z$ is involved. Since $h_2$ is $\cos\gamma$ times a function of $\cos^2\gamma$, it is a sum of odd harmonics $\cos[(2\,n+1)\,\gamma]$. The explicit form of $h_2$ and its expansion in powers of $B_z$ are given in Eq.~\eqref{eqh-hopf} below. The argument that follows uses only the parity of $h_2$ and does not depend on its harmonic content. The integral of $h_2$ against any weight that is even under $\gamma\to\pi-\gamma$ vanishes to all orders in $B_z$. This includes the pumping rate and the moments $M_0$ and $M_2$ defined below, in analogy with the RTA \cite{ips-firdous-vnr}. With the ansatz above, the left-hand side of Eq.~\eqref{eqlambdamu} equals $\lambda+a\cos\gamma+b\cos(2\gamma)$, since $h_2$ cancels. This cancellation is exact and independent of the harmonic content of $h_2$. The right-hand side contains the three harmonics of the overlap function, and these alone determine the closure of the ansatz, which gives
\begin{align}
\label{eqharm-bcd}
\lambda=\frac{3}{8}\,\beta_{\rm intr}\,M_0\,, \quad
a=\frac{1}{2}\,\beta_{\rm intr}\,M_1\,, \quad
b=\frac{1}{8}\,\beta_{\rm intr}\,M_2\,,
\end{align}
where $M_n=\int d\gamma^\prime\,\mathcal W(\gamma^\prime)\,\cos(n\,\gamma^\prime)\,\Lambda^z_2(\mu,\gamma^\prime)$. Here $\mathcal W=\mathcal J\,\mathcal D^{-1}/|\partial\xi/\partial\mathfrak{R}^\prime|$ is the weight of Eq.~\eqref{eqlambdamu} at $k^\prime=k_F$, which is even under $\gamma\to\pi-\gamma$, and so is $\tau_2$. The moments $M_0$ and $M_2$ involve only the even part of $\Lambda^z_2$, which is built from $\lambda$ and $b$ alone. They are therefore homogeneous and $\lambda=b=0$ solves them. This solution is the one selected by the number constraint, which removes the zero mode of the even sector, since an odd $\Lambda^z_2$ integrates to zero. In contrast, $M_1$ receives the contribution $-\int d\gamma^\prime\,\mathcal W\,\tau_2\,\cos\gamma^\prime\,h_2$ from the projection of $h_2$ onto $\cos\gamma$ with the weight $\mathcal W\,\tau_2$. At $B_z=0$, this contribution reduces to $-\,v_0\int d\gamma^\prime\,\mathcal W\,\tau_2\,\cos^2\gamma^\prime$ and is nonzero, so that $a$ is sourced. It is the vertex correction of the odd sector, generated by the in-scattering part of the collision integral. In the RTA, this part is dropped, which corresponds to $a=\lambda=b=0$ and $\Lambda^z_2=-\,\tau\,h_2$ with a constant $\tau$. The odd-in-$B_z$ terms are already present in that case through $h_2$ \cite{ips-firdous-vnr}, and $a$ renormalises them.

Reading off $\bs m_2$ from Eq.~\eqref{eqbcommhopf}, the Zeeman-like shift of Eq.~\eqref{eqmodi} modifies the dispersion of the $s=2$ band to
$\xi_2 (\boldsymbol k) = v_0 \, k + e \, v_0 \, B_z \, \cos^2 \gamma /(2\, k)$, and solving $\xi_2 = \mu $ for the Fermi momentum gives
\begin{align}
k_F =  \frac{ \mu + \sqrt{\mu^2- 2\, e\, v_0^2\, B_z \cos^2 \gamma } } {2\, v_0}\,.
\end{align}
We keep the root carrying the plus sign before the square-root, since it is the one which reduces to the known result $k_F = \mu /v_0 $ at $B_z=0$. The other root, ${ \big ( \mu - \sqrt{\mu^2- 2\, e\, v_0^2\, B_z \cos^2 \gamma } \big ) } /(2\, v_0)$, vanishes at $B_z=0$ and, for the weak fields $e \,B_z\, v_0^2 \ll \mu^2$ considered here, scales as $e \,v_0\, B_z \cos^2\gamma/(2 \,\mu)$. It is therefore parametrically smaller than $\mu/v_0$ by the weak-field parameter, $e \,B_z\, v_0^2/\mu^2$, and lies outside the range of validity of the low-energy expansion. Reality of the retained root requires $2\, e\, v_0^2\, B_z \le \mu^2$ for $B_z>0$, with no such restriction for $B_z<0$. With the retained root, the quantity $h_2$ can be written in closed form. We define $u=e\,B_z/k_F^2(\gamma)$, where $k_F$ depends on $\gamma$ only through $\cos^2\gamma$. For $\bs B=B_z\,\hat{\bs z}$, the definition of $h_2$ in Eq.~\eqref{eq_lambda_mu}, together with the BC and the OMM of Eq.~\eqref{eqbcommhopf}, then gives
\begin{align}
\label{eqh-hopf}
h_2(\mu,\gamma)=v_0\,\cos\gamma\;\frac{1+2\,u-\frac{1}{2}\,u\,(3+u)\cos^2\gamma}{1+u\,\cos^2\gamma}\,.
\end{align}
Expanding $h_2$ in powers of $B_z$ at fixed $\mu$, with the dependence of $k_F(\gamma)$ on $B_z$ included, the coefficients of $B_z^0$, $B_z^1$ and $B_z^2$ are odd polynomials in $\cos\gamma$ of degrees 1, 3, and 3, respectively. The degree for $B_z^2$ is 3 and not 5, because the $B_z$-dependence of $k_F(\gamma)$ cancels the $\cos^5\gamma$ term that $u$ alone would generate at this order. Those of $B_z^3$, $B_z^4$ and $B_z^5$ are odd polynomials of degrees 7, 9 and 11. The harmonics $\cos\gamma$ and $\cos (3\gamma)$ thus describe $h_2$ exactly through $\mathcal O(B_z^2)$. The harmonics $\cos (5 \gamma)$, $\cos (7 \gamma)$ and higher-degree odd harmonics enter from $\mathcal O(B_z^3)$ onwards.

Fig.~\ref{figexact}(b) illustrates the conductivity behaviour setting $v_0 = 0.0004$ and $\beta_{\rm intr}=1 $. For the chosen parameters, the response contains terms linear in $B_z$, and the linear parts dominate. On increasing the value of $\mu$, the magnitude of the response is seen to decrease.

\subsection{Two nodes of Hopf semimetal with oppositely-oriented dipole axes}

For a pair of conjugate Hopf nodes, the key geometric distinction lies in the orientation of BC-dipole axis. Taking the so-called valley-HSMs~\cite{graf-hopf} (containing an even number of Hopf nodes), the dipole-axis vectors ${\boldsymbol{\xi}}_\chi$ ($\chi=\pm1$) associated with two nodal points, point in opposite directions, i.e., $ {\boldsymbol{\xi}}_\chi=-{\boldsymbol{\xi}}_{-\chi} $. Consequently, the BC distribution around one node is the negative of that around its conjugate partner. In the valley-Hopf terminology of Ref.~\cite{graf-hopf}, this corresponds to the Berry dipole pointing upward in one valley and downward in the other. The OMM inherits the same reversal property [cf. Eq.~\eqref{eqbcommhopf}].  Taking the node with $\chi=+1$ as the reference node and choosing $ {\boldsymbol{\xi}}_{+1}=\hat{\boldsymbol z}, $ we write $ {\boldsymbol \Omega}_{\chi, s}=\chi\,{\boldsymbol \Omega}_s. $ The conjugate node then has $ {\boldsymbol{\xi}}_{-1}=-\hat{\boldsymbol z}, $ which implies
\begin{align} 
{\boldsymbol \Omega}_{\chi, s} = -\,{\boldsymbol \Omega}_{-\chi, s}, 
\quad {\boldsymbol m}_{\chi, s} = -\,{\boldsymbol m}_{-\chi, s}. 
\end{align}

\begin{figure*}[t!]
\includegraphics[width= 1 \textwidth]{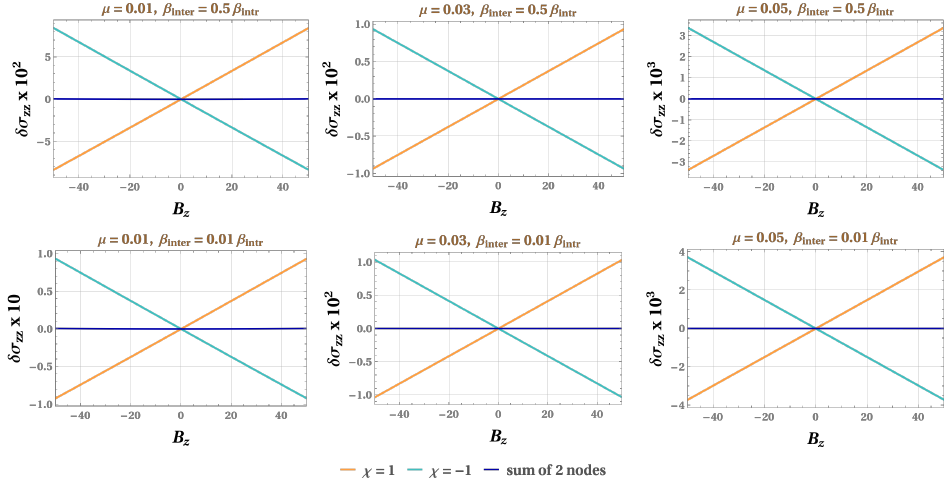}
\caption{\label{figexact2}
Behaviour of $\delta \sigma_{zz}$ for a two-node Hopf semimetal with oppositely-oriented Berry-curvature dipoles. The flat-band does not contribute, and only the $s=2$ band at each node participates in transport for $\mu>0$. The orange and cyan curves show the separate contributions from the two nodes, and the dark blue curve shows their sum. The columns correspond to $\mu=0.01$, $0.03$ and $0.05$, and the rows to $\beta_{\rm inter}/\beta_{\rm intr}=0.5$ and $0.01$.}
\end{figure*}

Using the eigenspinors of Eq.~\eqref{eqbcd2}, the $\phi$- and $\phi^\prime$-averaged overlap function for the $s=2$ bands (considering $\mu > 0$) of the two nodes is obtained as
\begin{align}
\label{eqoverlap-bcd-2node}
\mathcal T_{2,2}(\gamma,\gamma^\prime)
& = \frac{3 + 4\, \cos\gamma \cos\gamma^\prime
+ \cos(2\gamma) \cos(2\gamma^\prime)}{8} \,.
\end{align}
This expression is independent of $\chi $, unlike the monopole-node cases studied earlier \cite{timm, ips-exact-spin1, ips-exact-kwn, ips-exact-rsw}. For this system, we need to formulate the ansatz as
\begin{align}
\label{eqansatz-2node}
\Lambda^z_{\chi,2}(\mu,\gamma) & = \tau_{\chi,2}(\mu,\gamma)
\left[ -\, h_{\chi,2}(\mu,\gamma) + \lambda_\chi + a_\chi \cos\gamma + b_\chi \cos(2\gamma) \right],
\end{align}
where
\begin{align}
h_{\chi,2}(\mu,\gamma)
& = \mathcal D_{\chi,2}(\boldsymbol k)
\left[ w^z_{\chi,2}(\boldsymbol k)+ e \,B \left\{ \boldsymbol\Omega_{\chi,2}(\boldsymbol k)
\cdot \boldsymbol w_{\chi,2}(\boldsymbol k) \right\} \right].
\end{align}
Plugging this ansatz into Eq.~\eqref{eqlambdamu} furnishes six linear equations for the unknown coefficients $\{\lambda_\chi, a_\chi, b_\chi\}$, which can be written as the matrix equation
\begin{align}
\label{eqmatrix-2node}
\mathcal A \, \mathcal C = \mathcal H \,, \quad
\mathcal C =
\begin{bmatrix}
\lambda_+ & a_+ & b_+ & \lambda_- & a_- & b_-
\end{bmatrix}^{\rm T} \,.
\end{align}
The electron-number conservation,
\begin{align}
\sum_{\chi=\pm1} \int_k \delta f_{\chi,2}(\boldsymbol k)=0 \,,
\end{align}
must be included for consistency.

As for the single node, $h_{\chi,2}$ is odd under $\gamma\to\pi-\gamma$ for the full expression. The $\chi$-dependence enters only through the product $\chi\,B_z$ in quantities that are even under $\bs k\to-\bs k$, and it does not affect this parity. The node $\chi=+1$ coincides with the single node, and the relations $\bs\Omega_{\chi,2}=-\bs\Omega_{-\chi,2}$ and $\bs m_{\chi,2}=-\bs m_{-\chi,2}$ then imply that $h_{\chi,2}(\mu,\gamma)$ equals the single-node expression of Eq.~\eqref{eqh-hopf} with $B_z$ replaced by $\chi\,B_z$. Its harmonic content is therefore that of the single node, with $B_z^n$ replaced by $\chi^n\,B_z^n$. Its integral against any weight that is even under $\gamma\to\pi-\gamma$ vanishes to all orders in $B_z$. This is the analogue of the vanishing of the charge-pumping vector $\bs{\mathcal I}^\chi$ in the RTA \cite{ips-firdous-vnr}. The overlap function of Eq.~\eqref{eqoverlap-bcd-2node} preserves this parity, since only the term $4\,\cos\gamma\,\cos\gamma^\prime$ is odd, and it does not mix with the even terms. The even sector $\{\lambda_\chi,b_\chi\}$ is thus never sourced and vanishes, whereas $a_\chi$ is sourced by the projection of $h_{\chi,2}$ onto $\cos\gamma$ with the weight of Eq.~\eqref{eqlambdamu} multiplied by $\tau_{\chi,2}$.

For the collision integral, we parametrise the intranode and internode scattering strengths as
\begin{align}
\big|{\mathcal V}^{\chi,\chi}_{2,2}\big|^2
= \frac{2\,\pi}{\rho_{\rm imp}}\,\beta_{\rm intr} \,,\quad
\big|{\mathcal V}^{\chi,-\chi}_{2,2}\big|^2
= \frac{2\,\pi}{\rho_{\rm imp}}\,\beta^{\rm inter}_{2,2} \,.
\end{align}
We write $\beta_{\chi\tilde\chi}$ compactly for these strengths, with $\beta_{\chi\chi}=\beta_{\rm intr}$ and $\beta_{\chi,-\chi}=\beta^{\rm inter}_{2,2}$. The relations of Eq.~\eqref{eqharm-bcd} generalise to
\begin{align}
\lambda_\chi=\frac{3}{8}\sum_{\tilde\chi}\beta_{\chi\tilde\chi}\,M^{\tilde\chi}_0\,, \quad
a_\chi=\frac{1}{2}\sum_{\tilde\chi}\beta_{\chi\tilde\chi}\,M^{\tilde\chi}_1\,, \quad
b_\chi=\frac{1}{8}\sum_{\tilde\chi}\beta_{\chi\tilde\chi}\,M^{\tilde\chi}_2\,,
\end{align}
where $M^{\tilde\chi}_n$ is defined as in the single-node case for the node $\tilde\chi$. The coefficients $a_+$ and $a_-$ are coupled through $\beta^{\rm inter}_{2,2}$, which thus acts as an ordinary additional scattering channel for the vertex correction.

The vanishing of $\{\lambda_\chi,b_\chi\}$ also follows from the continuity equation of each node. Multiplying Eq.~\eqref{eqlambdamu} by $\mathcal W_\chi(\gamma)$ and integrating over $\gamma$ gives the rate equation for the number of electrons at the node $\chi$. The driving term contributes the pumping rate $\int d\gamma\,\mathcal W_\chi\,h_{\chi,2}$, which vanishes by parity. The intranode part of the collision integral drops out, because the scattering is elastic and $\mathcal T_{2,2}(\gamma,\gamma^\prime)$ is symmetric under $\gamma\leftrightarrow\gamma^\prime$. The internode part gives the rate of transfer to the other node, which is proportional to $\beta^{\rm inter}_{2,2}\int d\gamma\,d\gamma^\prime\,\mathcal W_\chi(\gamma)\,\mathcal W_{-\chi}(\gamma^\prime)\,\mathcal T_{2,2}(\gamma,\gamma^\prime)\left[\Lambda^z_{\chi,2}(\gamma)-\Lambda^z_{-\chi,2}(\gamma^\prime)\right]$. Since $\mathcal W$ is even and the term $4\,\cos\gamma\,\cos\gamma^\prime$ of $\mathcal T_{2,2}$ averages to zero against it, the odd part of $\Lambda^z_{\chi,2}$ does not contribute to this rate. Only the even part $\tau_{\chi,2}\left[\lambda_\chi+b_\chi\cos(2\gamma)\right]$ can transfer electrons between the nodes. Since the pumping rate vanishes, the homogeneous equations of the even sector have only the trivial solution once the number constraint is imposed. The continuity equation of each node is thus satisfied by itself, with neither pumping nor transfer. Consequently, the number constraint holds node by node and not only for the sum, and no charge is pumped between the nodes. The net $\delta\sigma_{zz}$ still depends on the ratio $\beta^{\rm inter}_{2,2}/\beta_{\rm intr}$, as seen in Fig.~\ref{figexact2}.

The matrix $\mathcal A$ of Eq.~\eqref{eqmatrix-2node} is singular as well, with rank 5 for $\beta^{\rm inter}_{2,2}\neq0$ and rank 4 for $\beta^{\rm inter}_{2,2}=0$, as discussed in Appendix~\ref{appsing}.
Reading off $\boldsymbol m_{\chi,2}=\chi\,\boldsymbol m_2$ from Eq.~\eqref{eqbcommhopf}, the Zeeman-like shift of Eq.~\eqref{eqmodi} modifies the dispersion of the $s=2$ band at the $\chi$-node to
\begin{align}
\label{eqxi-2node}
\xi_{\chi,2}(\boldsymbol k)
= v_0\,k + \frac{\chi\, e\, v_0\, B_z \cos^2\gamma}{2\,k} \,.
\end{align}
Solving $\xi_{\chi,2}=\mu$ for the Fermi momentum gives
\begin{align}
\label{eqkF-2node}
k_F^{\chi}
= \frac{\mu + \sqrt{\mu^2 - 2\,\chi\, e\, v_0^2\, B_z \cos^2\gamma}}{2\,v_0} \,.
\end{align}
We retain the root carrying the plus sign before the square-root, since it is the one which reduces to the known result $k_F=\mu/v_0$ at $B_z=0$. Reality of the retained root now requires $2\,e\,v_0^2\,|B_z| \le \mu^2$, restricting the allowed values of $B_z$ in both directions of the $B_z$-axis, because of the presence of the $\chi \, B_z$ term inside the square-root.
Finally, the net longitudinal magnetoconductivity for the two-node Hopf semimetal is
obtained from Eq.~\eqref{eqsigmazzexact} by summing over $\chi$.
The resulting behaviour of $\delta \sigma_{zz}$ is shown in Fig.~\ref{figexact2} for some distinct values of $\mu$ and $\beta_{\rm inter}$, with $v_0=0.0004$ and $\beta_{\rm intr} = 1$ held fixed. The curves have been illustrated for the $s=2$ bands at the 2 nodes as well as for their sum. For the chosen parameters, the curves of the individual nodes contain linear-in-$B_z$ terms with opposite slopes, and the linear parts dominate. The two curves are mirror images of each other, with the $\chi=+1$ curve at $B_z$ coinciding with the $\chi=-1$ curve at $-B_z$. This follows since $\xi_{\chi,2}$ depends on $\chi$ and $B_z$ only through the product $\chi\,B_z$ and the overlap function is independent of $\chi$. The sum over the two nodes is consequently even in $B_z$, so that the odd-in-$B_z$ terms cancel in the net response. The net response is therefore much smaller than the contribution of each node for these parameters. The ratio $\beta_{\rm inter}/\beta_{\rm intr}$ changes the linear terms of each node only slightly. Analogous to the single-node scenario, the magnitude of the response decreases for each case as we crank up $\mu$.

It is instructive to contrast these findings with the BC-monopole nodes of Refs.~\cite{timm, ips-exact-spin1, ips-exact-kwn, ips-exact-rsw}. For a BC-monopole node, the overlap function is a polynomial in $\cos\theta$ and $\cos\theta^\prime$, whose degree grows with the pseudospin. It is $\left(1+\chi\,\chi^\prime\cos\theta\cos\theta^\prime\right)/2$ for a pair of Weyl nodes \cite{timm} and for a Kramers-Weyl node \cite{ips-exact-kwn}, where the corresponding band index replaces $\chi$ in the latter. It contains terms up to $\cos^3\theta\,\cos^3\theta^\prime$ for the fourfold RSW node \cite{ips-exact-rsw}, where the intranode amplitudes $\beta^{1,1}_{\rm intr}$, $\beta^{3,3}_{\rm intr}$ and the internode amplitude $\beta_{\rm inter}$ multiply three different polynomials. In every one of these cases, the overlap function contains a $\theta$-independent term, as well as harmonics of both parities. The ansatz for $\Lambda^z$ therefore has to carry the constant $\lambda$ along with the full tower of harmonics, and the accompanying matrix is singular for the reason discussed in Appendix~\ref{appsing}, where the role of the continuity equation in the monopole and dipole cases is also compared. Moreover, the BC-monopole nodes have $\bs\Omega\propto\hat{\bs k}/k^2$ and $\bs m\propto\hat{\bs k}$, which are odd under $\bs k\to-\bs k$. The term in $h$ that is linear in $B_z$ then has components of both parities under $\theta\to\pi-\theta$. Consequently, the charge-pumping vector $\bs{\mathcal I}^\chi$ is nonzero and sources the even harmonics, in particular $\lambda_\chi$. How the intranode and internode amplitudes combine in the in-scattering term depends on the structure of the overlap function. For the Weyl pair, the coefficient of the $\cos\theta$ harmonic is multiplied by $\chi\,\chi^\prime$, and it is therefore proportional to $\beta_{\rm intr}-\beta_{\rm inter}$. At equal amplitudes, the odd harmonics drop out, the corresponding coefficients vanish, and the in-scattering term reduces to the RTA form \cite{timm}. For unequal amplitudes, these coefficients are nonzero, and the ratio of the internode to the intranode amplitude controls the sign of the quadratic magnetoconductivity once the OMM is included \cite{timm, ips-exact-spin1}. For the RSW node, there is no such cancellation at a special ratio of the amplitudes, since the three polynomials in the overlap function are all different, and the coefficients depend on the ratios $\beta_{\rm inter}/\beta^{1,1}_{\rm intr}$ and $\beta^{3,3}_{\rm intr}/\beta^{1,1}_{\rm intr}$ throughout \cite{ips-exact-rsw}. The HSM differs from all of these cases in two respects. First, the overlap function $\mathcal T_{2,2}(\gamma,\gamma^\prime)$ is independent of $\chi$. The two scattering channels therefore enter every harmonic with the same sign, and the in-scattering term for the node $\chi$ involves $\beta_{\rm intr}\,\Lambda^z_{\chi,2}+\beta^{\rm inter}_{2,2}\,\Lambda^z_{-\chi,2}$. No special ratio therefore removes the vertex correction $a_\chi$, which remains nonzero for all values of $\beta^{\rm inter}_{2,2}/\beta_{\rm intr}$. Second, $\bs\Omega$ and $\bs m$ are even in $\bs k$ for the BC-dipole, and $h_{\chi,2}$ is odd under $\gamma\to\pi-\gamma$ at every order in $B_z$. Hence $\bs{\mathcal I}^\chi=\bs0$, the even sector is never sourced, and $\lambda_\chi=b_\chi=0$. The internode scattering enters only as an additional channel for the vertex correction and, unlike in the BC-monopole case, it cannot reverse the sign of the response through charge pumping.

\subsection{\texorpdfstring{Limit of vanishing chemical potential}{Limit of vanishing chemical potential}}
\label{secmu0}

Let us comment on the limit $\mu\to0$ at fixed $B_z$. All the results of this subsection refer to $T=0$. The Boltzmann equation used in this work is applicable only for low $|B_z|$, where the formation of Landau levels can be ignored. We separate the conclusions that follow from the calculation within this regime from the features that arise only on extending the formulas beyond it. For the HSM node, the Zeeman-like shift gives $\xi_2(\bs k)=v_0\,k+e\,v_0\,B_z\cos^2\gamma/(2\,k)$. The only scales are $\mu$ and $v_0\sqrt{e\,B_z}$, so that the ratio of $\delta\sigma_{zz}$ to the conductivity at $B_z=0$ is, at fixed ratios of the scattering amplitudes, a function of the single parameter
\begin{align}
\zeta_{\rm h}=\frac{e\,B_z\,v_0^2}{\mu^2}\,.
\end{align}
The coefficient of the $n^{\rm th}$ power of $B_z$ in this ratio, whether odd or even, therefore scales as $\mu^{-2n}$ at fixed $B_z$, and the weak-field expansion is controlled only for $\zeta_{\rm h}\ll1$. The same condition governs the applicability of the Boltzmann equation. The parameter $\zeta_{\rm h}=\left(v_0\sqrt{e\,B_z}/\mu\right)^2$ is the only dimensionless combination of $\mu$, $B_z$ and $v_0$, and the neglect of Landau levels can therefore only be a condition on $\zeta_{\rm h}$. This is the regime of weak fields, $e\,B_z\,v_0^2\ll\mu^2$, considered in Sec.~\ref{secsinglehopf}, and it is the only regime in which our results are controlled. For $B_z>0$, the minimum of $\xi_2$ along the direction $\gamma$ equals $v_0\sqrt{2\,e\,B_z}\,|\cos\gamma|$. This coincides with the reality condition of the retained root of $k_F$ in Sec.~\ref{secsinglehopf}. The condition holds for all $\gamma$ only if $\zeta_{\rm h}\le1/2$. For $\zeta_{\rm h}>1/2$, the radicand of $k_F$ is negative for the directions with $\cos^2\gamma>\mu^2/(2\,e\,v_0^2\,B_z)$, and the retained root does not exist near the poles. At the boundary of this region, the two roots of the quadratic equation $\xi_2=\mu$ coincide. The argument of Sec.~\ref{secsinglehopf} for discarding the second root rests on its being parametrically small for weak fields, and it does not apply once $\zeta_{\rm h}$ is no longer small. The opening of the Fermi surface near the poles in the extended formulas is therefore a feature of the extension beyond the controlled regime, and not a result of the weak-field calculation. It occurs at $\mu\sim v_0\sqrt{e\,B_z}$, where Landau quantisation cannot be neglected. A square root in $k_F$ does not by itself fix the behaviour of $\delta\sigma_{zz}$ at $\zeta_{\rm h}=1/2$. The conductivity is built from integrals over $\gamma$, and their weight contains the inverse of the derivative of $\xi_2$ with respect to $k$ at fixed $\gamma$, evaluated at $k_F$. This derivative equals $\sqrt{\mu^2-2\,e\,v_0^2\,B_z\cos^2\gamma}\,/\,k_F$ and vanishes together with the radicand. We make no claim on the behaviour of $\delta\sigma_{zz}$ at $\zeta_{\rm h}=1/2$, since the Boltzmann treatment itself is not applicable there. No such restriction on the reality of $k_F$ exists for $B_z<0$, since the radicand is then positive for every $\gamma$. Which sign of $B_z$ is restricted depends on the sign conventions for $e$, the BC and the OMM that fix the Zeeman-like shift of Eq.~\eqref{eqmodi}, and in the two-node case on $\chi$, since only the product of $B_z$ with these signs enters the radicand. For either sign of $B_z$, the weak-field expansion is controlled only for $|\zeta_{\rm h}|\ll1$.

For the VNR, the Zeeman-like shift reduces at leading order in $\kappa/k_0$ to $e\,v_0\,B_z\sin^2\gamma/(2\,k_0)$, which does not depend on $\kappa$. The Fermi momentum becomes $\kappa_F=\mu/v_0-e\,B_z\sin^2\gamma/(2\,k_0)$, and both the OMM and the BC corrections organise in powers of
\begin{align}
\zeta_{\rm v}=\frac{e\,v_0\,B_z}{2\,k_0\,\mu}\,.
\end{align}
The expansion coefficients then grow as $\mu^{-n}$, which is milder than for the HSM. These expressions are controlled only for $|\zeta_{\rm v}|\ll1$. For $B_z>0$, the expression for $\kappa_F$ vanishes at $\gamma=\pm\pi/2$ when $\zeta_{\rm v}=1$, where the tube of the toroidal Fermi surface would pinch off in the extended formula. This lies outside the regime $|\zeta_{\rm v}|\ll1$ in which the expression was derived, and we draw no conclusion for the response from it.

At $T=0$, the BC and the OMM enter the Fermi-surface integrals only through their values at $k_F$ and $\kappa_F$, and their divergences at the node and at the ring are therefore not integrated over. The growth of the coefficients, together with the changes of the Fermi surface in the extended formulas, shows that the weak-field expansion ceases to be controlled once $|\zeta_{\rm h}|$ or $|\zeta_{\rm v}|$ is no longer small. Our calculation makes no prediction for the response beyond this regime, and the growth should not be read as a prediction of a divergence. A Landau-level description would be needed to assess the response when $|\zeta_{\rm h}|$ is not small, and it lies outside the scope of the present work. For the VNR, the scale $k_0$ additionally enters. At $\mu=v_0\sqrt{e\,B_z}$, the parameter $\zeta_{\rm v}$ equals $\sqrt{e\,B_z}/(2\,k_0)$, and we have not examined the validity of the semiclassical description for the VNR at such chemical potentials. For the PTNR, the BC and OMM are supported only on the ring, and no odd-in-$B_z$ term arises on account of time-reversal symmetry.

\section{Nonlinear response coefficients}
\label{secnonlinear}

In this section, we will focus on computing a nonlinear response in the form of the ECR. Contrary to $\sigma_{zz}$ computed in Sec.~\ref{secexact}, here we will restrict ourselves to the RTA since the exact method will be challenging to implement and solve for the purpose of computing the ECR. The main purpose of computing the associated response coefficients is to shed light on how the anisotropic BC shows up in transport, despite the Chern number being zero in the systems under study.

The reader is referred to Ref.~\cite{ips-shreya} for a detailed derivation of the forms of the electric and thermal current densities, denoted by ${\bs J}_s$ and ${\bs J}_s^{\rm th}$, respectively, starting from the Boltzmann equations (which we do not repeat here). The final expressions are obtained by solving for $ \delta  f_s (\bs r,\bs k)$, which is the deviation of the distribution of the quasiparticles from the equilibrium, up to second order in $\epsilon $. Here, $\epsilon \in [0, 1 ] $ is a perturbative parameter, which quantifies the smallness of the magnitude of the probe fields which usually comprise $\bs E  $, $\nabla_{\bs r} T $, and $\nabla_{\bs r} \mu $. In other words, we work in the regime where $ |\bs E| \propto \epsilon  $, $ |\nabla_{\bs r} T |  \propto \epsilon$, and $ |\nabla_{\bs r} \mu |  \propto \epsilon$, and use the expression,
\begin{align}
f_s (\bs r,\bs k)
=  f_0  (\varepsilon_s) + \epsilon \, f_s^{(1)}  (\varepsilon_s) 
+ \epsilon^2 \, f_s^{(2)}  (\varepsilon_s) 
+ \mathcal{O}(\epsilon^3)\,,
\end{align}
in a perturbative expansion in $\epsilon$. Finally, we set $\epsilon = 1 $ at the end of our computations.

For computing the ECR, we set $\bs B $ and $\nabla_{\bs r} T $ to zero. The part of the total electrical current density ($ {\bs J}_s $), which shows a quadratic dependence on the probe fields $\bs E$ and $\nabla_{\bs r} \mu $, is denoted by ${\bar {\bs J}}^s $. This is the electrochemical current density, with its components being proportional to the quadratic combinations of the form $ E_a  \, \partial_{b} \mu $. Its $a^{\rm th}$ component is given by
\begin{align}
\label{eqelectrocur}
{\bar J}^s_{a}   = \vartheta^s_{a  b c} \, \partial_b\mu \, E_c  \,,
\end{align}
where $\vartheta^s_{a b c }$ is a rank-three conductivity tensor for the $s^{\rm th}$ band. Since the dependence of the electric current on the probe fields is at second order, we call the resulting response nonlinear.
We divide it up as
\begin{align}
{\bar {\bs J}}^s  &  
 =  \bs{J}^{(s, 1)}   + \bs{J}^{(s, 2)}   \,,
 \quad \vartheta_{a  b c}^{s} = 
\vartheta_{a  b c}^{(s,1)} + \vartheta_{a  b c}^{(s,2)} \,,
\label{eqsecorderf} 
\end{align}
where
\begin{align}
\bs{J}^{(s, 1)} &=  e^2 \,\tau 
\int \frac{ d^3 \bs k}{(2\, \pi)^3 }  
\left( 
\nabla_{\bs r} \mu \cross \bs{\Omega}_{s}    \right ) 
\left(  {\boldsymbol v}^{(0,s)} \cdot \bs{E}   \right )
f^\prime_{0} ( \varepsilon_s) \,, \quad
\bs{J}^{(s, 2)}  = -\, e^2 \, \tau
\int \frac{ d^3 \bs k}{(2\, \pi)^3 }  
\, {\boldsymbol v}^{(0,s)}  \left[ 
\left( \bs{E} \cross \bs{\Omega}_s \right ) \cdot  
{\nabla}_{\bs r} \mu   \right]    f^\prime_0 (\varepsilon_s) \,,
\end{align}
\begin{align}
\label{eqtheta}
& \vartheta_{a  b c}^{(s,1)} =  e^{2} \,\tau
\sum_{d}  \epsilon_{a b d }\, t^s_{cd} \,, 
\quad \vartheta_{a  b c}^{(s,2)}  = - \,
e^2 \, \tau \sum_{d}  \epsilon_{b c d}\, t^s_{ad} \,,  \text{ and }
t^s_{ab} = \int \frac {d^3 \bs{k}} {(2 \, \pi)^3 } 
v^{(0,s)}_a \left( \Omega_s \right)_b \, f^{\prime}_0 (\varepsilon_s)   \,.
\end{align}
We would like to point out that a term arising from ${\nabla}_{\bs r} \mu$ was missed in Ref.~\cite{flores}, as noted in our earlier work \cite{ips-shreya}.

To express some of the above coefficients in a compact way, let us define the second-rank Berry-curvature dipole (BCD) tensor \cite{bcd-inti},
\begin{align}
\label{eqbcd}
& D^s_{ab} = - \,\epsilon_{ab} \int_k  \partial_a f_0 (\varepsilon_s) \left( \Omega_s \right)_b
= - \,\epsilon_{ab} \, t^s_{ab}\,.
\end{align}
It captures the dipole-moment of the BC over the occupied state. Here, $\epsilon_{ab}$ follows the cyclic convention $\epsilon_{xy}=\epsilon_{yz}=\epsilon_{zx}=+1$ (and antisymmetric otherwise), consistent with the three-index Levi-Civita symbol used throughout. We will find that, in certain cases, the nonzero components of $D^s_{ab}$ will be related to those of the $\vartheta$'s.

The vanishing and survival pattern of these coefficients, worked out
explicitly below for both the VNR and the HSM, follows directly from the
symmetries identified in Sec.~\ref{secmodels}. Since $\bs \Omega_s$ (and hence
$\boldsymbol m$) carries an overall factor of $k_z$ in both models, its
in-plane components are odd under the mirror $k_z \rightarrow -k_z$, while its
$z$-component is even; the band-velocity components $v_x$ and $v_y$ are even
and $v_z$ is odd under the same mirror. Every diagonal entry of
$t^s_{ab}$ is therefore built from a product of opposite $k_z$-parity, making
it odd overall, so it vanishes upon integration over the mirror-symmetric
Fermi surface of the untilted models. At the same time, the continuous
$C_{\infty z}$ symmetry about $\hat{\bs z}$ forbids any component mixing the
axial index $z$ with an equatorial index $x$ or $y$. Together, these two
symmetries force every component of $t^s_{ab}$, and hence every
$\vartheta^{(s,1)}_{abc}$ and $\vartheta^{(s,2)}_{abc}$, to vanish identically
at zero tilt. A tilt $\bs \eta = \eta_x \, \hat{\bs x}$ confined to the equatorial plane
breaks the $C_{\infty z}$ symmetry while leaving the mirror intact, so only
the $z$-mixing components, which are even in overall $k_z$-parity and were
previously forbidden by $C_{\infty z}$, can become nonzero, namely those
captured by $D^s_{zx}$. A tilt $\bs \eta = \eta_z \, \hat{\bs z}$ along the
axis instead preserves the $C_{\infty z}$ symmetry while breaking the mirror,
so only the diagonal components allowed by $C_{\infty z}$, namely
$t^s_{xx}=t^s_{yy}$ and $t^s_{zz}$, can become nonzero.
The same $\Xi = T \,M_y$ action that underlies
Eq.~\eqref{eq:theta_covariance_vnr} propagates to any Fermi-surface integral
of $\bs \Omega_s$, imposing the same selection rules on $D^s_{ab}$.

\subsection{Vortex nodal-ring}

All the nonlinear-response coefficients evaluate to zero for an untilted VNR. For the tilted ones, we discuss two distinct cases: $\bs \eta = \eta_x \,\bs{\hat x} $ and $ \bs \eta = \eta_z \,\bs{\hat z} $.

\subsubsection{Case I: Tilt with respect to $ k_x $-axis}

The antisymmetric tensor $ D_{ab}^s $ [cf. Eq.~\eqref{eqbcd}] evaluates to a nonzero value only for a nonzero tilt along the $k_x k_y$-plane, which we choose to be $\eta_x \,\bs{\hat x} $. In particular, $D^s_{zx}$ (equivalently $t^s_{zx}$) survives for a nonzero $\eta_x$, and we find that its transpose $t^s_{xz}$ is independently nonzero and, in fact, larger in magnitude than $t^s_{zx}$ over the whole range of $\eta_x$.
For $ \bs \eta = \eta_x \,\bs{\hat x} $, $\eta_x >0$, and $\mu \geq 0 $, we obtain
\begin{align}
 D^2_{zx} =\begin{cases}
-\frac{1}{8 \,\pi \, \eta_x} 
+  \frac{\eta_x^2 \,K\left(\eta_x^2\right)-K\left(\eta_x^2\right)
+ E\left(\eta_x^2\right)}{2\, \pi ^2 \,\eta_x^3}   & \text{ for }  \mu \geq\tilde \eta \\ & \\
-\frac{1}{8 \,\pi \, \eta_x} 
+ \frac{\sqrt{\tilde{\eta }^2-\mu ^2} \left(\sqrt{\tilde{\eta }^2-\mu ^2  \,\eta_x^2}-\tilde{\eta }\right)}
{4\, \pi ^2 \,\mu \,\tilde{\eta } \,\eta_x^3}
+ \frac{\cos ^{-1}\left(\frac{\mu }{\tilde{\eta }}\right)} {8 \,\pi ^2 \,\eta_x}
+
\frac{E\left(\sin ^{-1}\left(\frac{\mu \, \eta_x}{\tilde{\eta }}\right)| \eta_x^{-2}\right)
+E\left(\sin^{-1} \eta_x | \eta_x^{-2}\right)} {4\, \pi ^2 \,\eta_x^2}
 & \text{ for }  \mu < \tilde \eta  
 \end{cases}
\end{align}
and
\begin{align}
 D^1_{zx} =\begin{cases}
 0 & \text{ for }  \mu \geq\tilde \eta \\ & \\
\frac{\sqrt{\tilde{\eta }^2-\mu ^2}} {4 \,\pi ^2 \, \mu \, \eta_x^3}
-\frac{\cos ^{-1}\left(\frac{\mu }{\tilde{\eta }}\right)} {8 \,\pi ^2\, \eta_x}
-\frac{\sqrt{\left(\tilde{\eta }^2-\mu ^2\right) 
\left(\tilde{\eta }^2
-\mu ^2 \,\eta_x^2\right)}}{4 \,\pi ^2 \,\mu \, \tilde{\eta } \,\eta_x^3}
+ \frac{E\left(\sin ^{-1} \eta_x | \eta_x^{-2}\right)
-E\left(\sin ^{-1}\left(\frac{\mu \, \eta_x}{\tilde{\eta}}\right)| \eta_x^{-2}\right)}{4\, \pi ^2 \,\eta_x^2}
 & \text{ for }  \mu < \tilde \eta  
 \end{cases}
\end{align}
The transpose component $t^s_{xz}$ obeys an analogous case split. For $\mu \geq \tilde \eta$, only the $s=2$ band contributes,
\begin{align}
\label{eqtxz2}
t^2_{xz} = \frac{\pi \,\eta_x^2 + 4 \left(\eta_x^2-1\right)\left[\left(\eta_x^2-1\right) K\left(\eta_x^2\right) + E\left(\eta_x^2\right)\right]}{8 \,\pi^2 \,\eta_x^3} \,,
\end{align}
while for $\mu < \tilde \eta$ both bands contribute,
\begin{align}
\label{eqtxz1}
t^2_{xz} &= \frac{1}{8\, \pi \,\eta_x} - \frac{\cos^{-1}\left(\mu/\tilde \eta\right)}{8\, \pi^2 \,\eta_x} - \frac{1-\eta_x^2}{4\, \pi^2 \,\eta_x^3} \left[ \mathcal J(\theta_m) - \left(1-\eta_x^2\right) K\left(\eta_x^2\right) + E\left(\eta_x^2\right) \right] \,, \\
t^1_{xz} &= -\frac{\cos^{-1}\left(\mu/\tilde \eta\right)}{8 \,\pi^2 \,\eta_x} + \frac{1-\eta_x^2}{4\, \pi^2 \,\eta_x^3} \left[ \left(\eta_x^2-1\right) K\left(\eta_x^2\right) + E\left(\eta_x^2\right) - \mathcal J(\theta_m) \right] ,
\end{align}
where $\theta_m \equiv \sin^{-1}\left(\mu/\tilde \eta\right)$ and
\begin{align}
\label{eqJdef}
\mathcal J(\theta) \equiv \cot \theta \left[\sqrt{1-\eta_x^2 \sin^2 \theta} \,-1\right] + \left(\eta_x^2-1\right) F\left(\theta \,| \,\eta_x^2\right) + E\left(\theta \,| \,\eta_x^2\right) .
\end{align}
As with $D^s_{zx}$, both branches are independent of $\mu$ only in the $\mu \geq \tilde \eta$ case. For $\mu < \tilde \eta$ the sum $t^1_{xz}+t^2_{xz}$ must be taken. Taking $\eta_x \rightarrow 0$ from the $\mu \geq \tilde \eta$ side gives $t^2_{xz} \rightarrow 7 \, \eta_x/(64\,\pi)$, smooth and vanishing at $\eta_x=0$.
The second case for each band includes the scenarios with $\mu=0 $.
Here, $K (m)$ and $ E (m)$ give the complete elliptic integrals of the first and second kinds, respectively, with the parameter $m$ (which equals $\eta_x^2$ in the complete integrals above) as the argument. Similarly,
$F(\varphi\,|\,m)$ and $ E (\varphi\,|\,m)$ denote the incomplete elliptic integrals of the first and second kinds, respectively, with amplitude $\varphi$ and parameter $m$. Also, for the second case, we need to compute the sum $\left( D^1_{zx} +D^2_{zx} \right) $ as both the bands contribute for a given $\mu$. For checking the $\eta_x \to 0$ limit at fixed $\mu>0$, we note that, since $\mu/\tilde\eta \to \infty$, the relevant branch is $\mu \geq \tilde\eta$, where $D^1_{zx} = 0$ and $D^2_{zx} \to \eta_x/(64\pi)$. Hence $D^1_{zx}+D^2_{zx}$ has a smooth limit for $\eta_x \to 0$, vanishing exactly at $\eta_x = 0$.

All the diagonal components, which are proportional to $ \lbrace t^s_{aa} \rbrace $, vanish.
Feeding the above into Eq.~\eqref{eqtheta}, we find that the nonzero components are captured by
\begin{align}
\label{eqtheta1}
& \vartheta_{y z z}^{(s,1)} = - \, \vartheta_{ z y z}^{(s,1)}
=  -\,e^2 \,\tau \, D^s_{zx}  \,,\quad
\vartheta_{z  yz}^{(s,2)}  = -\, \vartheta_{zzy}^{(s,2)} 
=  e^2 \, \tau\, D^s_{zx}\,,
\end{align}
sourced by $t^s_{zx}$, together with a second independent pair, 
\begin{align}
\label{eqtheta1b}
\vartheta_{x y x}^{(s,1)} = - \, \vartheta_{y x x}^{(s,1)} = e^2\, \tau \, t^s_{xz} \,,\quad
\vartheta_{x x y}^{(s,2)} = - \, \vartheta_{x y x}^{(s,2)} = -\, e^2 \, \tau \, t^s_{xz} \,,
\end{align}
sourced by $t^s_{xz}$.
Here, $t^s_{xz}$ is given in Eqs.~\eqref{eqtxz2}--\eqref{eqJdef}.

\subsubsection{Case II: Tilt with respect to $ k_z $-axis}

For a nonzero tilt with respect to the $k_z$-axis, setting $\bs \eta = \eta_z \,\bs{\hat z}$, we find that only the diagonal terms, viz. $ \lbrace t^s_{aa} \rbrace $, survive. In this scenario, there is no horn-cyclide formation. For $\mu > 0$ and $\eta_z < 1$, only the $s=2$ band contributes: the $s=1$ band has dispersion $\varepsilon_1 = -v_0\,\kappa + v_0\,\eta_z\,k_z$, and since $\eta_z\,|k_z| < \sqrt{\kappa_\perp^2 + k_z^2}$ for $\eta_z < 1$, we have $\varepsilon_1 < 0$ for every allowed $(\kappa_\perp, k_z)$. Hence it cannot reach the positive chemical potential $\mu$. Only the $s=2$ band therefore has a Fermi surface at positive $\mu$. With all these details in hand, we obtain
\begin{align}
t^2_{xx} = t^2_{yy} = -\, \frac{t^2_{zz}} {2}
= \frac{\eta_z^2+2 \,\sqrt{1-\eta_z^2}-2} {16 \, \pi  \,\eta_z^3} \,.
\end{align}
Intriguingly, these are completely independent of the magnitude of $\mu$, and have a smooth limit for $\eta_z \rightarrow 0 $ when they vanish exactly at $\eta_z = 0 $. This behaviour is similar to the features observed in Ref.~\cite{ips-shreya}, where a node with a nonzero BC-monopole was considered. Finally, we list the nonzero components as follows:
\begin{align}
\label{eqtheta3}
& \vartheta_{a  b c}^{(s,1)} =  e^{2} \,\tau
\,\epsilon_{a b c }\, t^s_{cc} \,, 
\quad \vartheta_{a  b c}^{(s,2)}  = - \,e^2 \, \tau \, \epsilon_{ab c}\,t^s_{aa}   \,.
\end{align}

 
\subsection{${\mathcal P} {\mathcal T}$-symmetric nodal-ring}

Analogous to the VNR case, an untilted PTNR too has vanishing nonlinear-response coefficients. Hence, we discuss the tilted scenarios with $\bs \eta = \eta_x \,\bs{\hat x} $ and $ \bs \eta = \eta_z \,\bs{\hat z} $. 

\subsubsection{Case I: Tilt with respect to $ k_x $-axis}

Here, no diagonal components, viz. $\lbrace t^s_{aa} \rbrace $, survive. There is a nonzero nonlinear response only for the $D^s_{xy}$-component of the BCD. This can be seen by starting with
\begin{align}
& D^s_{ab} =  - \,\delta_{a x} \,\delta_{b y} \, v_0\,\eta_x 
 \int \frac{ d^3 \bs k}{(2\, \pi)^3 }  \left( \Omega_s \right)_y \, f^{\prime}_0 (\varepsilon_s)\,.
\end{align} 

Setting $\mu \geq 0$, the $s=2$ band contributes with
\begin{align}
D^2_{xy} & = - \,\pi \, v_0\,\eta_x
\int \frac{ d^3 \bs k}{(2\, \pi)^3 }   
 \cos \phi  \,\delta(k_z)\,\delta(k_\perp-k_0)\,\delta \big( \tilde \eta \cos \phi  -\mu \big)
=  \frac{-1 } {(2\, \pi)^2} \,\int_{-1}^{\min[1, \mu/\tilde \eta ] } \frac{du  \, u} { \sqrt{1-u^2} }  
\, \delta \Big( u -\frac{\mu} {\tilde \eta} \Big )
\nn & = \frac{ -1 } {(2\, \pi)^2} \,\frac{\mu} {\sqrt{ {\tilde \eta}^2 -\mu^2}} 
\,\Theta(1-\mu/{\tilde \eta})\,,
\end{align}
which is nonzero only if $\mu/\tilde \eta \leq 1$. The $s=1$ band also contributes, whose value is
\begin{align}
D^1_{xy} & =  \frac{1 } {(2\, \pi)^2} \,\int^{1}_{\mu/\tilde \eta} \frac{du  \, u} { \sqrt{1-u^2} }  
\, \delta \Big( u -\frac{\mu} {\tilde \eta} \Big ) \,\Theta(1-\mu/{\tilde \eta})
 = \frac{ 1 } {(2\, \pi)^2} \,\frac{\mu} {\sqrt{ {\tilde \eta}^2 -\mu^2}} \,\Theta(1-\mu/{\tilde \eta})\,.
\end{align}
However, if we add up the two contributions, the net value is zero.

Using Eq.~\eqref{eqcross} at $T=0$, we find that
\begin{align}
\frac{ J^{(s, 1)}_x } { e^2 \,\tau } & = (-1)^s \,\pi \,\partial_z \mu
\int \frac{ d^3 \bs k}{(2\, \pi)^3 }   
 \cos \phi  \,\delta(k_z)\,\delta(k_\perp-k_0)\,v_0\,\eta_x \, E_x \,
\delta \big( \tilde \eta \cos \phi  -\mu \big)
= \, \partial_z \mu\, E_x \,D^s_{xy} \,,\nn
\frac{ J^{(s , 1)}_y } { e^2 \,\tau } & =  (-1)^s \, \pi \,\partial_z \mu
\int \frac{ d^3 \bs k}{(2\, \pi)^3 }   
 \sin \phi  \,\delta(k_z)\,\delta(k_\perp-k_0) \,v_0\,\eta_x \, E_x \,\delta \big( \tilde \eta \cos \phi  -\mu \big)
= 0\,,
\nn \frac{ J^{(s, 1)}_z } { e^2 \,\tau } & = -(-1)^s \,  \pi 
\int \frac{ d^3 \bs k}{(2\, \pi)^3 }   
(\partial_x \mu \cos \phi + \partial_y \mu \sin \phi) \,\delta(k_z)\,\delta(k_\perp-k_0)
\,v_0\,\eta_x \, E_x \,
\delta \big( \tilde \eta \cos \phi  -\mu \big)
 =  -\,\partial_x \mu \,E_x \, D^s_{xy}  \,.
\end{align}
Now, for $\bs{J}^{(s, 2)}$, only the $x$-component survives, which evaluates to
\begin{align}
\frac{ J^{(s, 2)}_x } { e^2 \,\tau } & =- (-1)^s \, \pi \,v_0\,\eta_x
\int \frac{ d^3 \bs k}{(2\, \pi)^3 }  
  \delta(k_z)\,\delta(k_\perp-k_0)
\left[\partial_x \mu\, E_z \cos \phi + \partial_y \mu\,  E_z \sin \phi
- \partial_z \mu\,( E_x \cos \phi + E_y \sin \phi ) \right ]
\delta \big( \tilde \eta \cos \phi  -\mu \big)
\nn & = \left (\partial_z \mu\, E_x - \partial_x \mu \, E_z   \right )D^s_{xy}\,.
\end{align}

From the above expressions, we can easily infer that the only nonzero individual-band-coefficients are captured by
\begin{align}
-\,\vartheta_{x z x}^{(s,1)} = \vartheta_{zxx}^{(s,1)}  = \vartheta_{x xz}^{(s,2)}  = -\,\vartheta_{xzx}^{(s,2)}
= -\,e^2 \,\tau\,D^s_{xy} \,.
\end{align}
These coefficients, however, do not survive the band sum. The physical electrochemical current is obtained by summing over the band contributions, and since $D^1_{xy}$ and $D^2_{xy}$ are equal in magnitude and opposite in sign, we have $\vartheta^{(1)}_{abc}+\vartheta^{(2)}_{abc}=0$ for the coefficients listed above. The net ECR of the PTNR therefore vanishes for an in-plane tilt.

\subsubsection{Case II: Tilt with respect to $ k_z $-axis}

A nonzero tilt $ \bs \eta = \eta_z \,\bs{\hat z} $ does not give any nonzero nonlinear response, and this vanishing is exact rather than approximate. Since the BC of Eq.~\eqref{eqbcpnr} is confined entirely to the nodal-ring, any nonzero $t^s_{ab}$ requires the Fermi surface to intersect the ring itself, i.e., the locus $\kappa=0$. On the ring, the dispersion reduces to $\varepsilon_s|_{\kappa=0} = v_0\,\bs \eta \cdot \bs k|_{\kappa=0}$, and for a tilt along $\bs{\hat z}$ this is $\eta_z\, v_0 \, k_z|_{\kappa=0}=0$ identically, because $k_z$ vanishes everywhere on the ring by definition. The ring therefore stays pinned at zero energy irrespective of $\eta_z$, so it never coincides with a Fermi surface at $\mu>0$. This is in sharp contrast with the in-plane tilt of case~I, where the ring's energy varies with $\phi$ and does cross $\mu$ for $\mu \le \tilde \eta$, which is why that case alone produces a nonzero $D^s_{xy}$.

\subsection{Three-band Hopf semimetal}

For the BC-dipole nodal point, keeping with the trend of the nodal-ring systems, there is zero nonlinear response for no tilt. Therefore, we will consider the two cases: $\bs \eta = \eta_x \,\bs{\hat x} $ and $ \bs \eta = \eta_z \,\bs{\hat z} $. It turns out that the $s=0$ band does not lead to a nonzero ECR even for the tilted cases. We therefore do not discuss it any further.

\subsubsection{Case I: Tilt with respect to $ k_x $-axis}

In analogy with the VNR and PTNR cases, all diagonal components, viz. $\lbrace t^s_{aa} \rbrace $, vanish.
As for the off-diagonal components proportional to the BCD, $D^s_{zx}$ survives just like the VNR system. As in the VNR, its transpose $t^s_{xz}$ is independently nonzero as well, and turns out to be larger in magnitude. Explicit calculations lead to
\begin{align}
D^s_{zx} = \frac{\eta _x \left(2 \,\eta _x^2-3\right)
+ 3 \left( 1- \eta _x^2 \right) \tanh ^{-1} \eta _x}   {12 \,\pi ^2\, \eta _x^4}\;
\text{ for } s=1 \text{ and } s=2\,,
\end{align}
together with
\begin{align}
\label{eqhopftxz}
t^s_{xz} = \frac{3\,\eta_x - 5\,\eta_x^3 - 3\left(1-\eta_x^2\right)^2 \tanh^{-1} \eta_x}{12\, \pi^2\, \eta_x^4}\;
\text{ for } s=1 \text{ and } s=2\,.
\end{align}
Intriguingly, the results are completely independent of $\mu$. Eqs.~\eqref{eqtheta1} and \eqref{eqtheta1b} are both applicable due to the similarities in the surviving terms, with $D^s_{zx}$ feeding the former and $t^s_{xz}$ of Eq.~\eqref{eqhopftxz} feeding the latter. The $\eta_x \rightarrow 0$ limit is smooth for both: $D^s_{zx} \rightarrow -\,\eta _x
/ (30 \,\pi ^2 ) + \order{\eta_x^3}$, while $t^s_{xz} \rightarrow -\,2\,\eta_x/(15\,\pi^2) + \order{\eta_x^3}$.

\subsubsection{Case II: Tilt with respect to $ k_z $-axis}

Again, agreeing with the VNR results, we get a nonzero answer for the diagonal components $\propto t^s_{aa}$, while all the off-diagonal components proportional to the BCD vanish. As in the VNR case, for $\mu > 0$ and $\eta_z < 1$, only the $s=2$ band contributes. The $s=1$ band has $\varepsilon_1 = -v_0\,k + v_0\,\eta_z\,k_z < 0$ for every allowed $\bs k$, and the flat-band $s=0$ has vanishing group-velocity. For the diagonal ones, we get
\begin{align}
& t^s_{xx} = t^s_{yy} = 
\frac{\eta _z \; _2F_1\left(1,\frac{3}{2};\frac{7}{2};\eta _z^2\right)} {30\, \pi ^2}
\text{ and }
t^s_{zz} = \frac{\eta _z \left(2 \,\eta _z^2-3\right)+3 \left( 1- \eta _z^2\right) 
\tanh ^{-1} \eta _z} {6 \, \pi ^2 \, \eta _z^4}
 \text{ for } s=1 \text{ and } s=2\,.
\end{align}
Here, $ _2F_1(a, b; c; z) $ is the Gaussian or ordinary hypergeometric function. Again, the results are completely independent of $\mu$ and have a smooth limit for $\eta_z \rightarrow  0 $ (vanishing exactly at $\eta_z  =  0$). As with the VNR, this independence follows from an exact cancellation, here between the $1/k^2$ carried by the BC of Eq.~\eqref{eqbcommhopf} and the $k^2$ supplied by the Jacobian.
The two diagonal answers above satisfy the identity $t^s_{zz}=-\,2\,t^s_{xx}$ for all $\eta_z$. The identity is proved by substituting $u = \cos \gamma$ in both momentum integrals. The two angular integrals reduce to
\begin{align}
t^s_{xx} = -\,\frac{1}{2\,(2\,\pi)^2}
\int_{-1}^{1} \frac{u\,(1-u^2)}{1+\eta_z \,u}\,du\,, \quad
t^s_{zz} = -\,\frac{1}{(2\,\pi)^2}
\int_{-1}^{1} \frac{u^3 + \eta_z \,u^2}{1+\eta_z \,u}\,du\,,
\end{align}
so that
\begin{align}
t^s_{zz} + 2\,t^s_{xx} = -\,\frac{1}{(2\,\pi)^2}
\int_{-1}^{1} \frac{u^3 + \eta_z \,u^2 + u\,(1-u^2)}{1+\eta_z \,u}\,du
= -\,\frac{1}{(2\,\pi)^2} \int_{-1}^{1} u\,du = 0\,.
\end{align}
The last step uses that the integrand $u$ is odd over the symmetric interval $[-1,1]$. This proves the identity $t^s_{zz}=-\,2\,t^s_{xx}$ for every $\eta_z$. Eq.~\eqref{eqtheta3} is applicable to get the final forms of the three-index tensors.

\subsection{\texorpdfstring{Combined in-plane and axial tilt}{Combined in-plane and axial tilt}}
\label{seccombined}

Let us finally consider $\bs\eta=\eta_x\,\hat{\bs x}+\eta_z\,\hat{\bs z}$. We define $\eta\equiv|\bs\eta|=\sqrt{\eta_x^2+\eta_z^2}$ and $\hat\eta_{x,z}\equiv\eta_{x,z}/\eta$, and we restrict ourselves to $\mu>0$ and $\eta<1$. Eq.~\eqref{eqtheta} remains valid for an arbitrary tilt, and only the evaluation of $t^s_{ab}$ is required. The intermediate steps of the calculation are collected in Appendix~\ref{appcomb}.

This tilt breaks both the $C_{\infty z}$ rotation and the $k_z$ mirror. The antiunitary symmetry $\Xi$ survives since the tilt has no $y$ component. For the VNR and the HSM, the components $\Omega^{x,z}_s$ and $v_{x,z}$ are even under $k_y\to-k_y$, whereas $\Omega^y_s$ and $v_y$ are odd, and $f_0^\prime$ is even. Hence $t^s_{xy}$, $t^s_{yx}$, $t^s_{yz}$ and $t^s_{zy}$ vanish, while $t^s_{xx}$, $t^s_{yy}$, $t^s_{zz}$, $t^s_{xz}$ and $t^s_{zx}$ are allowed. The mirror and the rotation by $\pi$ about $\hat{\bs z}$ further show that the diagonal components are even in $\eta_x$ and odd in $\eta_z$, whereas $t^s_{xz}$ and $t^s_{zx}$ are odd in $\eta_x$ and even in $\eta_z$. We now present the explicit forms of these five components, starting with the HSM.

For the HSM, only the $s=2$ band has a Fermi surface at $\mu>0$. Defining the three functions
\begin{align}
\label{eqFhopf}
& \mathcal F_1 (\eta) = \frac{3\,(1-\eta^2)\,\tanh^{-1}\eta - 3\,\eta + 2\,\eta^3}{3\,\eta^4}\,, \quad
\mathcal F_2 (\eta) = \frac{3\,(1-\eta^2)^2\,\tanh^{-1}\eta - 3\,\eta + 5\,\eta^3}{3\,\eta^4}\,, \nn &
\mathcal F_3 (\eta) = \frac{3\,(1-\eta^2)\,(\eta^2-5)\,\tanh^{-1}\eta + 15\,\eta - 13\,\eta^3}{3\,\eta^4}\,,
\end{align}
each of which vanishes linearly as $\eta\to0$, we find
\begin{align}
\label{eqhopfcomb}
& t^2_{xx} = -\,\frac{\hat\eta_z}{4\,\pi^2} \left[ \mathcal F_1 + \mathcal F_3\,\hat\eta_x^2 \right] , \quad
t^2_{yy} = -\,\frac{\hat\eta_z}{4\,\pi^2}\, \mathcal F_1 \,, \quad
t^2_{zz} = -\,\frac{\hat\eta_z}{4\,\pi^2} \left[ 2\,\mathcal F_1 + \mathcal F_2 + \mathcal F_3\,\hat\eta_z^2 \right] , \nn &
t^2_{xz} = -\,\frac{\hat\eta_x}{4\,\pi^2} \left[ \mathcal F_2 + \mathcal F_3\,\hat\eta_z^2 \right] , \quad
t^2_{zx} = -\,\frac{\hat\eta_x}{4\,\pi^2} \left[ \mathcal F_1 + \mathcal F_3\,\hat\eta_z^2 \right] .
\end{align}
For $\eta_x=0$ these reduce to the results of Case~II, and for $\eta_z=0$ to those of Case~I, including the relation $D^2_{zx}=-\,t^2_{zx}$. The results are independent of $\mu$, as for the single tilts. The three exact relations
\begin{align}
\label{eqhopfrel}
& t^2_{xx}+t^2_{yy}+t^2_{zz} = 0 \,, \quad
t^2_{xx}-t^2_{yy} = -\,\frac{\hat\eta_x^2\,\hat\eta_z}{4\,\pi^2}\,\mathcal F_3 \,, \quad
t^2_{xz}-t^2_{zx} = -\,\frac{\hat\eta_x}{4\,\pi^2\,\eta^2} \left[ \eta - (1-\eta^2)\,\tanh^{-1}\eta \right]
\end{align}
follow directly. The vanishing of the trace holds for any tilt, since $(\bs n+\bs\eta)\cdot\bs n=1+\bs\eta\cdot\bs n$, with $\bs n=\hat{\bs k}$, cancels the denominator of the integrand and leaves an integral of $n_z$ over the sphere. Hence, as soon as both tilts are present, the isotropy $t^2_{xx}=t^2_{yy}$ is lifted, and the relation $t^2_{zz}=-2\,t^2_{xx}$ of Case~II is replaced by $t^2_{zz}+2\,t^2_{xx}=t^2_{xx}-t^2_{yy}$. At small tilt, the expansions
\begin{align}
\label{eqhopfser}
& t^2_{xx} = \frac{1}{\pi^2} \left[ \frac{\eta_z}{30} + \frac{\eta_z^3}{70} - \frac{\eta_x^2\,\eta_z}{42} \right] + \mathcal O(\eta^5)\,, \quad
t^2_{yy} = \frac{1}{\pi^2} \left[ \frac{\eta_z}{30} + \frac{\eta_z^3}{70} + \frac{\eta_x^2\,\eta_z}{70} \right] + \mathcal O(\eta^5)\,, \nn &
t^2_{zz} = \frac{1}{\pi^2} \left[ -\,\frac{\eta_z}{15} - \frac{\eta_z^3}{35} + \frac{\eta_x^2\,\eta_z}{105} \right] + \mathcal O(\eta^5)\,, \nn &
t^2_{xz} = \frac{1}{\pi^2} \left[ -\,\frac{2\,\eta_x}{15} - \frac{2\,\eta_x^3}{105} - \frac{2\,\eta_x\,\eta_z^2}{35} \right] + \mathcal O(\eta^5)\,, \quad
t^2_{zx} = \frac{1}{\pi^2} \left[ \frac{\eta_x}{30} + \frac{\eta_x^3}{70} - \frac{\eta_x\,\eta_z^2}{42} \right] + \mathcal O(\eta^5)
\end{align}
show that the response is additive to linear order in the tilt, with $\eta_x$ feeding only $t^2_{xz}$ and $t^2_{zx}$, and $\eta_z$ only the diagonal components. The interference between the two tilt directions is captured by $\Delta_{ab}\equiv t^2_{ab}(\eta_x,\eta_z)-t^2_{ab}(\eta_x,0)-t^2_{ab}(0,\eta_z)$. It starts at third order, with
\begin{align}
\label{eqhopfdelta}
& \Delta_{xx} = -\,\frac{\eta_x^2\,\eta_z}{42\,\pi^2}\,, \quad
\Delta_{yy} = \frac{\eta_x^2\,\eta_z}{70\,\pi^2}\,, \quad
\Delta_{zz} = \frac{\eta_x^2\,\eta_z}{105\,\pi^2}\,, \quad
\Delta_{xz} = -\,\frac{2\,\eta_x\,\eta_z^2}{35\,\pi^2}\,, \quad
\Delta_{zx} = -\,\frac{\eta_x\,\eta_z^2}{42\,\pi^2}\,,
\end{align}
up to corrections of order $\eta^5$. No component is proportional to the product $\eta_x\,\eta_z$, and the interference is therefore of the type $\eta_x^2\,\eta_z$ for the diagonal components and $\eta_x\,\eta_z^2$ for the off-diagonal ones.

For the VNR, only the $s=2$ band contributes if $\mu\ge\tilde\eta$, where $\tilde\eta=v_0\,k_0\,\eta_x$. In that regime, the results again do not depend on $\mu$ and are given by
\begin{align}
\label{eqvnrcomb}
t^2_{ab} = \frac{1}{16\,\pi^3} \int_0^{2\,\pi} d\phi \;\mathscr S_{ab}(\phi) \,,
\end{align}
where the integrands $\mathscr S_{ab}(\phi)$ are elementary functions of $R(\phi)=\sqrt{\eta_z^2+\eta_x^2\cos^2\phi}$ and $\sqrt{1-R^2(\phi)}$, listed in Eqs.~\eqref{eqSab} and \eqref{eqJpq} of Appendix~\ref{appcomb}. The condition $\eta<1$ is precisely the requirement that $1-R^2(\phi)>0$ for all $\phi$, which ensures that the toroidal Fermi surface remains closed.

For $\mu\ge\tilde\eta$, the remaining $\phi$-integral can be performed in closed form. With $\varrho=\sqrt{1-\eta_z^2}$, $\varsigma^2=\eta_x^2/\varrho^2$ and $\hat\Pi\equiv\Pi\!\left(-\eta_z^2/\varrho^2\,\big|\,\varsigma^2\right)$, where $K$, $E$ and $\Pi$ denote the complete elliptic integrals of the first, second and third kind, we obtain
\begin{align}
\label{eqvnrclosed}
t^2_{ab} = \frac{1}{4\,\pi^2} \left[ \mathcal A_{ab}\,K(\varsigma^2) + \mathcal B_{ab}\,E(\varsigma^2) + \mathcal C_{ab}\,\hat\Pi \right] + \mathcal E_{ab}\,,
\end{align}
with the coefficients $\mathcal A_{ab}$, $\mathcal B_{ab}$, $\mathcal C_{ab}$ and $\mathcal E_{ab}$ listed in Eq.~\eqref{eqvnrcoef} of Appendix~\ref{appcomb}. For $\eta_x\to0$ and $\eta_z\to0$, Eq.~\eqref{eqvnrclosed} reduces to the results of Case~II and Case~I, respectively. The single-tilt results of the VNR and the HSM are thus the limits $\eta_z\to0$ and $\eta_x\to0$ of Eqs.~\eqref{eqvnrclosed} and \eqref{eqhopfcomb}, where the former holds for $\mu\ge\tilde\eta$. The regime $\mu<\tilde\eta$ of the VNR is covered by Eq.~\eqref{eqvnrcombmu}.

 The trace identity holds for the same reason as in the HSM. The expansions,
\begin{align}
\label{eqvnrser}
& t^2_{xx} = \frac{1}{\pi} \left( -\,\frac{\eta_z}{64} - \frac{\eta_z^3}{128} 
+ \frac{7\,\eta_x^2\,\eta_z}{512} \right ) + \mathcal O(\eta^5)\,, \quad
t^2_{yy} = \frac{1}{\pi} \left ( -\,\frac{\eta_z}{64} - \frac{\eta_z^3}{128} 
- \frac{3\,\eta_x^2\,\eta_z}{512} \right) + \mathcal O(\eta^5)\,, \nn &
t^2_{zz} = \frac{1}{\pi} \left ( \frac{\eta_z}{32} + \frac{\eta_z^3}{64} - \frac{\eta_x^2\,\eta_z}{128} \right ) + \mathcal O(\eta^5)\,, \nn &
t^2_{xz} = \frac{1}{\pi} \left ( \frac{7\,\eta_x}{64} + \frac{5\,\eta_x^3}{512} + \frac{9\,\eta_x\,\eta_z^2}{128} \right ) + \mathcal O(\eta^5)\,, \quad
t^2_{zx} = \frac{1}{\pi} \left ( -\,\frac{\eta_x}{64} - \frac{3\,\eta_x^3}{512} 
+ \frac{\eta_x\,\eta_z^2}{128} \right) + \mathcal O(\eta^5)\,,
\end{align}
lead to the same structure as for the HSM. The response is additive to linear order and the interference terms are
\begin{align}
\label{eqvnrdelta}
& \Delta_{xx} = \frac{7\,\eta_x^2\,\eta_z}{512\,\pi}\,, \quad
\Delta_{yy} = -\,\frac{3\,\eta_x^2\,\eta_z}{512\,\pi}\,, \quad
\Delta_{zz} = -\,\frac{\eta_x^2\,\eta_z}{128\,\pi}\,, \quad
\Delta_{xz} = \frac{9\,\eta_x\,\eta_z^2}{128\,\pi}\,, \quad
\Delta_{zx} = \frac{\eta_x\,\eta_z^2}{128\,\pi}\,.
\end{align}
For $\mu<\tilde\eta$, the $\phi$-integration is restricted, as in the single-tilt cases. With $\phi_m\equiv\cos^{-1}\left[\min(1,\mu/\tilde\eta)\right]$, we obtain
\begin{align}
\label{eqvnrcombmu}
t^2_{ab} = \frac{1}{16\,\pi^3} \int_{\phi_m}^{2\,\pi-\phi_m} d\phi \;\mathscr S_{ab}(\phi) \,, \quad
t^1_{ab} = -\,\frac{1}{16\,\pi^3} \int_{-\phi_m}^{\phi_m} d\phi \;\mathscr S_{ab}(\phi) \,,
\end{align}
The band sum is then given by $t^1_{ab}+t^2_{ab}=t^2_{ab}\big|_{\mu\ge\tilde\eta}-\frac{1}{8\,\pi^3}\int_{-\phi_m}^{\phi_m} d\phi \;\mathscr S_{ab}(\phi)$. The trace vanishes for every $\mu$, since the $\phi$-restriction does not affect the identity $\int_0^{2\pi} d\gamma\,\sin\gamma=0$. In both systems, the condition $\eta<1$ keeps the effective tilt in each cross-section of the Fermi surface below unity.

For the PTNR, all the contributions are supported on the ring, where the energy is $\varepsilon_s=\tilde\eta\cos\phi$ irrespective of $\eta_z$. As in Case~I, only the tilt part $v_0\,\bs\eta$ of the velocity survives on the ring, and the BC is $\propto\hat{\bs\phi}$. Hence $t^s_{ab}$ is nonzero only for $b=y$ and $a\in\{x,z\}$, with
\begin{align}
\label{eqptnrcomb}
t^s_{ay} = (-1)^s\,\frac{\eta_a}{\eta_x}\,\frac{\mu}{4\,\pi^2\,\sqrt{\tilde\eta^2-\mu^2}}\;\Theta(\tilde\eta-\mu) \,, \quad a\in\{x,z\} \,.
\end{align}
The component $t^s_{xy}$ coincides with $-D^s_{xy}$ of Case~I, and $t^s_{zy}=(\eta_z/\eta_x)\,t^s_{xy}$ is new. It exists only if $\tilde\eta\ge\mu$, which explains why a pure axial tilt gives no response. Since $t^1_{ay}=-\,t^2_{ay}$, the band-summed ECR of the PTNR continues to vanish for the combined tilt.

Finally, the nonzero three-index coefficients follow from Eq.~\eqref{eqtheta}. For the VNR and the HSM, they are
\begin{align}
\label{eqthetacomb}
& \vartheta^{(s,1)}_{xyx} = -\,\vartheta^{(s,1)}_{yxx} = e^2\,\tau\,t^s_{xz}\,, \quad
\vartheta^{(s,1)}_{xyz} = -\,\vartheta^{(s,1)}_{yxz} = e^2\,\tau\,t^s_{zz}\,, \quad
\vartheta^{(s,1)}_{yzx} = -\,\vartheta^{(s,1)}_{zyx} = e^2\,\tau\,t^s_{xx}\,, \nn &
\vartheta^{(s,1)}_{yzz} = -\,\vartheta^{(s,1)}_{zyz} = e^2\,\tau\,t^s_{zx}\,, \quad
\vartheta^{(s,1)}_{zxy} = -\,\vartheta^{(s,1)}_{xzy} = e^2\,\tau\,t^s_{yy}\,, \nn &
\vartheta^{(s,2)}_{axy} = -\,\vartheta^{(s,2)}_{ayx} = -\,e^2\,\tau\,t^s_{az}\,, \quad
\vartheta^{(s,2)}_{ayz} = -\,\vartheta^{(s,2)}_{azy} = -\,e^2\,\tau\,t^s_{ax}\,, \quad a\in\{x,z\}\,, \nn &
\vartheta^{(s,2)}_{yzx} = -\,\vartheta^{(s,2)}_{yxz} = -\,e^2\,\tau\,t^s_{yy}\,.
\end{align}
The sets of nonzero components of Eqs.~\eqref{eqtheta1}, \eqref{eqtheta1b} and \eqref{eqtheta3} are thus active simultaneously. For the PTNR, the individual-band coefficients are $\vartheta^{(s,1)}_{zxc}=-\,\vartheta^{(s,1)}_{xzc}=e^2\,\tau\,t^s_{cy}$ and $\vartheta^{(s,2)}_{axz}=-\,\vartheta^{(s,2)}_{azx}=e^2\,\tau\,t^s_{ay}$ with $a,c\in\{x,z\}$, which cancel in the sum over $s$.

\section{Summary and outlook}
\label{secsum}

In this paper, we have examined how a dipole-textured BC, rather than the familiar monopolar texture $\bs\Omega_s \propto \hat{\bs k}$ of an isotropic Weyl node, shapes the transport response of three exotic semimetals: the VNR, the PTNR, and the three-band HSM. The VNR and the PTNR carry a toroidal Fermi surface, while the HSM carries a spherical one. Saliently, both the VNR and the HSM host anisotropic dipole-like BC and OMM profiles from their underlying nodal structure, the former spreading it continuously along the nodal-ring and the latter localising it at the isolated Hopf point [cf. Fig.~\ref{figbc}]. This shared feature, rather than any similarity in their dispersions, is what drives the qualitative parallels we find between the VNR and the HSM throughout our analysis of the longitudinal magnetoconductivity ($\sigma_{zz}$) and the ECR, and it is consistent with the anisotropic collective-mode and transport response already reported for the HSM elsewhere \cite{hopf-plasmon}.

For the linear response, obtained from an exact solution of the Boltzmann equation rather than the RTA, we found that $\sigma_{zz}$ for the VNR and the HSM contains odd-in-$B_z$ terms alongside the usual even-in-$ B_z $ ones [cf. Eq.~\eqref{eqsigmazzexact} and Fig.~\ref{figexact}]. We traced this feature to the absence of standalone time-reversal symmetry in these two models. Their vortex-like pseudospin winding around the nodal locus is compensated only by the weaker $\Xi$-covariant relation identified in Sec.~\ref{secmodels}, rather than by the ordinary Onsager--Casimir relation \cite{onsager1, onsager2, onsager3}. This makes the VNR and a single HSM node, or an HSM with imbalanced nodes, the natural candidates for observing odd-in-$B_z$ terms in the topology-induced magnetoconductance in experiments. Extending the exact treatment to a two-node HSM, with conjugate nodes carrying oppositely-oriented BC-dipole axes, sharpens this picture further [cf. Fig.~\ref{figexact2}]. The net $\delta \sigma_{zz}$ is the sum of two mirror-image single-node contributions, whose odd-in-$B_z$ terms cancel and leave a net response that is even in $B_z$ and, for the chosen parameters, much smaller than the individual contributions. The scattering ratio changes the linear terms of each node only slightly. No charge is pumped between the nodes, in agreement with the RTA result of Ref.~\cite{ips-firdous-vnr}, and the internode term acts as an ordinary additional scattering channel. As with the single node, the magnitude of the response falls off as $\mu$ is increased.

The ECR exhibits a related but distinct pattern. For the VNR and the HSM, the two tilt directions considered separately, $\eta_x\,\bs{\hat x}$ and $\eta_z\,\bs{\hat z}$, activate mutually exclusive sets of response components. Axial tilts and equatorial tilts probe non-overlapping entries of $\vartheta^{(s,1)}_{abc}$ and $\vartheta^{(s,2)}_{abc}$, with the surviving components fixed entirely by symmetry. A tilt along $\bs{\hat x}$ breaks the $C_{\infty z}$ rotation while leaving the mirror intact, allowing the off-diagonal BC-dipole (BCD) components to become nonzero. A tilt along $\bs{\hat z}$ does the reverse, preserving $C_{\infty z}$ while breaking the mirror, and permits only the diagonal components of $t^s_{ab}$ instead. The VNR and the HSM again track each other closely, sharing an identical mapping from the surviving BCD or diagonal terms onto the nonzero components of $\vartheta^{(s,1)}_{abc}$ and $\vartheta^{(s,2)}_{abc}$ [cf. Eqs.~\eqref{eqtheta1} and \eqref{eqtheta3}]. The PTNR departs from this pattern: its BC is confined entirely to the nodal-ring rather than spread throughout the Fermi sea, and its ECR vanishes identically for both tilt directions. An in-plane tilt does produce nonzero coefficients for the individual bands ($s=1$ and $s=2$), but their contributions are equal and opposite and cancel in the band-summed current. An axial tilt leaves the ring pinned at zero energy and yields nothing at all. This null result is the natural counterpart, on the ECR side, of the PTNR's retention of both $T$ and the mirror. In every case, the response vanishes smoothly as the tilt is switched off, confirming that it is the tilt-induced mixing of the anisotropic BC into the Fermi-surface average, and not the tilt by itself, that drives the effect. When both tilt directions are present simultaneously, the VNR and the HSM are described by closed-form expressions that contain the single-tilt results as limiting cases [cf. Eqs.~\eqref{eqhopfcomb} and \eqref{eqvnrclosed}]. The two directions act additively to linear order in the tilt. Their interference first appears at third order, through terms $\propto\eta_x^2\,\eta_z$ in the diagonal components and $\propto\eta_x\,\eta_z^2$ in the off-diagonal ones [cf. Eqs.~\eqref{eqhopfdelta} and \eqref{eqvnrdelta}], and it lifts the isotropy of the diagonal components in the $xy$-plane. For the PTNR, the individual-band coefficients acquire additional components, which still cancel in the band sum [cf. Eq.~\eqref{eqptnrcomb}].

Taken together, our results indicate that a spatially extended or dipolar BC texture, rather than a nonzero Chern number, is the key ingredient behind both the anomalous linear-in-$B_z$ magnetoconductance and the anisotropic ECR reported here. Several extensions follow naturally from this picture. The most immediate one is to relax the collinearity of $\bs E$ and $\bs B$. Doing so would activate the Hall and planar-Hall components of the conductivity tensor, which would prove to be sensitive experimental probes of an anisotropic Fermi surface and BC texture in tilted Weyl and Dirac semimetals \cite{burkov17_giant, ma19_planar, li18_giant}, and would be a natural transport signature to look for in the VNR and the HSM. A second direction is to restore the temperature gradient $\nabla_{\bs r} T$ that we set to zero in Sec.~\ref{secnonlinear}. The same Boltzmann-transport machinery would then furnish the thermal and thermoelectric analogues of the ECR, including a planar thermal-Hall response \cite{nandy_thermal_hall, ips-ruiz, ips-spin1-ph, ips-rsw-ph} and an anomalous Nernst signal \cite{watzman18_dirac}, both again controlled by the BC and the OMM of the underlying bands.
Next, it will be worthwhile to investigate how the regime of strong magnetic fields affects the transport features \cite{gusynin06_magneto, yadav23_magneto, dirac-qtm, deng2019_quantum, ips-sanskar-2dgnr}. Finally, it would be worth asking how the picture developed here is modified for a non-Hermitian dissipative realisation of a nodal-ring or a Hopf semimetal \cite{ips-hessen}, where various symmetries have been shown to protect exceptional points.

\section*{Data availability}

All data generated or analysed during this study are included within the article.

\appendix

\section{Limits of angular integrals for tilted nodal-rings}
\label{appcyclides}

Clearly, if $  \mu < \tilde \eta $, we have to restrict the upper (lower) limit of $\cos \phi $ to $  \mu / \tilde \eta$ for the $s=2$ ($s=1$) band. Overall, let us spell out the step-by-step strategy when we apply a chemical potential $\mu \geq 0 $: 
\begin{enumerate}
\item Implementing the Dirac-delta function, we get rid of the $\kappa $-integral and replace $\kappa $ by $ (\mu - \tilde \eta \cos \phi ) /(v_0\, \zeta_s) $ in the integrand. Say, the integral is now $I_s = \int_0^{2\pi} d\gamma \int_{-\pi}^{\pi} d \phi \,\mathcal{I}_s (\gamma, \phi)$.

\item Perform the $\gamma$-integral first to get the form $I_s =  \int_{-\pi}^{\pi} d \phi \,\tilde{\mathcal{I}_s } (\phi)$.

\item We divide up the integral as $I_s = I_1^{(s)} + I_2^{(s)} $, where $ I_1^{(s)} = 
\int_{-\pi}^{ - \phi_m } d \phi \, \tilde{\mathcal{I}_s} (\phi) $,
$ I_2^{(s)} =  \int_{\phi_m}^{\pi} d \phi \,\tilde{\mathcal{I}_s} (\phi)$, and $\phi_{m}= \max[0, \cos^{-1}(\mu/\tilde \eta ) ]$. If $\tilde{\mathcal{I}_s} (\phi)$ is even (odd) in $\phi$, then $I_s = 2\, I_2^{(s)}$ ($I_s =  0$).

\item Set $s= 2$ and change variables as $ u = \cos \phi $, which lead to $ I_2^{(2)} =  \int_{-1}^{0} 
\frac{d u}{\sqrt {1-u^2}} \,\tilde {\mathcal{I}_2} (\cos^{-1} u) + \int_{0}^{u_m} 
\frac{d u}{\sqrt {1-u^2}} \,\tilde {\mathcal{I}_2} (\cos^{-1} u) $, where $u_m$ is the upper limit determined by $ \min [1, {\mu} /{ \tilde \eta } \big]$.

\item If $ {\mu} < \tilde \eta  $, a section of the $s=1 $ band is cut by $\mu \geq 0$ as well, and the conductivity gets contributions both from $s=1$ and $s=2$. For the $s=1$ band, we need to employ a strategy similar to the $s=2$ band, using the appropriate limits for the $\phi$-integrals.

\end{enumerate}

\section{Intermediate steps for the combined in-plane and axial tilts}
\label{appcomb}

In this appendix, we derive the results of Sec.~\ref{seccombined} for $\bs\eta=\eta_x\,\hat{\bs x}+\eta_z\,\hat{\bs z}$ with $\eta=\sqrt{\eta_x^2+\eta_z^2}<1$ and $\mu>0$. We start with the HSM, whose angular structure is the simplest, and continue with the VNR and the PTNR.

\subsection{Three-band Hopf semimetal}

Let $\bs n\equiv\hat{\bs k}=(\sin\gamma\cos\phi,\,\sin\gamma\sin\phi,\,\cos\gamma)$. From Eqs.~\eqref{eqevhopf} and \eqref{eqbcommhopf}, the dispersive bands have $\varepsilon_s=v_0\,k\,[(-1)^s+\bs\eta\cdot\bs n]$, $\bs v^{(0,s)}=v_0\,[(-1)^s\,\bs n+\bs\eta]$ and $\bs\Omega_s=n_z\,\bs n/k^2$. Since $\eta<1$, the factor $(-1)^s+\bs\eta\cdot\bs n$ has the sign of $(-1)^s$. Hence the $s=1$ band lies below zero energy everywhere, and only the $s=2$ band has a Fermi surface at $\mu>0$, with $k_F=\mu/[v_0\,(1+\bs\eta\cdot\bs n)]$. Using $f_0^\prime=-\,\delta(\varepsilon_2-\mu)$ in Eq.~\eqref{eqtheta}, we have
\begin{align}
t^2_{ab} & = -\,\frac{1}{(2\,\pi)^3} \int d\Omega \int_0^\infty dk\,k^2\; v_0\,(n_a+\eta_a)\,\frac{n_z\,n_b}{k^2}\; \frac{\delta (k-k_F)}{v_0\,(1+\bs\eta\cdot\bs n)}\nn
& = -\,\frac{1}{8\,\pi^3} \int d\Omega\; \frac{(n_a+\eta_a)\,n_z\,n_b}{1+\bs\eta\cdot\bs n}\,.
\end{align}
The factor $k^2$ of the Jacobian cancels against the $1/k^2$ of the BC, which is why the result is independent of $\mu$. Setting $b=a$ and summing over $a$, we use $(\bs n+\bs\eta)\cdot\bs n=1+\bs\eta\cdot\bs n$ to obtain $\sum_a t^2_{aa}=-\frac{1}{8\pi^3}\int d\Omega\,n_z=0$.

To evaluate the integral for a general tilt, we expand the tensors in terms of $\hat{\bs\eta}=\bs\eta/\eta$ and use $c=\hat{\bs\eta}\cdot\bs n$ as the polar variable about the tilt axis. The moments
\begin{align}
P_j = 2\,\pi \int_{-1}^{1} \frac{c^j\,dc}{1+\eta\,c}
\end{align}
are obtained from the recursion $P_j=\frac{2\pi}{\eta}\left[\int_{-1}^1 c^{j-1}dc-\frac{P_{j-1}}{2\pi}\right]$, starting from $P_0=4\,\pi\,\tanh^{-1}\eta/\eta$. They read
\begin{align}
& P_1 = \frac{4\,\pi}{\eta} \left[ 1 - \frac{\tanh^{-1}\eta}{\eta} \right], \quad
P_2 = \frac{4\,\pi}{\eta^2} \left[ \frac{\tanh^{-1}\eta}{\eta} - 1 \right], \quad
P_3 = \frac{4\,\pi}{\eta} \left[ \frac13 + \frac{1}{\eta^2} - \frac{\tanh^{-1}\eta}{\eta^3} \right].
\end{align}
By covariance, the tensor moments of $\bs n$ take the forms
\begin{align}
& \int d\Omega\, \frac{n_a\,n_b}{1+\bs\eta\cdot\bs n} = \mathscr B_0\,\delta_{ab} + \mathscr B_2\,\hat\eta_a\,\hat\eta_b \,, \nn &
\int d\Omega\, \frac{n_a\,n_b\,n_c}{1+\bs\eta\cdot\bs n} = \mathscr C_1 \left( \delta_{ab}\,\hat\eta_c + \delta_{ac}\,\hat\eta_b + \delta_{bc}\,\hat\eta_a \right) + \mathscr C_3\,\hat\eta_a\,\hat\eta_b\,\hat\eta_c \,.
\end{align}
The coefficients are fixed by contracting with $\delta_{ab}$ and with $\hat\eta_a\hat\eta_b$ (and likewise for the rank-three tensor), which gives $\mathscr B_0=(P_0-P_2)/2$, $\mathscr B_2=(3\,P_2-P_0)/2$, $\mathscr C_1=(P_1-P_3)/2$ and $\mathscr C_3=(5\,P_3-3\,P_1)/2$. Explicitly,
\begin{align}
& \mathscr B_0 = \frac{2\,\pi \left[ \eta - (1-\eta^2)\,\tanh^{-1}\eta \right]}{\eta^3} \,, \quad
\mathscr B_2 = \frac{2\,\pi \left[ (3-\eta^2)\,\tanh^{-1}\eta - 3\,\eta \right]}{\eta^3} \,, \nn &
\mathscr C_1 = 2\,\pi\,\mathcal F_1 \,, \quad
\mathscr C_3 = \frac{2\,\pi \left[ 3\,(3\,\eta^2-5)\,\tanh^{-1}\eta + 15\,\eta - 4\,\eta^3 \right]}{3\,\eta^4}\,,
\end{align}
and the combinations that appear below satisfy $\mathscr C_1+\eta\,\mathscr B_0=2\,\pi\,\mathcal F_2$ and $\mathscr C_3+\eta\,\mathscr B_2=2\,\pi\,\mathcal F_3$, with $\mathcal F_{1,2,3}$ given in Eq.~\eqref{eqFhopf}. Inserting the tensor moments, with $\eta_a=\eta\,\hat\eta_a$, yields
\begin{align}
-\,8\,\pi^3\,t^2_{ab} = \mathscr C_1 \left( \delta_{az}\,\hat\eta_b + \delta_{ab}\,\hat\eta_z + \delta_{bz}\,\hat\eta_a \right) + \mathscr C_3\,\hat\eta_a\,\hat\eta_z\,\hat\eta_b + \eta\,\hat\eta_a \left( \mathscr B_0\,\delta_{bz} + \mathscr B_2\,\hat\eta_z\,\hat\eta_b \right) .
\end{align}
For $\hat\eta_y=0$, the components with $y$ and one other index vanish, and the five remaining components are
\begin{align}
& -\,8\,\pi^3\,t^2_{xx} = \hat\eta_z \left[ \mathscr C_1 + (\mathscr C_3+\eta\,\mathscr B_2)\,\hat\eta_x^2 \right] , \quad
-\,8\,\pi^3\,t^2_{yy} = \hat\eta_z\,\mathscr C_1 \,, \nn &
-\,8\,\pi^3\,t^2_{zz} = \hat\eta_z \left[ 3\,\mathscr C_1 + \eta\,\mathscr B_0 + (\mathscr C_3+\eta\,\mathscr B_2)\,\hat\eta_z^2 \right] , \nn &
-\,8\,\pi^3\,t^2_{xz} = \hat\eta_x \left[ \mathscr C_1 + \eta\,\mathscr B_0 + (\mathscr C_3+\eta\,\mathscr B_2)\,\hat\eta_z^2 \right] , \quad
-\,8\,\pi^3\,t^2_{zx} = \hat\eta_x \left[ \mathscr C_1 + (\mathscr C_3+\eta\,\mathscr B_2)\,\hat\eta_z^2 \right] .
\end{align}
Using the combinations quoted above, together with $3\,\mathscr C_1+\eta\,\mathscr B_0=2\,\pi\,(2\,\mathcal F_1+\mathcal F_2)$, we arrive at Eq.~\eqref{eqhopfcomb}. The relations in Eq.~\eqref{eqhopfrel} follow by inspection, with $\mathcal F_2-\mathcal F_1=[\eta-(1-\eta^2)\,\tanh^{-1}\eta]/\eta^2$.

For $\hat\eta_x=0$, we have $\eta=\eta_z$, and $t^2_{zz}=-\frac{1}{4\pi^2}\left(2\,\mathcal F_1+\mathcal F_2+\mathcal F_3\right)=\frac{\eta_z\,(2\,\eta_z^2-3)+3\,(1-\eta_z^2)\,\tanh^{-1}\eta_z}{6\,\pi^2\,\eta_z^4}$, which is the result of Case~II. For $\hat\eta_z=0$, we obtain $t^2_{xz}=-\mathcal F_2/(4\,\pi^2)$ and $t^2_{zx}=-\mathcal F_1/(4\,\pi^2)$, which reproduce Eq.~\eqref{eqhopftxz} and $D^2_{zx}=-\,t^2_{zx}$ of Case~I. The small-tilt expansions of Eq.~\eqref{eqhopfser} follow by expanding $\tanh^{-1}\eta$ in powers of $\eta$ in Eq.~\eqref{eqFhopf} up to third order in the tilt.

\subsection{Vortex nodal-ring}

In the toroidal coordinates of Eq.~\eqref{eqtrsf}, we define the unit vector $\bs\nu=(\cos\gamma\cos\phi,\,\cos\gamma\sin\phi,\,\sin\gamma)$. From Eqs.~\eqref{eqv} and \eqref{eqbcomm}, we have $\bs v^{(0,s)}=v_0\,[(-1)^s\,\bs\nu+\bs\eta]$ and $\bs\Omega_s=(-1)^{s+1}\,\sin\gamma\;\bs\nu/(2\,k_\perp\,\kappa)$, where $k_\perp=k_0+\kappa\cos\gamma$. The Jacobian $\kappa\,k_\perp$ cancels the factor $1/(k_\perp\,\kappa)$ of the BC exactly, which is again the reason for the independence of $\mu$. The energy is $\varepsilon_s-\mu=v_0\,\kappa\,\zeta_s+\tilde\eta\cos\phi-\mu$, with
\begin{align}
\zeta_s=(-1)^s+\bs\eta\cdot\bs\nu\,, \quad \bs\eta\cdot\bs\nu = \eta_x\cos\gamma\cos\phi+\eta_z\sin\gamma\,, \quad
\kappa_F = \frac{\mu-\tilde\eta\cos\phi}{v_0\,\zeta_s}\,.
\end{align}
For fixed $\phi$, the maximum of $\bs\eta\cdot\bs\nu$ over $\gamma$ is $R(\phi)=\sqrt{\eta_z^2+\eta_x^2\cos^2\phi}$, which is smaller than unity for all $\phi$ if and only if $\eta<1$. Hence $\zeta_2>0$ and $\zeta_1<0$, and the condition $\kappa_F>0$ reduces to $\cos\phi\le\mu/\tilde\eta$ for $s=2$ and to $\cos\phi>\mu/\tilde\eta$ for $s=1$, independently of $\gamma$. Performing the $\kappa$-integral with the $\delta$-function, we obtain
\begin{align}
t^s_{ab} = \frac{(-1)^s}{16\,\pi^3} \int d\phi\; \Theta(\kappa_F) \int_0^{2\,\pi} d\gamma\; \sin\gamma\; \frac{\left[(-1)^s\,\nu_a+\eta_a\right]\nu_b}{|\zeta_s|}\,.
\end{align}
For $s=2$, the $\gamma$-integral defines
\begin{align}
\label{eqSab}
\mathscr S_{ab}(\phi) \equiv \int_0^{2\,\pi} d\gamma\; \sin\gamma\; \frac{(\nu_a+\eta_a)\,\nu_b}{1+\bs\eta\cdot\bs\nu}\,.
\end{align}
For $s=1$, the substitution $\gamma\to\gamma+\pi$ at fixed $\phi$ reverses the sign of $\sin\gamma$ and of $\bs\nu$. It maps the $\gamma$-integral onto $\mathscr S_{ab}(\phi)$, and the prefactor $(-1)^s$ then gives $t^1_{ab}=-\frac{1}{16\pi^3}\int_{\cos\phi>\mu/\tilde\eta}d\phi\,\mathscr S_{ab}(\phi)$. This leads to Eqs.~\eqref{eqvnrcomb} and \eqref{eqvnrcombmu}. The trace is $\sum_a\mathscr S_{aa}=\int_0^{2\pi}d\gamma\,\sin\gamma=0$, because $(\bs\nu+\bs\eta)\cdot\bs\nu=1+\bs\eta\cdot\bs\nu$.

Let $a\equiv\eta_x\cos\phi$ and $b\equiv\eta_z$, so that $\bs\eta\cdot\bs\nu=a\cos\gamma+b\sin\gamma$. Using $\nu_x=\cos\gamma\cos\phi$, $\nu_y=\cos\gamma\sin\phi$ and $\nu_z=\sin\gamma$, the nonzero integrands are
\begin{align}
\label{eqSabexp}
& \mathscr S_{xx} = \cos^2\phi\;J_{21}+\eta_x\cos\phi\;J_{11} \,, \quad
\mathscr S_{yy} = \sin^2\phi\;J_{21} \,, \quad
\mathscr S_{zz} = J_{03}+\eta_z\,J_{02} \,, \quad
\mathscr S_{xz} = \cos\phi\;J_{12}+\eta_x\,J_{02} \,, \nn &
\mathscr S_{zx} = \cos\phi \left( J_{12}+\eta_z\,J_{11} \right) ,\quad
J_{pq}(a,b) = \int_0^{2\,\pi} d\gamma
\frac{\cos^p\gamma\,\sin^q\gamma}{1+a\cos\gamma+b\sin\gamma}\,.
\end{align}
The components $\mathscr S_{xy}$, $\mathscr S_{yx}$, $\mathscr S_{yz}$ and $\mathscr S_{zy}$ vanish because they are odd under $\phi\to-\phi$.

To evaluate $J_{pq}$, we write $a\cos\gamma+b\sin\gamma=R\cos\theta$ with $\theta=\gamma-\gamma_0$, $\cos\gamma_0=a/R$, $\sin\gamma_0=b/R$ and $R=\sqrt{a^2+b^2}$. Then $\cos\gamma$ and $\sin\gamma$ become linear combinations of $\cos\theta$ and $\sin\theta$, and monomials with odd powers of $\sin\theta$ integrate to zero. With $W\equiv\sqrt{1-R^2}$ and $M_m\equiv\int_0^{2\pi}d\theta\,\cos^m\theta/(1+R\cos\theta)$, we have the recursion $M_m=\frac1R\left[\int_0^{2\pi}d\theta\,\cos^{m-1}\theta-M_{m-1}\right]$, with
\begin{align}
M_0=\frac{2\,\pi}{W}\,, \quad M_1=-\,\frac{2\,\pi\,(1-W)}{R\,W}\,, \quad M_2=\frac{2\,\pi\,(1-W)}{R^2\,W}\,, \quad M_3=\frac{1}{R}\left[\pi-M_2\right]\,,
\end{align}
and $\sin^2\theta=1-\cos^2\theta$. Collecting the terms, and using $W^2=1-R^2$ to simplify, we find
\begin{align}
\label{eqJpq}
& J_{11} = \frac{2\,\pi\,a\,b\,(1-W)^2}{R^4\,W} \,, \quad
J_{02} = \frac{2\,\pi\,(1-W)\,(a^2\,W+b^2)}{R^4\,W} \,, \quad
J_{21} = \frac{\pi\,b\,(1-W)^2 \left[ a^2\,(W-2)-b^2\,W \right]}{R^6\,W} \,, \nn &
J_{03} = -\,\frac{\pi\,b\,(1-W)^2 \left[ 3\,a^2\,W+b^2\,(W+2) \right]}{R^6\,W} \,, \quad
J_{12} = \frac{\pi\,a\,(1-W)^2 \left[ b^2\,(W-2)-a^2\,W \right]}{R^6\,W}\,.
\end{align}
These are regular at $R\to0$, since $1-W=R^2/(1+W)$. Inserting them into Eq.~\eqref{eqSabexp}, with $a=\eta_x\cos\phi$ and $b=\eta_z$, gives the integrands $\mathscr S_{ab}(\phi)$ of Eq.~\eqref{eqvnrcomb}.

Two limits serve as checks. For $\eta_x=0$, we have $a=0$ and $R=\eta_z$, and hence $\mathscr S_{xx}=\cos^2\phi\;J_{21}$ and $\mathscr S_{yy}=\sin^2\phi\;J_{21}$ with $J_{21}\big|_{a=0}=-\,\pi\,(1-W)^2/\eta_z^3$. Since $\int_0^{2\pi}\cos^2\phi\,d\phi=\int_0^{2\pi}\sin^2\phi\,d\phi=\pi$ and $(1-W)^2=2-2\,W-\eta_z^2$, we obtain $t^2_{xx}=t^2_{yy}=\frac{\eta_z^2+2\,\sqrt{1-\eta_z^2}-2}{16\,\pi\,\eta_z^3}$, as in Case~II. For $\eta_z=0$, we have $b=0$ and therefore $J_{11}=J_{21}=J_{03}=0$, which means that all the diagonal components vanish. The remaining components $\mathscr S_{xz}$ and $\mathscr S_{zx}$ involve $R=\eta_x\,|\cos\phi|$ and $\sqrt{1-\eta_x^2\cos^2\phi}$, whose $\phi$-integrals produce the complete elliptic integrals $K(\eta_x^2)$ and $E(\eta_x^2)$ of Case~I. For a general tilt, the result is a single quadrature over $\phi$ of elementary functions of $R(\phi)$ and $W(\phi)$, where $W(\phi)=\sqrt{1-\eta_z^2}\,\sqrt{1-\varsigma^2\cos^2\phi}$ and $\varsigma^2=\eta_x^2/(1-\eta_z^2)$.

Let us now show how the $\phi$-integral in Eq.~\eqref{eqvnrcomb} is carried out for $\mu\ge\tilde\eta$. We exchange the order of integration in Eq.~\eqref{eqSab}, so that the $\phi$-integral is performed first at fixed $\gamma$. We define
\begin{align}
P(\gamma)=1+\eta_z\,\sin\gamma\,, \quad \mathcal W(\gamma)=\sqrt{P^2-\eta_x^2\,\cos^2\gamma}\,.
\end{align}
The quantity $\mathcal W^2$ is the product of the values of $1+\bs\eta\cdot\bs\nu$ at $\phi=0$ and $\phi=\pi$, and it is therefore positive for $\eta<1$. With $\bs\eta\cdot\bs\nu=\eta_x\cos\gamma\cos\phi+\eta_z\sin\gamma$, the elementary integrals
\begin{align}
\int_0^{2\,\pi}\frac{d\phi}{1+\bs\eta\cdot\bs\nu}=\frac{2\,\pi}{\mathcal W}\,, \quad
\int_0^{2\,\pi}\frac{\cos\phi\;d\phi}{1+\bs\eta\cdot\bs\nu}=\frac{2\,\pi}{\eta_x\cos\gamma}\left(1-\frac{P}{\mathcal W}\right), \quad
\int_0^{2\,\pi}\frac{\cos^2\phi\;d\phi}{1+\bs\eta\cdot\bs\nu}=\frac{2\,\pi\,P}{\eta_x^2\cos^2\gamma}\left(\frac{P}{\mathcal W}-1\right)
\end{align}
together with $\sin^2\phi=1-\cos^2\phi$ give, after the factors of $\cos\gamma$ from $\nu_x$ and $\nu_y$ have cancelled against the denominators,
\begin{align}
\label{eqvnrgamma}
& t^2_{xx}=\frac{1}{8\,\pi^2}\int_0^{2\,\pi}d\gamma\,\sin\gamma \left[\frac{P^2/\eta_x^2-P}{\mathcal W}+1-\frac{P}{\eta_x^2}\right], \quad
t^2_{yy}=\frac{1}{8\,\pi^2}\int_0^{2\,\pi}d\gamma\,\sin\gamma \left[\frac{1-\sin^2\gamma-P^2/\eta_x^2}{\mathcal W}+\frac{P}{\eta_x^2}\right], \nn &
t^2_{zz}=\frac{1}{8\,\pi^2}\int_0^{2\,\pi}d\gamma\,\frac{\sin^2\gamma\,(\sin\gamma+\eta_z)}{\mathcal W}\,, \quad
t^2_{xz}=\frac{1}{8\,\pi^2}\int_0^{2\,\pi}d\gamma\,\sin^2\gamma \left[\frac{\eta_x-P/\eta_x}{\mathcal W}+\frac{1}{\eta_x}\right], \nn &
t^2_{zx}=\frac{1}{8\,\pi^2\,\eta_x}\int_0^{2\,\pi}d\gamma\,\sin\gamma\,(\sin\gamma+\eta_z)\left(1-\frac{P}{\mathcal W}\right).
\end{align}
The terms without $\mathcal W$ are trigonometric polynomials. Using $\int_0^{2\pi}d\gamma\,\sin^2\gamma=\pi$ and the vanishing of the odd moments, they give the elementary contributions $\mathcal E_{ab}$ of Eq.~\eqref{eqvnrcoef}.

To evaluate the remaining integrals, we introduce the angle $\psi$ through $\sin\psi=(\sin\gamma+\eta_z)/P$, which is the analogue of a boost along $\hat{\bs z}$. Then $\cos\psi=\varrho\,\cos\gamma/P$, and we obtain the relations
\begin{align}
\sin\gamma=\frac{\sin\psi-\eta_z}{\lambda}\,, \quad P=\frac{\varrho^2}{\lambda}\,, \quad \mathcal W=\frac{\varrho^2\,\Delta}{\lambda}\,, \quad d\gamma=\frac{\varrho\;d\psi}{\lambda}\,,
\end{align}
with $\lambda(\psi)=1-\eta_z\sin\psi$ and $\Delta(\psi)=\sqrt{1-\varsigma^2\cos^2\psi}$. The parameter $\varsigma^2=\eta_x^2/\varrho^2$ arises from $P^2-\eta_x^2\cos^2\gamma=P^2\,(1-\varsigma^2\cos^2\psi)$. Hence $\int_0^{2\pi}d\gamma\,F/\mathcal W=\varrho^{-1}\int_0^{2\pi}d\psi\,F/\Delta$, and the integrand $F$ becomes a Laurent polynomial in $\lambda$ with powers from $\lambda^{-3}$ to $\lambda^0$. All the integrals are therefore combinations of
\begin{align}
\mathcal K_j=\int_0^{2\,\pi}d\psi\,\frac{\lambda^j}{\Delta}\,, \quad j=0,-1,-2,-3\,.
\end{align}
The first one is $\mathcal K_0=4\,K(\varsigma^2)$. For $j=-1$, the part of $1/\lambda$ that is odd in $\sin\psi$ integrates to zero, and the remainder is $1/(1-\eta_z^2\sin^2\psi)=\varrho^{-2}\left[1+(\eta_z^2/\varrho^2)\cos^2\psi\right]^{-1}$ in terms of $\cos\psi$, which gives $\mathcal K_{-1}=4\,\hat\Pi/\varrho^2$. The integrals with $j=-2$ and $j=-3$ follow from the total derivatives
\begin{align}
\int_0^{2\,\pi}d\psi\,\frac{d}{d\psi}\left[\cos\psi\;\Delta\;\lambda^m\right]=0\,,
\end{align}
which relate $\mathcal K_{m-1},\dots,\mathcal K_{m+3}$ for every integer $m$. Taking $m=-1$ and $m=-2$, and using $\mathcal K_1=\mathcal K_0$ and $\mathcal K_2=4\,K+4\,\eta_z^2\left[E-(1-\varsigma^2)\,K\right]/\varsigma^2$, we find
\begin{align}
& \mathcal K_{-2}=-\frac{4\,K}{\varrho^2}+\frac{4\,\eta_z^2\,E}{\eta^2\,\varrho^2}+\frac{4\,(2\,\eta_x^2+\eta_z^2)\,\hat\Pi}{\eta^2\,\varrho^4}\,, \nn &
\mathcal K_{-3}=-\frac{2\,(4\,\eta_x^2+3\,\eta_z^2)\,K}{\eta^2\,\varrho^4}+\frac{6\,\eta_z^2\,(2\,\eta_x^2+\eta_z^2)\,E}{\eta^4\,\varrho^4}+\frac{2\,\mathcal N\,\hat\Pi}{\eta^4\,\varrho^6}\,,
\end{align}
with $\mathcal N=2\,\eta_x^4\,\eta_z^2+6\,\eta_x^4+3\,\eta_x^2\,\eta_z^4+5\,\eta_x^2\,\eta_z^2+\eta_z^6+2\,\eta_z^4$, where $K=K(\varsigma^2)$ and $E=E(\varsigma^2)$. Inserting these into Eq.~\eqref{eqvnrgamma} yields Eq.~\eqref{eqvnrclosed} with the coefficients
\begin{align}
\label{eqvnrcoef}
& \mathcal A_{xx}=\frac{\varrho\left[2\,\eta_x^2\,(\eta^2-1)-\eta_z^2\right]}{\eta_x^2\,\eta_z\,\eta^2}\,, \quad
\mathcal B_{xx}=-\,\frac{\eta_z\,\varrho\left[2\,\eta_x^2\,(\eta^2-2)-\eta_z^2\right]}{\eta_x^2\,\eta^4}\,, \nn &
\mathcal C_{xx}=-\,\frac{2\,\eta_x^6-2\,\eta_x^4-3\,\eta_x^2\,\eta_z^4+\eta_x^2\,\eta_z^2-\eta_z^6}{\eta_x^2\,\eta_z\,\varrho\,\eta^4}\,, \quad
\mathcal E_{xx}=-\,\frac{\eta_z}{8\,\pi\,\eta_x^2}\,, \nn &
\mathcal A_{yy}=\frac{\varrho}{\eta_x^2\,\eta_z}\,, \quad
\mathcal B_{yy}=-\,\frac{\eta_z\,\varrho}{\eta_x^2\,\eta^2}\,, \quad
\mathcal C_{yy}=-\,\frac{\eta_x^2\,\eta_z^2+\eta_x^2+\eta_z^4}{\eta_x^2\,\eta_z\,\varrho\,\eta^2}\,, \quad
\mathcal E_{yy}=\frac{\eta_z}{8\,\pi\,\eta_x^2}\,, \nn &
\mathcal A_{zz}=-\,\frac{\varrho\,(2\,\eta^2-1)}{\eta_z\,\eta^2}\,, \quad
\mathcal B_{zz}=\frac{\eta_z\,\varrho\,(2\,\eta^2-3)}{\eta^4}\,, \quad
\mathcal C_{zz}=\frac{2\,\eta_x^4+\eta_x^2\,\eta_z^2-\eta_x^2-\eta_z^4+2\,\eta_z^2}{\eta_z\,\varrho\,\eta^4}\,, \quad
\mathcal E_{zz}=0\,, \nn &
\mathcal A_{xz}=\frac{2\,\eta_x^4+2\,\eta_x^2\,\eta_z^2+\eta_z^2-1}{\eta_x\,\varrho\,\eta^2}\,, \quad
\mathcal B_{xz}=\frac{\varrho\left[2\,\eta_x^2\,(\eta^2-1)+\eta_z^2\right]}{\eta_x\,\eta^4}\,, \quad
\mathcal C_{xz}=-\,\frac{4\,\eta_x^4+5\,\eta_x^2\,\eta_z^2-3\,\eta_x^2+\eta_z^4}{\eta_x\,\varrho\,\eta^4}\,, \quad
\mathcal E_{xz}=\frac{1}{8\,\pi\,\eta_x}\,, \nn &
\mathcal A_{zx}=\frac{\varrho\,(2\,\eta^2-1)}{\eta_x\,\eta^2}\,, \quad
\mathcal B_{zx}=-\,\frac{\varrho\left[2\,\eta_x^2\,\eta_z^2+2\,\eta_x^2+2\,\eta_z^4-\eta_z^2\right]}{\eta_x\,\eta^4}\,, \quad
\mathcal C_{zx}=\mathcal C_{xz}\,, \quad
\mathcal E_{zx}=\frac{1}{8\,\pi\,\eta_x}\,.
\end{align}
The coefficients contain inverse powers of $\eta_x$ and $\eta_z$, but these cancel between the elliptic integrals and the elementary terms, and each $t^2_{ab}$ is regular at $\eta_x\to0$ and $\eta_z\to0$. The trace identity $t^2_{xx}+t^2_{yy}+t^2_{zz}=0$ holds term by term in $K$, $E$, $\hat\Pi$ and the elementary part. For $\eta_x\to0$, we recover $t^2_{xx}=t^2_{yy}=\left(\eta_z^2+2\,\varrho-2\right)/(16\,\pi\,\eta_z^3)$ and $t^2_{zz}=-\,2\,t^2_{xx}$. For $\eta_z\to0$, we have $\hat\Pi\to K(\eta_x^2)$, and $t^2_{xz}$ and $t^2_{zx}$ reduce to the expressions with $K(\eta_x^2)$ and $E(\eta_x^2)$ of Case~I.

For the small-tilt expansion, we expand $1/(1+\bs\eta\cdot\bs\nu)=\sum_{n}(-\bs\eta\cdot\bs\nu)^n$ in Eq.~\eqref{eqSab} up to $n=3$. The remaining $\gamma$- and $\phi$-integrals involve only trigonometric monomials. Summing the contributions order by order in the tilt gives Eq.~\eqref{eqvnrser}, and $\sum_a t^2_{aa}=0$ is satisfied at each order.

\subsection{$\mathcal{PT}$-symmetric nodal-ring}

From Eq.~\eqref{eqbcpnr}, we have $(\Omega_s)_b=(-1)^{s+1}\,\pi\,\delta(k_z)\,\delta(\kappa_\perp)\,\hat\phi_b$, which is nonzero only on the ring. There, $\varepsilon_s=\tilde\eta\cos\phi$, since $k_z=0$ and $\kappa_\perp=0$, and the tilt part of the velocity is $v_0\,\eta_a$. As in Case~I, we retain only the latter. Hence
\begin{align}
t^s_{ab} = (-1)^{s}\,\pi\,v_0\,\eta_a\,\frac{k_0}{(2\,\pi)^3} \int_0^{2\,\pi} d\phi\; \hat\phi_b\;\delta (\tilde\eta\cos\phi-\mu) \,,
\end{align}
with $\hat\phi_b=(-\sin\phi,\cos\phi,0)$. Only $b=y$ survives the $\phi$-integration, with $\hat\phi_y=\cos\phi$. The $\delta$-function has support at $\phi=\pm\cos^{-1}(\mu/\tilde\eta)$ for $\mu\le\tilde\eta$, and each root contributes $\cos\phi/(\tilde\eta\,|\sin\phi|)=\mu/(\tilde\eta\,\sqrt{\tilde\eta^2-\mu^2})$. The two roots together give
\begin{align}
\int_0^{2\,\pi} d\phi\,\cos\phi\;\delta (\tilde\eta\cos\phi-\mu) = \frac{2\,\mu}{\tilde\eta\,\sqrt{\tilde\eta^2-\mu^2}}\;\Theta(\tilde\eta-\mu)\,.
\end{align}
Using $\tilde\eta=v_0\,k_0\,\eta_x$ then yields Eq.~\eqref{eqptnrcomb}. The $a=x$ component reproduces $D^s_{xy}=-\,t^s_{xy}$ of Case~I. The ratio $t^s_{zy}/t^s_{xy}=\eta_z/\eta_x$ is independent of $\mu$, and $t^1_{ay}=-\,t^2_{ay}$ holds because of the factor $(-1)^{s+1}$ of the BC.

\subsection{Three-index coefficients}

Equation~\eqref{eqtheta} reads $\vartheta^{(s,1)}_{abc}=e^2\,\tau\,\epsilon_{abd}\,t^s_{cd}$ and $\vartheta^{(s,2)}_{abc}=-\,e^2\,\tau\,\epsilon_{bcd}\,t^s_{ad}$. For the VNR and the HSM, only $t^s_{xx}$, $t^s_{yy}$, $t^s_{zz}$, $t^s_{xz}$ and $t^s_{zx}$ are nonzero, and the sum over the repeated index $d$ contains at most one term for each ordered triple $(a,b,c)$. Listing the triples for which this term is nonzero gives Eq.~\eqref{eqthetacomb}. For the PTNR, the nonzero entries are $t^s_{xy}$ and $t^s_{zy}$, which gives the coefficients quoted at the end of Sec.~\ref{seccombined}.

\section{\texorpdfstring{Singularity of the collision matrix}{Singularity of the collision matrix}}
\label{appsing}

This appendix discusses the singularity of the matrix $\mathcal A$ of Eqs.~\eqref{eqmatrix0}, \eqref{eqmatrix} and \eqref{eqmatrix-2node} for all the cases treated in Sec.~\ref{secexact}, with $\chi$ labelling the nodes whenever there are two of them. The underlying reason is that elastic scattering moves electrons between states of equal energy and neither creates nor destroys them. A uniform shift of the distribution over the Fermi surface therefore cannot relax. We show below that this constant mode is a null vector of $\mathcal A$.

Consider an electron on the node $\chi$ at the angle $\theta$ of the Fermi surface, where $\theta$ stands for the angle $\gamma$ of Sec.~\ref{secexact}. Electrons on the node $\tilde\chi$ at the angle $\theta^\prime$ can scatter into this state. Let $\mathcal W_{\tilde\chi}(\theta^\prime)$ be the weight of the angle $\theta^\prime$ on the node $\tilde\chi$, which contains the Jacobian, the phase-space factor and the inverse radial derivative of the energy, but not the relaxation time. Let $\mathcal T_{\chi\tilde\chi}(\theta,\theta^\prime)$ be the overlap function of the spinors at the two angles, and let $\beta_{\chi\tilde\chi}=\beta_{\tilde\chi\chi}$ be the scattering strength between the two nodes. The inverse relaxation time of the node $\chi$ is then the total out-scattering rate,
\begin{align}
\frac{1}{\tau_\chi(\theta)}=\sum_{\tilde\chi}\beta_{\chi\tilde\chi}\int d\theta^\prime\,\mathcal W_{\tilde\chi}(\theta^\prime)\,\mathcal T_{\chi\tilde\chi}(\theta,\theta^\prime)\,.
\label{eqsingrate}
\end{align}
Let $\Lambda_\chi(\theta)$ be the mean free path of the node $\chi$, and let $h_\chi(\theta)$ be the driving term, which is generated by the electric field and does not depend on $\Lambda_\chi$. Eq.~\eqref{eqlambdamu} then takes the form
\begin{align}
\frac{\Lambda_\chi(\theta)}{\tau_\chi(\theta)}+h_\chi(\theta)=\sum_{\tilde\chi}\beta_{\chi\tilde\chi}\int d\theta^\prime\,\mathcal W_{\tilde\chi}(\theta^\prime)\,\mathcal T_{\chi\tilde\chi}(\theta,\theta^\prime)\,\Lambda_{\tilde\chi}(\theta^\prime)\,.
\label{eqsingbolt}
\end{align}
The first term on the left is the out-scattering term, and the right-hand side is the in-scattering term. We now set $h_\chi=0$ and $\Lambda_{\tilde\chi}=1$ on every node. The right-hand side then equals the right-hand side of Eq.~\eqref{eqsingrate}, and thus $1/\tau_\chi(\theta)$, and the left-hand side is $1/\tau_\chi(\theta)$ as well. A constant therefore solves the homogeneous equation. This is why the electron-number constraint is needed, since the collision operator alone cannot fix the amplitude of this mode.

The overlap function of every channel can be written as a sum of products of cosines. Let $t_{nn}$ be the real coefficients of this sum, which do not depend on $\chi$ and $\tilde\chi$. Their values for each system follow from the respective overlap function in Sec.~\ref{secexact}. We then have
\begin{align}
\mathcal T_{\chi\tilde\chi}(\theta,\theta^\prime)=\sum_n t_{nn}\cos(n\,\theta)\cos(n\,\theta^\prime)\,.
\end{align}
Let $u^{\tilde\chi}_n=\int d\theta^\prime\,\mathcal W_{\tilde\chi}(\theta^\prime)\cos(n\,\theta^\prime)$ be the cosine moments of the weight without the relaxation time, and let $r^\chi_n$ be the coefficient of $\cos(n\,\theta)$ in $1/\tau_\chi$. Inserting the expansion of $\mathcal T_{\chi\tilde\chi}$ into Eq.~\eqref{eqsingrate} gives
\begin{align}
\frac{1}{\tau_\chi(\theta)}=\sum_n r^\chi_n\cos(n\,\theta)\,,\qquad r^\chi_n=t_{nn}\sum_{\tilde\chi}\beta_{\chi\tilde\chi}\,u^{\tilde\chi}_n\,.
\label{eqsingr}
\end{align}
Let $\alpha^\chi_n$ be the unknown coefficients of the ansatz, which are the coefficients $\lambda$, $a$ and $b$ of Sec.~\ref{secexact}. The ansatz reads $\Lambda_\chi=\tau_\chi\,[-h_\chi+\sum_n\alpha^\chi_n\cos(n\,\theta)]$. In the homogeneous problem the term with $h_\chi$ is absent, and therefore $\Lambda_\chi/\tau_\chi=\sum_n\alpha^\chi_n\cos(n\,\theta)$. For $\Lambda_\chi=1$ the left-hand side equals $1/\tau_\chi$, which is $\sum_n r^\chi_n\cos(n\,\theta)$ by Eq.~\eqref{eqsingr}. The constant mode thus corresponds to $\alpha^\chi_n=r^\chi_n$, and the question is whether $\mathcal A$ annihilates this vector.

Let $\mathcal Q^{\tilde\chi}_{nm}=\int d\theta^\prime\,\mathcal W_{\tilde\chi}(\theta^\prime)\,\tau_{\tilde\chi}(\theta^\prime)\cos(n\,\theta^\prime)\cos(m\,\theta^\prime)$ be the cosine moments of the weight including the relaxation time. Inserting the ansatz into Eq.~\eqref{eqsingbolt} and projecting on $\cos(n\,\theta)$ gives the linear system $\mathcal A\,\alpha=\mathcal H$ of the main text, with the source $\mathcal H_{\chi n}=-\,t_{nn}\sum_{\tilde\chi}\beta_{\chi\tilde\chi}\int d\theta^\prime\,\mathcal W_{\tilde\chi}\,\tau_{\tilde\chi}\cos(n\,\theta^\prime)\,h_{\tilde\chi}$. Let $\delta$ denote the Kronecker delta. The matrix elements of $\mathcal A$ are
\begin{align}
\mathcal A_{(\chi n),(\tilde\chi m)}=\delta_{\chi\tilde\chi}\,\delta_{nm}-\beta_{\chi\tilde\chi}\,t_{nn}\,\mathcal Q^{\tilde\chi}_{nm}\,.
\label{eqsingA}
\end{align}
Since $\sum_m r^{\tilde\chi}_m\cos(m\,\theta^\prime)=\tau_{\tilde\chi}^{-1}(\theta^\prime)$ by Eq.~\eqref{eqsingr}, the factor $\tau_{\tilde\chi}$ in $\mathcal Q^{\tilde\chi}_{nm}$ cancels, and we obtain
\begin{align}
\sum_m\mathcal Q^{\tilde\chi}_{nm}\,r^{\tilde\chi}_m=u^{\tilde\chi}_n\,.
\label{eqsingQr}
\end{align}
Let $r$ be the vector with components $r^{\tilde\chi}_m$. Applying $\mathcal A$ to it, and using Eq.~\eqref{eqsingQr} together with the expression of $r^\chi_n$ in Eq.~\eqref{eqsingr}, we find
\begin{align}
\sum_{\tilde\chi,\,m}\mathcal A_{(\chi n),(\tilde\chi m)}\,r^{\tilde\chi}_m=r^\chi_n-t_{nn}\sum_{\tilde\chi}\beta_{\chi\tilde\chi}\,u^{\tilde\chi}_n=0\,.
\label{eqsingnull}
\end{align}
The vector $r$ is therefore a null vector of $\mathcal A$. The identity holds for any weights $\mathcal W_\chi$, and hence for any $B_z$ and any ratio $\beta^{\rm inter}/\beta^{\rm intr}$.

The identity of Eq.~\eqref{eqsingnull} provides a test of any numerical implementation of $\mathcal A$ that does not rely on a reference matrix. The test can be implemented in the following steps. We first evaluate the moments $u^\chi_n$ of every node over the full range of $\theta$. The coefficients $r^\chi_n$ then follow from Eq.~\eqref{eqsingr} and define the inverse relaxation time, and the relaxation time is thereby built from exactly these numbers. Next, we evaluate the moments $\mathcal Q^\chi_{nm}$ with this relaxation time and assemble $\mathcal A$ from its matrix elements in Eq.~\eqref{eqsingA}. We then compute the vector $\mathcal A\,r$ and its relative norm $\lVert\mathcal A\,r\rVert/\lVert r\rVert$ together with its individual components. This residual should vanish up to the accuracy of the angular integrals, whereas a residual of the order of unity signals an inconsistency in the normalisation. Finally, the rank is obtained from the singular values of $\mathcal A$, and not from its determinant, which does not carry a scale. Singular values below a tolerance that is set by the integration accuracy are counted as zero, and their number should equal the deficit given below.

The zero mode is unique, which follows from the symmetry of $\mathcal W_\chi\,\mathcal M_{\chi\tilde\chi}\propto\mathcal W_\chi\,\mathcal W_{\tilde\chi}\,\beta_{\chi\tilde\chi}\,\mathcal T_{\chi\tilde\chi}$ under the exchange of $(\chi,\theta)$ and $(\tilde\chi,\theta^\prime)$. It gives
\begin{align}
\sum_\chi\int d\theta\,\mathcal W_\chi\,\Lambda_\chi\,\mathcal L\Lambda_\chi
= -\,\frac{1}{2}\sum_{\chi,\tilde\chi}\beta_{\chi\tilde\chi}\int d\theta\,d\theta^\prime\,\mathcal W_\chi(\theta)\,\mathcal W_{\tilde\chi}(\theta^\prime)\,\mathcal T_{\chi\tilde\chi}(\theta,\theta^\prime)\left[\Lambda_\chi(\theta)-\Lambda_{\tilde\chi}(\theta^\prime)\right]^2 \,,
\end{align}
which vanishes only if $\Lambda$ is constant on every set of nodes connected by a nonzero $\beta_{\chi\tilde\chi}$. The overlap functions are squared spinor overlaps, which are nonnegative and vanish only at isolated angles. For a single node, the kernel of $\mathcal L$ is thus spanned by the constant function. The same holds for a pair of nodes with $\beta^{\rm inter}\neq0$, since the internode channel connects them. The kernel of $\mathcal A$ is then one-dimensional, and the deficit is 1. For $\beta^{\rm inter}=0$, the two nodes decouple, and the constants on the two nodes can be chosen independently, so that the deficit is 2. The ranks are therefore $1$ out of $2$ for the VNR, $2$ out of $3$ for the single-node HSM, and $5$ out of $6$ or $4$ out of $6$ for the two-node HSM, with $\beta^{\rm inter}\neq0$ or $\beta^{\rm inter}=0$, respectively.

The number of null vectors equals the number of independent electron numbers that are conserved by the elastic collision operator, since each of them corresponds to one constant mode within the span of the ansatz. A single node conserves its own electron number, which gives a deficit of 1. For two nodes with $\beta^{\rm inter}=0$, the electron number at each node is separately conserved, which gives a deficit of 2. For $\beta^{\rm inter}\neq0$, only the total electron number is conserved, and the deficit is again 1. This count reproduces the ranks quoted above.

The vector $u$ with components $u^\chi_n$ is a left null vector of $\mathcal A$, which yields the solvability condition of the linear system. To see this, let $\mathcal P_\chi=\int d\theta\,\mathcal W_\chi\,h_\chi$ be the pumping rate of the node $\chi$. Since $\beta_{\chi\tilde\chi}=\beta_{\tilde\chi\chi}$ and $\mathcal Q^{\tilde\chi}_{nm}$ is symmetric in $n$ and $m$, Eqs.~\eqref{eqsingr} and \eqref{eqsingQr} give $\sum_{\chi,n}u^\chi_n\,\mathcal A_{(\chi n),(\tilde\chi m)}=u^{\tilde\chi}_m-\sum_n r^{\tilde\chi}_n\,\mathcal Q^{\tilde\chi}_{nm}=0$. The contraction of $\mathcal A\,\alpha=\mathcal H$ with $u$ therefore reads, with the help of Eq.~\eqref{eqsingr} for the source,
\begin{align}
0=\sum_{\chi,n}u^\chi_n\,\mathcal H_{\chi n}=-\sum_\chi\mathcal P_\chi\,.
\end{align}
The system is thus solvable only if the total pumping rate of every set of nodes connected by a nonzero $\beta_{\chi\tilde\chi}$ vanishes. This is the continuity equation, and its node-resolved form follows from Eq.~\eqref{eqlambdamu}.

Multiplying Eq.~\eqref{eqlambdamu} by $\mathcal W_\chi(\theta)$ and integrating over $\theta$, the intranode part of the collision integral drops out, and one obtains the node-resolved equation $\mathcal P_\chi=\beta^{\rm inter}\,\mathcal R_\chi$ between the pumping rate $\mathcal P_\chi=\int d\theta\,\mathcal W_\chi\,h_\chi$ and the internode transfer rate $\mathcal R_\chi$, derived in Appendix~\ref{appcont}. Summed over the nodes, it reproduces the solvability condition above. For a BC-monopole, $\mathcal P_\chi\neq0$ for each node, and the node-resolved equation relates the transfer rate to the pumping, thereby constraining the even harmonics of $\Lambda^z$. For the HSM, both rates vanish by parity, as shown in the two HSM subsections of Sec.~\ref{secexact}, and the node-resolved equation reduces to $0=0$. It gives neither a condition on the coefficients nor a relation between the nodes, and the amplitude of the zero mode is fixed solely by the electron-number conservation constraint. Since this amplitude multiplies a constant function, it vanishes for the odd source of the HSM. For the VNR, the mirror symmetry $k_z\to-k_z$ makes $\mathcal W$, $\tau_2$ and the harmonics $1$ and $\cos\gamma$ even under $\gamma\to-\gamma$, whereas $h_2$ is odd. Both components of the source in Eq.~\eqref{eqmatrix0} therefore vanish, and the number constraint then sets $\lambda=a=0$, which leaves $\Lambda^z_2=-\,\tau_2\,h_2$ with the anisotropic $\tau_2(\mu,\gamma)$.

\section{\texorpdfstring{Continuity equation and internode transfer rate}{Continuity equation and internode transfer rate}}
\label{appcont}

In this appendix, we derive the node-resolved continuity equation used in Sec.~\ref{secexact}, for a pair of nodes labelled by $\chi=\pm$, and we evaluate the internode transfer rate for a pair of BC-monopoles and for the HSM. We write the angular variable generically as $\theta$, with the understanding that it is $\gamma$ for the HSM. Let $\mathcal W_\chi(\theta)$ denote the weight of Eq.~\eqref{eqlambdamu} at the Fermi surface of the node $\chi$, and let $\beta_{\chi\tilde\chi}$ be the intranode and internode strengths introduced above. Eq.~\eqref{eqlambdamu} then takes the form
\begin{align}
\label{eqappcont1}
h_\chi(\theta) + \frac{\Lambda^z_\chi(\theta)}{\tau_\chi(\theta)}
= \sum_{\tilde\chi}\beta_{\chi\tilde\chi}\int d\theta^\prime\,\mathcal W_{\tilde\chi}(\theta^\prime)\,\mathcal T_{\chi\tilde\chi}(\theta,\theta^\prime)\,\Lambda^z_{\tilde\chi}(\theta^\prime) \,, \quad
\frac{1}{\tau_\chi(\theta)}
= \sum_{\tilde\chi}\beta_{\chi\tilde\chi}\int d\theta^\prime\,\mathcal W_{\tilde\chi}(\theta^\prime)\,\mathcal T_{\chi\tilde\chi}(\theta,\theta^\prime)\,.
\end{align}
We multiply the first of these equations by $\mathcal W_\chi(\theta)$ and integrate over $\theta$. Inserting the second equation in the term containing $\Lambda^z_\chi/\tau_\chi$, we obtain
\begin{align}
\label{eqappcont2}
\mathcal P_\chi \equiv \int d\theta\,\mathcal W_\chi\,h_\chi
= \sum_{\tilde\chi}\beta_{\chi\tilde\chi}\int d\theta\,d\theta^\prime\,\mathcal W_\chi(\theta)\,\mathcal W_{\tilde\chi}(\theta^\prime)\,\mathcal T_{\chi\tilde\chi}(\theta,\theta^\prime)\left[\Lambda^z_{\tilde\chi}(\theta^\prime)-\Lambda^z_\chi(\theta)\right] .
\end{align}
The overlap function of the intranode channel is symmetric under $\theta\leftrightarrow\theta^\prime$, and the weights of the two arguments are the same. The $\tilde\chi=\chi$ term is therefore odd under the exchange of $\theta$ and $\theta^\prime$ and vanishes. What remains is the continuity equation of the node $\chi$,
\begin{align}
\label{eqappcont3}
\mathcal P_\chi = \beta^{\rm inter}\,\mathcal R_\chi \,, \quad
\mathcal R_\chi \equiv \int d\theta\,d\theta^\prime\,\mathcal W_\chi(\theta)\,\mathcal W_{-\chi}(\theta^\prime)\,\mathcal T_{\chi,-\chi}(\theta,\theta^\prime)\left[\Lambda^z_{-\chi}(\theta^\prime)-\Lambda^z_\chi(\theta)\right] ,
\end{align}
where $\mathcal R_\chi$ is the internode transfer rate. Since $\mathcal T_{\chi,-\chi}(\theta,\theta^\prime)=\mathcal T_{-\chi,\chi}(\theta^\prime,\theta)$, we have $\mathcal R_++\mathcal R_-=0$, and the sum of Eq.~\eqref{eqappcont3} over the two nodes states that the total pumping rate vanishes.

To make $\mathcal R_\chi$ explicit, we expand the internode overlap function as $\mathcal T_{\chi,-\chi}(\theta,\theta^\prime)=\sum_{n,m}t_{nm}\cos^n\theta\,\cos^m\theta^\prime$, with $t_{nm}=t_{mn}$, and define the moments
\begin{align}
N^\chi_n=\int d\theta\,\mathcal W_\chi\,\cos^n\theta \,, \quad
L^\chi_n=\int d\theta\,\mathcal W_\chi\,\cos^n\theta\;\Lambda^z_\chi \,.
\end{align}
Then
\begin{align}
\label{eqappcont4}
\mathcal R_\chi = \sum_{n,m}t_{nm}\left[N^\chi_n\,L^{-\chi}_m - L^\chi_n\,N^{-\chi}_m\right] .
\end{align}

For a pair of BC-monopoles \cite{timm}, $\mathcal T_{\chi,-\chi}=(1-\cos\theta\,\cos\theta^\prime)/2$, so that $t_{00}=1/2$, $t_{11}=-1/2$ and
\begin{align}
\mathcal R_\chi = \frac{1}{2}\left(N^\chi_0\,L^{-\chi}_0 - N^{-\chi}_0\,L^\chi_0\right)
-\frac{1}{2}\left ( N^\chi_1\,L^{-\chi}_1 - N^{-\chi}_1\,L^\chi_1\right ).
\end{align}
For $B_z=0$, the weights are even under $\theta\to\pi-\theta$, so that $N^\chi_1=0$, and $N^+_0=N^-_0\equiv N_0$. In this limit, $\mathcal R_\chi=N_0\left(L^{-\chi}_0-L^\chi_0\right)/2$ is proportional to the difference of the zeroth moments of $\Lambda^z$, which involve only its even part. At nonzero $B_z$, the OMM-induced distortion of the Fermi surface gives the weights an odd part, with $N^\chi_1\neq0$, and the first moments, which involve the odd part of $\Lambda^z$, also contribute. The pumping rate $\mathcal P_\chi$ is nonzero, since $h_\chi$ has an even part. Eq.~\eqref{eqappcont3} then relates the difference of the moments of $\Lambda^z$ on the two nodes to the pumping rate, and this constrains the coefficients of the even sector.

For the HSM, $\mathcal T_{2,2}$ of Eq.~\eqref{eqoverlap-bcd-2node} has $t_{00}=3/8$, $t_{11}=1/2$ and $t_{22}=1/8$ in the basis $\{1,\cos\gamma,\cos(2\gamma)\}$, which gives
\begin{align}
\mathcal R_\chi = \sum_{n=0}^{2}c_n\left ( N^\chi_n\,L^{-\chi}_n - N^{-\chi}_n\,L^\chi_n \right ) , \quad
c_0=\frac{3}{8}\,, \quad c_1=\frac{1}{2}\,, \quad c_2=\frac{1}{8}\,,
\end{align}
where $N_n$ and $L_n$ are now moments in $\cos(n\gamma)$. The weights $\mathcal W_\chi$ are even under $\gamma\to\pi-\gamma$ at every order in $B_z$, so that $N^\chi_1=0$. The term with $n=1$ drops out, and only the zeroth and second moments of $\Lambda^z$, built from $\lambda_\chi$ and $b_\chi$, enter. Since $\mathcal P_\chi=0$ by parity, Eq.~\eqref{eqappcont3} is a homogeneous equation for these coefficients, as stated in Sec.~\ref{secexact}.

 
\bibliography{ref_nl}


\end{document}